\documentclass{aa}  

\usepackage[version=3]{mhchem}   
\usepackage[nottoc, notlof, notlot]{tocbibind}   
\usepackage{enumitem}
\usepackage{graphicx}
\usepackage{color}
\usepackage{colortbl}
\usepackage{txfonts}
\begin{document}

   \title{Exploring molecular complexity with ALMA (EMoCA): \\ Detection of three new hot cores in Sagittarius~B2(N)}


  \author{M. Bonfand\inst{1}, A. Belloche\inst{1}, K. M. Menten\inst{1}, R.~T. Garrod\inst{2}, H.~S.~P. M\"{u}ller\inst{3}
          }
  \authorrunning{M. Bonfand et al.}
  \titlerunning{Detection of three new hot cores in Sgr~B2(N)}

   \institute{\inst{1} Max-Planck Institut f\"{u}r Radioastronomie, Auf dem H\"{u}gel 69, Bonn, Germany \\
              \inst{2} Departments of Chemistry and Astronomy, University of Virginia, Charlottesville, VA 22904, USA \\
              \inst{3} I. Physikalishes Institut, Universit\"{a}t zu K\"{o}ln, Z\"{u}lpicher Str. 77, 50937 K\"{o}ln, Germany
             }

   \date{}

 
  \abstract
   {The Sagittarius~B2 molecular cloud contains several sites forming high-mass stars. Sgr~B2(N) is one of its main centers of activity. It hosts several compact and ultra-compact HII regions, as well as two known hot molecular cores (Sgr~B2(N1) and Sgr~B2(N2)) in the early stage of the high-mass star formation process, where complex organic molecules (COMs) are detected in the gas phase.}
   {Our goal is to use the high sensitivity of the Atacama Large Millimeter/submillimeter Array (ALMA) to characterize the hot core population in Sgr~B2(N) and thereby shed a new light on the star formation process in this star-forming region.}
   {We use a complete 3~mm spectral line survey conducted with ALMA to search for faint hot cores in the Sgr~B2(N) region. The chemical composition of the detected sources and the column densities are derived by modelling the whole spectra under the assumption of local thermodynamic equilibrium. Population diagrams are constructed to fit rotational temperatures. Integrated intensity maps are produced to derive the peak position and fit the size of each molecule's emission distribution. The kinematic structure of the hot cores is investigated by analyzing the line wing emission of typical outflow tracers. The H$_2$ column densities are computed from ALMA and SMA continuum emission maps.}
   {We report the discovery of three new hot cores in Sgr~B2(N) that we call Sgr~B2(N3), Sgr~B2(N4), and Sgr~B2(N5). The three sources are associated with class II methanol masers, well known tracers of high-mass star formation, and Sgr~B2(N5) also with a UCHII region. Their H$_2$ column densities are found to be $\sim$16 up to 36 times lower than the one of the main hot core Sgr~B2(N1). The spectra of these new hot cores have spectral line densities of 11 up to 31 emission lines per GHz above the 7$\sigma$ level, assigned to 22--25 molecules plus 13--20 less abundant isotopologs. We derive rotational temperatures around 140--180~K for the three new hot cores and mean source sizes of 0.4$\arcsec$ for Sgr~B2(N3) and 1.0$\arcsec$ for Sgr~B2(N4) and Sgr~B2(N5). The chemical composition of Sgr~B2(N3), Sgr~B2(N4), and Sgr~B2(N5) is very similar, but it differs from that of Sgr~B2(N2). Finally, Sgr~B2(N3) and Sgr~B2(N5) show high velocity wing emission in typical outflow tracers, with a bipolar morphology in their integrated intensity maps suggesting the presence of an outflow, like in Sgr~B2(N1). No sign of an outflow is found around Sgr~B2(N2) and Sgr~B2(N4). We derive statistical lifetimes of $4 \times 10^4$ yr for the class II methanol maser phase and 6 $\times 10^4$ yr for the hot core phase in Sgr~B2(N).}
   {The associations of the hot cores with class II methanol masers, outflows, and/or UCHII regions tentatively suggest the following age sequence: Sgr~B2(N4), Sgr~B2(N3), SgrB2(N5), Sgr~B2(N1). The status of Sgr~B2(N2) is unclear. It may contain two distinct sources, a UCHII region and a very young hot core.}

   \keywords{stars: formation -- ISM: individual objects: Sagittarius B2(N) -- astrochemistry -- ISM: molecules              }

   \maketitle

\newcommand{\Msol}{\ensuremath{M_{\odot}}}

\section{Introduction}
        \label{intro}

 The Sagittarius B2 molecular cloud (Sgr~B2 hereafter) is one of the most prominent regions forming high-mass stars in our Galaxy, with a mass of $\sim$10$^7 \Msol$ in a diameter of $\sim$40~pc \citep{lis1990}. It is located in the Central Molecular Zone, close to the Galactic center ($\sim$100~pc from the central supermassive black hole Sgr~A$^*$ in projection) and at a distance of 8.34~$\pm$~0.16~kpc from the Sun \citep{reid2014}. The cloud contains several sites of on-going high-mass star formation. Its two major centers of activity are called Sgr~B2(M)ain and Sgr~B2(N)orth. They both host a cluster of (ultra)compact HII regions \citep{mehringer1993, gaume1995, depree1998, depree2015} as well as class II methanol masers \citep{caswell1996}; both phenomena provide strong evidence of on-going high-mass star formation. The high density of molecular lines and the continuum emission detected toward this region highlight the presence of a large amount of material to form new stars. Sgr~B2(N) contains two dense and compact hot molecular cores, called Sgr~B2(N1) and Sgr~B2(N2) \citep{belloche2016}, separated by $\sim$5$\arcsec$ in the North-South direction \citep[corresponding to 0.2 pc in projection,][]{belloche2008, qin2011}. They are both in the early stage of star formation when a high-mass protostar has already formed and started to warm up its circumstellar envelope, and is producing ionizing radiation that creates an ultracompact HII region around it. The earliest stages of the formation process of high-mass stars are still poorly understood since, first, the high dust content of high-mass protostellar cores makes observations at infrared and shorter wavelengths impossible and, second, because of the small angular sizes of these faraway regions. However, it is in these stages that many molecules are formed, directly in the gas phase or at the surface of dust grains. These molecules can readily be studied by their rotational spectrum and now, with the Atacama Large Millimeter/submillimeter Array (ALMA), at exquisite angular resolution and sensitivity. While the newly ignited protostar increases the temperature of its surroundings, the molecules frozen in the ice mantles of the dust grains are released in the gas phase, explaining the large number of molecules detected toward Sgr~B2.

Over the past five decades, nearly 200 molecules have been discovered in the interstellar medium (ISM) or in circumstellar envelopes of evolved stars (see, e.g., http://www.astro.uni-koeln.de/cdms/molecules). Among them, about 60 are composed of six atoms or more and are called Complex Organic Molecules (COMs) in the field of astrochemistry \citep{herbst2009}. Most species have been discovered toward the warm and dense parts of star forming regions and many of the first detections of interstellar molecules at radio and (sub)millimeter wavelengths were made toward Sgr~B2, such as acetic acid \ce{CH3COOH} \citep{mehringer1997}, glycolaldehyde \ce{CH2(OH)CHO} \citep{hollis2000}, acetamide \ce{CH3CONH2} \citep{hollis2006}, and aminoacetonitrile \ce{NH2CH2CN} \citep{belloche2008}; see also the overview by \citet{Menten2004}. Sgr~B2(N) thus appears to be one of the best targets for studying COMs and searching for new molecules, and thus expanding our view of the chemical complexity of the ISM. 

Through the investigation of the chemical composition of Sgr~B2(N), we wish to characterize its hot core population to shed a new light on the star formation process in this region. To this aim we analyze a spectral line survey recently conducted with ALMA in its cycles 0 and 1 at high angular resolution. We take advantage of the high sensitivity of the EMoCA survey (standing for "Exploring Molecular Complexity with ALMA") to search for fainter hot cores in Sgr~B2(N) in order to extend our view of the distribution of active star forming regions in this outstanding cloud. The article is structured as follows. The observations and method of analysis are presented in Sect.~2. The results are given in Sect.~3 and discussed in Sect.~4. Finally the conclusions are presented in Sect.~5.

\section{Observations and method of analysis}
         \label{obs}

        \subsection{Atacama Large Millimiter/submillimeter Array (ALMA)}
                    \label{obs-ALMA}

The ALMA observations that provided the data used here target Sgr~B2(N) with the phase center located half way between Sgr~B2(N1) and Sgr~B2(N2), at $\alpha _{\rm J2000}$  = 17$^h$47$^m$19.87$^s$, $\delta _{\rm J2000}$ = -28$^{\rm o}$22$'$16$\arcsec$. It is a complete spectral line survey between 84.1~GHz and 114.4~GHz, conducted at high angular resolution ($\sim$1.6$\arcsec$) and with a high sensitivity $\sim$3~mJy/beam per 488~kHz (1.7 to 1.3~km~s$^{-1}$) wide channel. The spectral line survey is divided into five spectral setups, each one delivering four spectral windows. Details about the different setups and the data reduction are presented in \citet{belloche2016}. The size (HPBW) of the primary beam of the 12~m antennas varies between 69$\arcsec$ at 84~GHz and 51$\arcsec$ at 114~GHz \citep{remijan2015}.

        \subsection{Submillimeter Array (SMA)}
                    \label{obs-SMA}

\citet{qin2011} observed the Sgr~B2 region using the SMA in the compact and very extended configurations, reporting the first high-angular-resolution submillimeter continuum observations of this region. The continuum map used here was obtained at 342.883~GHz, with a synthesized beam of 0.4$\arcsec$ $\times$ 0.24$\arcsec$ and a position angle of 14.4$^{\rm o}$. The map has been corrected for the primary beam attenuation. More details can be found in \citet{qin2011}.

         \subsection{Radiative transfer modelling of the line survey}
                     \label{modelling}

Given the high densities observed in the Sgr~B2(N) region \citep[][]{belloche2008, bellochescience, qin2011}, it is appropriate to work under the local thermodynamic equilibrium (LTE) approximation. We use Weeds \citep{maret2011}, which is part of the CLASS software\footnote{See http://www.iram.fr/IRAMFR/GILDAS.}, to perform the line identification and modelling of the spectra, after correction for the primary beam attenuation. A synthetic spectrum is produced for each species solving the radiative transfer equation and taking into account the finite angular resolution of the interferometer, the line opacity, and line blending. Each molecule is modelled separately adjusting the following parameters: column density, rotational temperature, angular size of the emitting region (assumed to be Gaussian), velocity offset with respect to the systemic velocity of the source, and linewidth (FWHM).
For each species, population diagrams are plotted to derive the rotational temperature and 2D Gaussians are fitted to integrated intensity maps to measure the size of the emitting region. The linewidth and velocity offset are derived from 1D-Gaussian fits to the lines detected in the spectra. All parameters are then adjusted manually until a good fit to the data is obtained. Finally, contributions from all species are added to obtain the complete synthetic spectrum. Through this modelling process, a line is assigned to a given molecule only if all lines from this molecule emitted in the frequency range of the survey are detected with the right intensity predicted by the model and if no line is missing in the observed spectrum. More information about the modelling procedure can be found in \citet{belloche2016}. The spectroscopic predictions used to model the spectra are the same as in \citet{belloche2016, belloche2017}, and \citet{muller2016}. They originate mainly from the CDMS and JPL catalogs \citep{endres2016, pearson2010}.

\section{Results}
        \label{results}

        \subsection{Detection of three new sources}
                    \label{search-new-hot-cores}

\begin{figure}[!t]
       \begin{center}
            \includegraphics[width=\hsize]{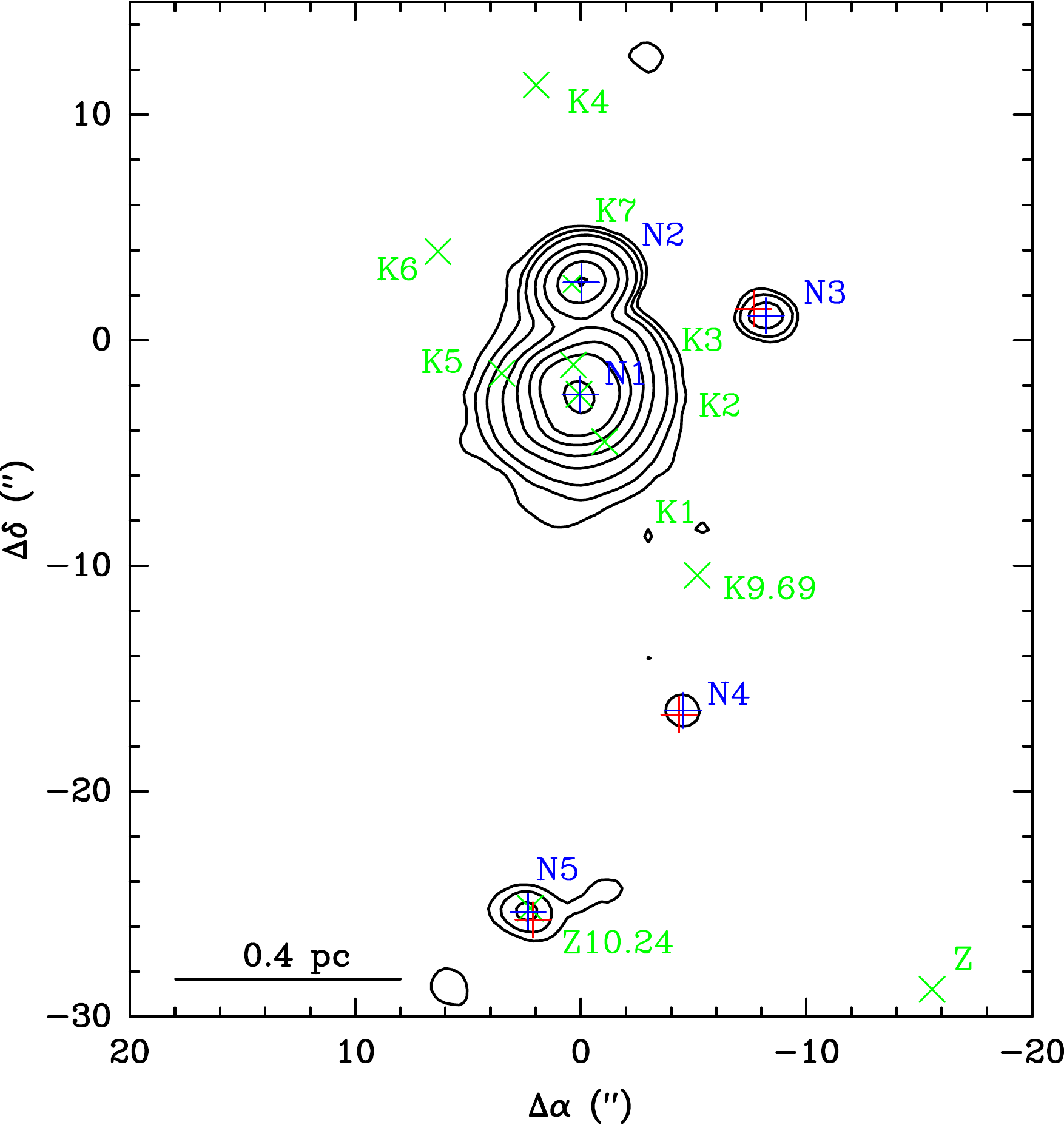} 
             \caption{\label{hot_cores_contour_map} Contour map of the number of channels with continuum-subtracted flux density above the 7$\sigma$ level (1$\sigma$ $\sim$3 mJy/beam). The contour levels are: 500, 1000, 2000, 5000, 10000, 20000, 30000, and 40000. The blue crosses mark the peaks of spectral line density. Red crosses represent the 6.7~GHz class II methanol masers \citep{caswell1996}. Green crosses represent the compact and ultra-compact HII regions \citep{gaume1995, depree2015}. The offsets are defined with respect to the phase center (see Sect.~\ref{obs-ALMA}).}
       \end{center}
       \end{figure}

 The high sensitivity of our ALMA data set allows us to search for fainter hot cores in the vicinity of Sgr~B2(N1) and Sgr~B2(N2). To this aim, we counted the channels with continuum-subtracted flux densities above the 7$\sigma$ level (1$\sigma$~$\sim$3~mJy/beam) over the whole frequency range for each pixel in the field of view. In this count we excluded setup 3 which has the lowest angular resolution \citep[HPBW $>$~2$\arcsec$, see Table~2 in][]{belloche2016} and we counted only once the channels located in frequency ranges overlapped by adjacent spectral windows. The analyzed ranges cover 47296 channels, $i.e.$ 23.1~GHz. The result is presented in Fig.~\ref{hot_cores_contour_map} as a contour map of the Sgr~B2(N) region. On this map, the contours represent the number of channels with flux density above the 7$\sigma$ threshold, $i.e$ higher contours reflect the presence of more emission lines. The figure clearly shows the two main hot cores Sgr~B2(N1) and Sgr~B2(N2). It also reveals high spectral line densities toward three other positions, unveiling the presence of three new sources that we call Sgr~B2(N3), Sgr~B2(N4), and Sgr~B2(N5). The region inside the contour located South-East of Sgr~B2(N5) shows emission lines only in setups 2 and 5 (4 spectral windows) with line intensities lower than toward Sgr~B2(N5) but a line content that is very similar. We think this structure is a deconvolution artifact.

We fit 2D Gaussians to the channel count map with the GILDAS procedure GAUSS-2D in order to derive the peak position of the five sources. The results of the fits are listed in Table~\ref{result-fit-channel-map} and the position of the spectral line density peaks of each source is marked with a blue cross in Fig.~\ref{hot_cores_contour_map}. These positions are adopted as reference positions of the hot cores. Known HII regions \citep{gaume1995, depree2015} and class II methanol masers \citep{caswell1996} are shown in Fig.~\ref{hot_cores_contour_map} as green and red crosses, respectively. The distance of each hot core to the closest UCHII region and the distance to the closest class II methanol maser are given in Table~\ref{result-fit-channel-map}. Sgr~B2(N1), Sgr~B2(N2), and Sgr~B2(N5) are associated with UCHII regions. All the new hot cores, Sgr~B2(N3), Sgr~B2(N4), and Sgr~B2(N5) happen to be associated with class II methanol masers.

\begin{table*}[!t]
\begin{center}
  \caption{\label{result-fit-channel-map} Position and spectral line density of the hot cores embedded in Sgr~B2(N), and distance to UCHII regions and class II methanol masers.} 
  \setlength{\tabcolsep}{1.2mm}
  \begin{tabular}{cccrrccc}
    \hline
Source &  $\Delta \alpha$ ; $\Delta \delta$ \tablefootmark{a}     & $\alpha _{\rm J2000}$ ; $\delta _{\rm J2000}$  \tablefootmark{b}  & \multicolumn{1}{c}{$N_{\rm channels}$ \tablefootmark{c}}  & \multicolumn{1}{c}{$n_{\rm l}$ \tablefootmark{d}}  & $d_{\rm l-maser}$ \tablefootmark{e} & $d_{\rm l-UCHII}$ \tablefootmark{e} & FWHM$_{\rm UCHII}$ \tablefootmark{f} \\ 
       &   ($\arcsec$)                                                      &    17$^h$47$^m$ ; -28$^{\rm o}$22$'$ &    & (GHz$^{-1}$)  & ($\arcsec$) & ($\arcsec$)   & ($\arcsec$)  \\
    \hline
    \hline 
N1 & +0.02(0.01) ; -2.40(0.01)   & 19.872(0.002)$^s$ ; 18.40(0.01)$\arcsec$ & 40710 &  438  &  \_         &  0.06(0.01) & 0.12 \\
N2 & -0.02(0.01) ; +2.58(0.01)   & 19.868(0.001)$^s$ ; 13.42(0.01)$\arcsec$ & 30546 &  460  &  \_         &  0.42(0.01) & 0.08 \\
N3 & -8.21(0.07) ; +1.09(0.06)   & 19.248(0.005)$^s$ ; 14.91(0.06)$\arcsec$ & 3008  &  45   &  0.63(0.41) &  \_         & \_   \\
N4 & -4.52(0.11) ; -16.41(0.05)  & 19.528(0.008)$^s$ ; 32.41(0.05)$\arcsec$ & 932   &  14   &  0.25(0.41) &  \_         & \_   \\
N5 & +2.34(0.01) ; -25.34(0.01)  & 20.047(0.007)$^s$ ; 41.34(0.01)$\arcsec$ & 2369  &  35   &  0.43(0.40) &  0.19(0.01) & $<$0.25 \\
    \hline
\end{tabular}
\end{center}
\tablefoot{\tablefoottext{a}{Equatorial offsets of the spectral line density peak with respect to the phase center (see Sect.~\ref{obs-ALMA}). The uncertainties in parentheses come from the 2D-Gaussian fit to the contour map. They are only statistical.}
\tablefoottext{b}{Same position given in J2000 Equatorial coordinates.}
\tablefoottext{c}{Number of channels with continuum-subtracted flux densities above 7$\sigma$.}
\tablefoottext{d}{Estimation of the spectral line density above 7$\sigma$ (excluding setup 3) assuming mean linewidths of $\sim$7~km~s$^{-1}$ for Sgr~B2(N1) and $\sim$5~km~s$^{-1}$ for the others hot cores \citep[see Sect.~\ref{kinematic-structure} and][]{belloche2016}.}
\tablefoottext{e}{Distance between the hot core and the closest class II methanol maser or UCHII region. The uncertainties given in parentheses are calculated based on the errors given by the Gaussian fits. In the case of the methanol masers, they also take into account the uncertainty on the maser positions (0.4$\arcsec$) given by \citet{caswell1996}.} 
\tablefoottext{f}{Deconvolved angular size of the UCHII region \citep{depree2015, gaume1995}.}}

\end{table*}

         \subsection{Continuum properties of the new sources}
                     \label{continuum}

In order to calculate H$_2$ column densities from the dust thermal emission arising from the hot cores embedded in Sgr~B2(N) we need to know the dust mass opacity coefficient $\kappa_{\nu}$, which depends on the dust emissivity exponent $\beta$. To derive these parameters we need observations at at least two frequencies. Here we use the ALMA and SMA data after processing them in the following way: the flux densities measured in the ALMA continuum maps need to be corrected for the contribution of the free-free emission and the SMA continuum map has to be smoothed to the ALMA resolution. 

                \subsubsection{ALMA data: correction for free-free emission}
                               \label{continuum-alma}

\begin{figure}[!t]
\begin{center}
 \includegraphics[width=\hsize]{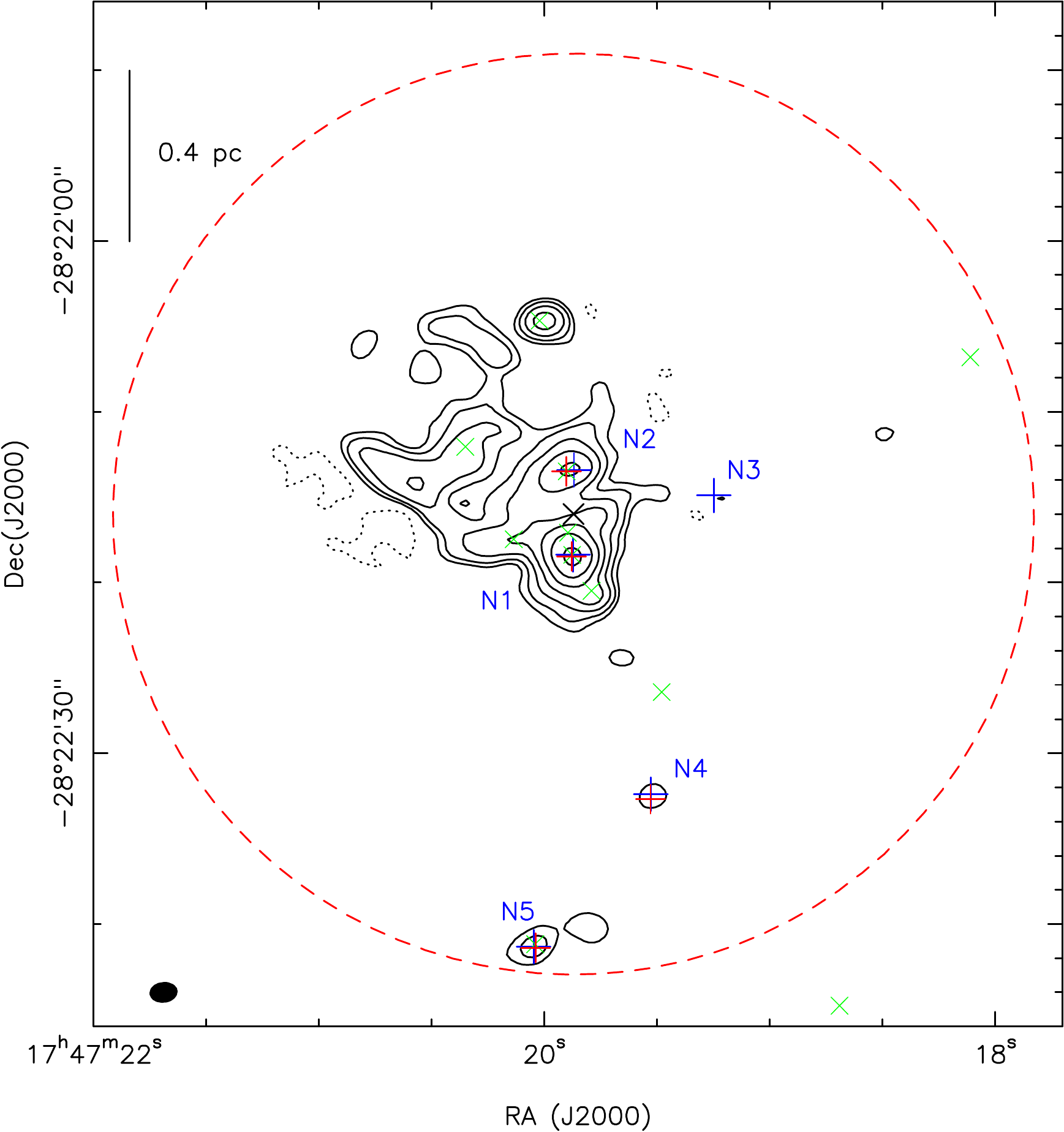} 
 \caption{\label{ALMA_continuum_map} Continuum map of the Sgr~B2(N) region obtained with ALMA at 108~GHz. Contour levels (positive in black solid line and negative in dashed line) start at 5 times the rms noise level, $\sigma$, of 3.0~mJy/beam and double in value up to 320$\sigma$. The filled ellipse shows the synthesized beam (1.65$\arcsec$ $\times$ 1.21$\arcsec$, PA=-83.4$^{\rm o}$). The black cross represents the phase center. The blue crosses mark the positions of the five hot cores embedded in Sgr~B2(N) derived from Fig.~\ref{hot_cores_contour_map}. The red crosses represent the average peak positions of the continuum emission derived from 2D-Gaussian fits to all ALMA continuum maps. The green crosses mark the compact and ultra-compact HII regions. The dotted red circle represents the size (HPBW) of the primary beam of the 12~m antennas at 108~GHz. The map is not corrected for the primary beam attenuation.}
\end{center}

\end{figure}

We investigate 20 continuum maps obtained at different frequencies over the whole frequency range of the ALMA survey. As an example, Fig.~\ref{ALMA_continuum_map} shows the continuum map obtained at 108~GHz with ALMA. Only four of the five hot cores embedded in Sgr~B2(N) are detected above the 5$\sigma$ level in this map. Over the whole frequency range, Sgr~B2(N1) and Sgr~B2(N2) show a strong continuum emission while Sgr~B2(N4) and Sgr~B2(N5) appear clearly weaker. Sgr~B2(N3) is not detected at all; for this source we can only derive an upper limit to its H$_2$ column density from the measurement of the noise level in the maps. In this case, noise histograms are plotted from all pixels using the command GO NOISE of the GILDAS software to derive the rms noise level in each map. For the other hot cores we measure the peak flux density $S_{\nu}^{\rm beam}$(Jy/beam) by fitting a 2D-Gaussian to the continuum maps at different frequencies (using the GAUSS-2D task of the GILDAS software). The derived flux density is then corrected for the primary beam attenuation. We note that the compact continuum source detected to the North of Sgr~B2(N2) corresponds to the UCHII region K4. It does not have a high spectral line density (see Fig.~\ref{hot_cores_contour_map}) and thus does not harbor a hot core.

The average peak position of the continuum emission associated to each hot core, derived from the 2D-Gaussian fit, is marked with a red cross in Fig.~\ref{ALMA_continuum_map}. These positions are also given in Table~\ref{continuum-pics-table} as well as the distances to the closest hot core and UCHII region. The offset between the continuum peak position and the hot core reference position is at most one fourth of the beam.

\begin{table*}[!t]
\begin{center}
  \caption{\label{continuum-pics-table} Position and size of the ALMA continuum sources, and distances to hot cores and UCHII regions.} 
  \setlength{\tabcolsep}{1.5mm}
\begin{tabular}{cccccc}
    \hline
Source & $\Delta \alpha$ ; $\Delta \delta$ \tablefootmark{a}     & $\alpha _{\rm J2000}$ ; $\delta _{\rm J2000}$  \tablefootmark{b}   & FWHM$_c$  \tablefootmark{c} & $d_{\rm c-l}$ \tablefootmark{d}   &  $d_{\rm c-UCHII}$ \tablefootmark{e} \\ 
       &  ($\arcsec$)                                                 &    17$^h$47$^m$ ; -28$^{\rm o}$22$'$                            &  ($\arcsec$)                     & ($\arcsec$)                          & ($\arcsec$)  \\
    \hline
    \hline 
N1 & +0.12(0.03);-2.48(0.05)   & 19.879(0.002)$^s$ ; 18.48(0.05)$\arcsec$  & 2.1(0.2) &  0.13(0.04)  & 0.10(0.05) \\
N2 & +0.42(0.11);2.50(0.15)    & 19.902(0.008)$^s$ ; 13.50(0.15)$\arcsec$  & 2.6(0.3) &  0.45(0.11)  & 0.02(0.11) \\
N4 & -4.50(0.31);-16.69(0.65)  & 19.529(0.050)$^s$ ; 32.69(0.65)$\arcsec$  & 0.7(0.1) &  0.28(0.65)  & \_ \\
N5 & +2.21(0.15);-25.38(0.18)  & 20.037(0.002)$^s$ ; 41.38(0.18)$\arcsec$  & 2.0(0.4) &  0.14(0.15)  & 0.20(0.18) \\
    \hline
\end{tabular}
\end{center}
\tablefoot{\tablefoottext{a}{Equatorial offsets of the continuum peak with respect to the phase center (see Sect.~\ref{obs-ALMA}). The uncertainties in parentheses represent the standard deviation weighted by the errors given by the 2D-Gaussian fit.}
\tablefoottext{b}{Same position given in J2000 Equatorial coordinates.}
\tablefoottext{c}{Average deconvolved angular size of the continuum source derived from 2D-Gaussian fits to the ALMA continuum maps. The uncertainty given in parentheses corresponds to the standard deviation.}
\tablefoottext{d}{Distance between the continuum peak position and the position of the closest hot core derived from Fig.~\ref{hot_cores_contour_map}. The uncertainty given in parentheses is calculated based on the errors given by the 2D-Gaussian fits.} 
\tablefoottext{e}{Distance between the continuum peak position and the peak position of the closest UCHII region \citep{gaume1995, depree2015}. The uncertainty given in parentheses is calculated based on the errors given by the 2D-Gaussian fit.}}
\end{table*}

Usually (sub)millimeter continuum emission observed toward Sgr~B2(N) is attributed to thermal emission from interstellar dust \citep{kuan1996}, while at wavelength~$\geq$~1~cm the continuum emission is dominated by free-free emission arising from HII regions. However in the ALMA frequency range considered in this paper, the free-free emission may significantly contribute to the 3~mm flux densities. Figure~\ref{ALMA_continuum_map} shows that Sgr~B2(N1), Sgr~B2(N2), and Sgr~B2(N5) are associated with UCHII regions. Furthermore the overall shape of the extended continuum emission detected with ALMA around Sgr~B2(N1) and Sgr~B2(N2) is similar to the shape of the 1.3~cm free-free emission reported by \citet{gaume1995}. The peak flux densities measured at 3~mm toward the hot cores need thus to be corrected for the contribution of the free-free emission. Details about this analysis are presented in Appendix~\ref{appendix-continuum}. The fraction of free-free emission estimated toward Sgr~B2(N1) varies between $\sim$17\% at 85~GHz and $\sim$9\% at 114~GHz. The free-free contribution is higher toward Sgr~B2(N2), from $\sim$75\% at 85~GHz, down to $\sim$32\% at 113.5~GHz, and it varies between 40\% and 60\% toward Sgr~B2(N5). 
The flux densities measured in the 20 ALMA continuum maps and corrected for the free-free contribution are presented in Tables~\ref{tab-continuum-n1}-\ref{tab-continuum-n5}. We do not make any correction for Sgr~B2(N3) and Sgr~B2(N4) because they are not associated with any known UCHII region.

                    \subsubsection{SMA data: smoothing to the ALMA resolution}
                                   \label{continuum-SMA}

\begin{figure}[!t]
\begin{center} 
 \includegraphics[width=\hsize]{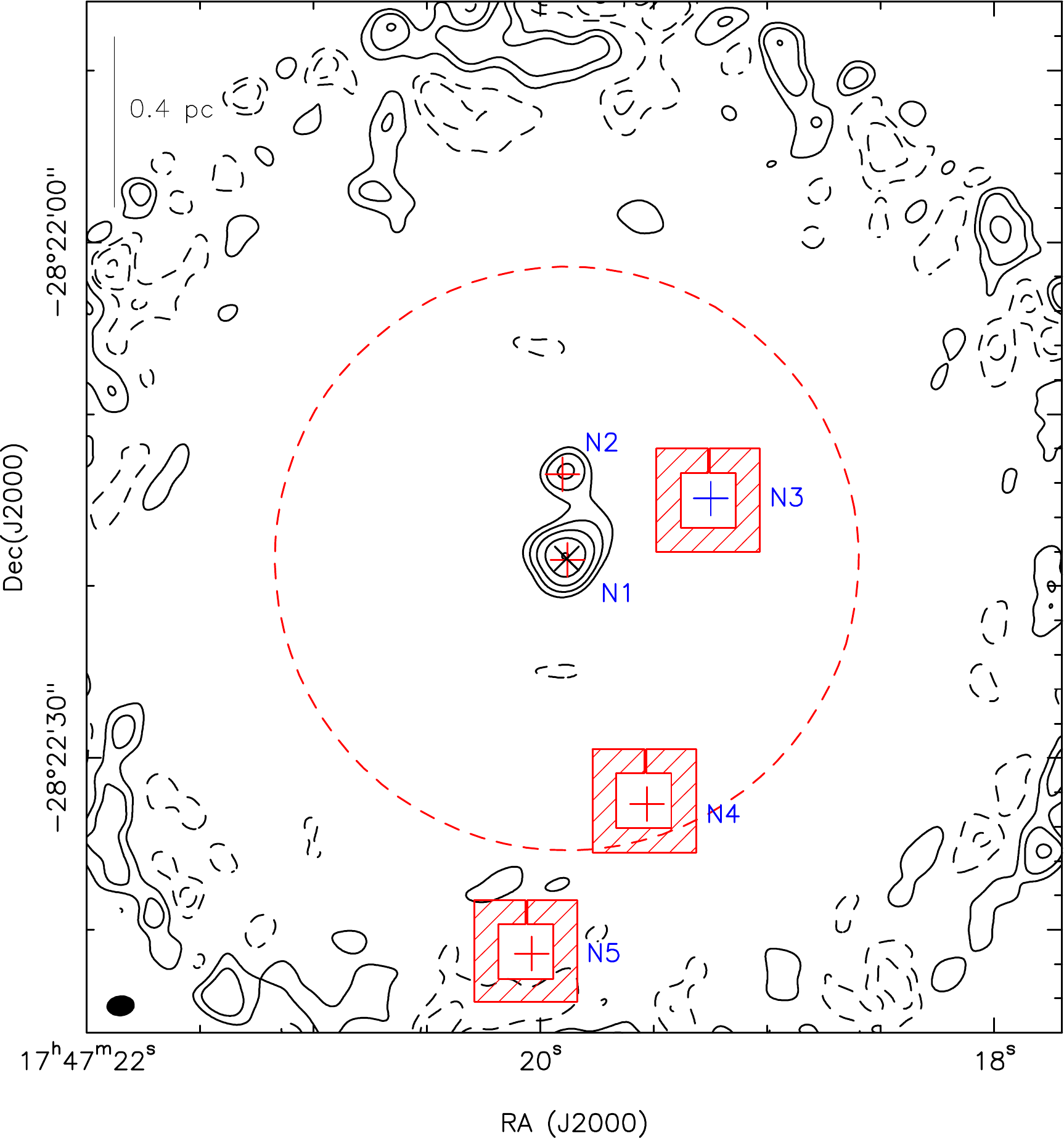} 
 \caption{\label{SMA_continuum_map} Continuum map of the Sgr~B2(N) region obtained with the SMA at 343~GHz \citep{qin2011} and smoothed to the ALMA resolution. The new beam is shown in the bottom left corner (1.65$\arcsec$ $\times$ 1.21$\arcsec$, PA=-83.4$^{\rm o}$). The contour levels (positive in solid line, negative in dashed line) start at 8$\sigma$ and double up to 128$\sigma$ with $\sigma$ = 0.15~Jy/beam, the noise level measured inside the polygon defined around Sgr~B2(N3). The black cross represents the SMA phase center. The red crosses mark the peak positions of the ALMA continuum emission derived from Fig.~\ref{ALMA_continuum_map}. The position of Sgr~B2(N3) (blue cross) is derived from Fig.~\ref{hot_cores_contour_map}. The red dashed circle represents the size (HPBW) of the primary beam of the SMA 6~m antennas at 343~GHz. The map is corrected for the primary beam attenuation.}
\end{center}
\end{figure}

 We use the SMA map obtained by \citet{qin2011} to derive the dust emissivity index $\beta$ in a joint ALMA/SMA analysis. To this aim we first need to smooth the SMA map to the ALMA resolution. We use the task GAUSS-SMOOTH of the GILDAS software. Figure~\ref{SMA_continuum_map} shows as an example the SMA map smoothed to the same resolution as the ALMA map shown in Fig.~\ref{ALMA_continuum_map}. We proceed in the same way to smooth the SMA map to the 20 different angular resolutions corresponding to the five ALMA setups \citep[20 spectral windows, 4 per setup; see][for details about the angular resolution]{belloche2016}. 

 Figure~\ref{SMA_continuum_map} shows that only the two main hot cores Sgr~B2(N1) and Sgr~B2(N2) are detected in the SMA map. To derive peak flux densities toward these two sources we fit 2D Gaussians to the smoothed maps. For the other hot cores that are not detected, we measure in each map the noise level inside the polygons plotted in red in Fig.~\ref{SMA_continuum_map}. Each square has a dimension of 6$\arcsec$ and an area of about twice the beam size is excluded in the middle. The flux densities or rms derived from the 20 maps are listed in Tables~\ref{tab-continuum-n1}-\ref{tab-continuum-n5}. At 343~GHz, the contribution of the free-free emission to the flux density is lower than 0.1\% and can safely be ignored \citep{qin2011}.

                   \subsubsection{SMA data: original resolution}
                                 \label{continuum-SMA-original}

To derive H$_2$ column densities from the SMA map at its original resolution ($\sim$0.3$\arcsec$), we measure the peak flux densities toward Sgr~B2(N1) and Sgr~B2(N2) with 2D-Gaussian fits to the map. For the fainter hot cores  Sgr~B2(N3), Sgr~B2(N4), and Sgr~B2(N5) not detected in the SMA map, we measure the average noise level inside a polygon of 6$''^2$ excluding an area of about twice the beam size in the middle.

                   \subsubsection{H$_2$ column densities}
                                 \label{h2-column-density}

We assume a dust temperature $T_{\rm d}$ $\sim T_{\rm gas}$~$\sim$150~K (see Sect.~\ref{Trot}). Neglecting the cosmic microwave background temperature, the radiative transfer equation can be written as follows:
\begin{equation}
\label{eq-radiative-transfer}
S_{\nu}^{\rm beam} = \Omega_{\rm beam}  B_{\nu}(T_d) (1 - \mathrm{e}^{-\tau}),
\end{equation}
with $\Omega_{\rm beam} =  \frac{\pi}{4 \ln 2} \times HPBW_{\rm max} \times HPBW_{\rm min}$ the solid angle of the synthesized beam, $B_{\nu}(T_{\rm d}$) the Planck function at the dust temperature, and $\tau$ the dust opacity. $S_{\nu}^{\rm beam}$ is the peak flux density measured in the continuum maps in Jy/beam. $S_{\nu}^{\rm beam}$ is corrected for the primary beam attenuation and, in the case of the ALMA data, for the free-free contribution.

From this equation we can calculate the dust opacity as follows: 
\begin{equation}
\label{eq-tau1}
\tau = -\ln \left( 1 - \frac{S_{\nu}^{\rm beam}}{\Omega_{\rm beam} B_{\nu}(T_d)}  \right).
\end{equation} 
The results of the opacity calculations for both the ALMA and SMA data are summarized in Table~\ref{tau-calculations}. The emission is optically thin for all cores except Sgr~B2(N1) at the SMA frequency. In the SMA map at its original resolution (0.3$\arcsec$), the dust emission of Sgr~B2(N1) is optically thick and inconsistent with a temperature of 150~K. From Eq.~\ref{eq-radiative-transfer} and assuming optically thick emission, we derive a lower limit to the dust temperature of 200~K at a scale of 0.3$\arcsec$. 

The dust opacity, $\tau$, is related to the H$_2$ column density, $N_{\rm H_2}$, via: 
\begin{equation}
\label{eq-tau2}
\tau = \mu_{\rm H_2}  m_{\rm H}  \kappa_{\nu}  N_{\rm H_2},
\end{equation}
with $\mu_{\rm H_2}$ = 2.8 the mean molecular weight per hydrogen molecule \citep{kauffmann2008}, $m_{\rm H}$ the mass of atomic hydrogen, and $\kappa_{\nu}$ the dust mass opacity (in cm$^2$ g$^{-1}$) given by the power law:
\begin{equation}
\label{eq-kappa}
\kappa_{\nu} = \frac{\kappa_0}{\chi_d} \left( \frac{\nu}{\nu _0}  \right) ^{\beta},
\end{equation}
with $\chi_d$ = 100 the standard gas-to-dust ratio, $\beta$ the dust emissivity exponent, and $\kappa_0$ the mass absorption coefficient at frequency $\nu_0$. 

We derive the dust emissivity index $\beta$ from the ALMA and SMA data. From Eqs.~\ref{eq-radiative-transfer}, \ref{eq-tau2}, and \ref{eq-kappa} we can write:  
\begin{equation}
S^{\rm ALMA}_{\nu} = \Omega_{\rm ALMA}   B_{\nu}^{\rm ALMA}(T_d) \left[1 - \mathrm{exp}{\left( -\mu_{\rm H_2}  m_{\rm H} \frac{\kappa_0}{\chi_d} \left( \frac{\nu_{\rm ALMA}}{\nu_0}  \right)^{\beta}   N_{\rm H_2} \right)} \right] \nonumber
\end{equation}
and
\begin{equation}
S^{\rm SMA}_{\nu} = \Omega_{\rm SMA}   B_{\nu}^{\rm SMA}(T_d) \left[ 1 - \mathrm{exp}{\left(-\mu_{\rm H_2}  m_{\rm H}  \frac{\kappa_0}{\chi_d} \left( \frac{\nu_{\rm SMA}}{\nu_0}  \right)^{\beta}  N_{\rm H_2}\right)} \right], \nonumber
\end{equation}
which can also be written as: 
\begin{equation}
 N_{\rm H_2} \left( \frac{\nu_{\rm ALMA}}{\nu_0} \right)^{\beta} = -\frac{1}{\mu_{\rm H_2}  m_{\rm H}  \frac{\kappa_0}{\chi_d}} \times \ln \left(1 - \frac{S^{\rm ALMA}_{\nu}}{\Omega_{\rm ALMA} B_{\nu}^{\rm ALMA}}  \right) \nonumber
\end{equation}
and
\begin{equation}
N_{\rm H_2} \left( \frac{\nu_{\rm SMA}}{\nu_0} \right)^{\beta} = -\frac{1}{\mu_{\rm H_2}  m_{\rm H}  \frac{\kappa_0}{\chi_d}} \times \ln \left(1 - \frac{S^{\rm SMA}_{\nu}}{\Omega_{\rm SMA} B_{\nu}^{\rm SMA}}  \right). \nonumber
\end{equation}
Then $\beta$ is given by:
\begin{equation}
\beta =  \ln \left[  \frac{ \ln \left(1 - \frac{S^{\rm ALMA}_{\nu}}{\Omega_{\rm ALMA} B_{\nu}^{\rm ALMA}}  \right)}{ \ln \left(1 - \frac{S^{\rm SMA}_{\nu}}{\Omega_{\rm SMA} B_{\nu}^{\rm SMA}}  \right)} \right] \times \frac{1}{\ln \left( \frac{\nu_{\rm ALMA}}{\nu_{\rm SMA}}  \right) },
\end{equation}
which only depends on the peak flux density measured on the continuum maps, the beam solid angle, the Planck function, and the frequency. We perform the calculations using the flux densities measured toward Sgr~B2(N1) and Sgr~B2(N2) in the ALMA maps and in the SMA maps smoothed to the ALMA resolution, excluding setup 3 (low angular resolution). We obtain $\beta$~=~0.8$\pm$0.1 and $\beta$~=~1.2$\pm$0.1 for Sgr~B2(N1) and Sgr~B2(N2), respectively.

As mentioned above, the continuum emission detected with SMA toward Sgr~B2(N1) is very optically thick. Even if the emission in the SMA map smoothed to the ALMA resolution should be less optically thick and our analysis does not assume optically thin emission, it is likely that the value derived for $\beta$ toward Sgr~B2(N1) is underestimated due to these high opacities. It could also be that we have underestimated the contribution of free-free emission toward Sgr~B2(N1) at the ALMA frequencies. Therefore, we consider the value obtained toward Sgr~B2(N2) as more reliable, and we adopt $\beta = 1.2$ for all sources. Such a flat index suggests a dust opacity spectrum intermediate between the models of dust grains without ice mantles and those with thin ice mantles of \citet[][see their Fig. 5]{ossenkopf1994}. We adopt a dust mass absorption coefficient $\kappa_0 = 1.6$~cm$^2$~g$^{-1}$ (of dust) at $\lambda_0 = 1.3$~mm which is intermediate between these models. With this normalization, we obtain a methanol abundance of $\sim 2\times 10^{-5}$ relative to H$_2$ for Sgr~B2(N2) (see Sect.~\ref{chemical-composition}), consistent with the peak gas-phase abundance of methanol predicted by our chemical models \citep[][]{garrod2009,muller2016}. A higher value of $\kappa_0$ would imply lower H$_2$ column densities and, in turn, higher methanol abundances that would not be realistic anymore.

\begin{table}[!t]
\begin{center}
  \caption{\label{tau-calculations} Dust opacity of the five hot cores for a dust temperature of 150~K.} 
  \begin{tabular}{clcc}
    \hline
  Source   &   $\tau _{\rm ALMA}$ \tablefootmark{a}  & $\tau _{\rm SMA(1.6\arcsec)}$ \tablefootmark{b} & $\tau _{\rm SMA(0.3\arcsec)}$ \tablefootmark{c}         \\
    \hline
    \hline 
N1 &  0.36    & 0.95    & \_ \tablefootmark{*}  \\
N2 &  0.05    & 0.19    & 0.61 \\
N3 &  $<$0.009 & $<$0.02 & $<$0.07 \\ 
N4 &  0.008    & $<$0.06 & $<$0.12 \\
N5 &  0.03    & $<$0.15 & $<$0.25 \\
    \hline
\end{tabular}
\end{center}
\tablefoot{\tablefoottext{a}{Dust opacity calculated based on the ALMA data. $S_{\nu}^{\rm beam}$ has been corrected for the primary beam attenuation and the free-free contamination.}
\tablefoottext{b}{Dust opacity calculated based on the SMA map smoothed to the ALMA resolution ($\sim$1.6$\arcsec$).}
\tablefoottext{c}{Dust opacity calculated based on the SMA map at its original resolution ($\sim$0.3$\arcsec$).}
\tablefoottext{*}{Optically thick and inconsistent with a temperature of 150~K.}
}
\end{table}

\begin{figure}[!t]
\begin{center}
 \includegraphics[width=\hsize]{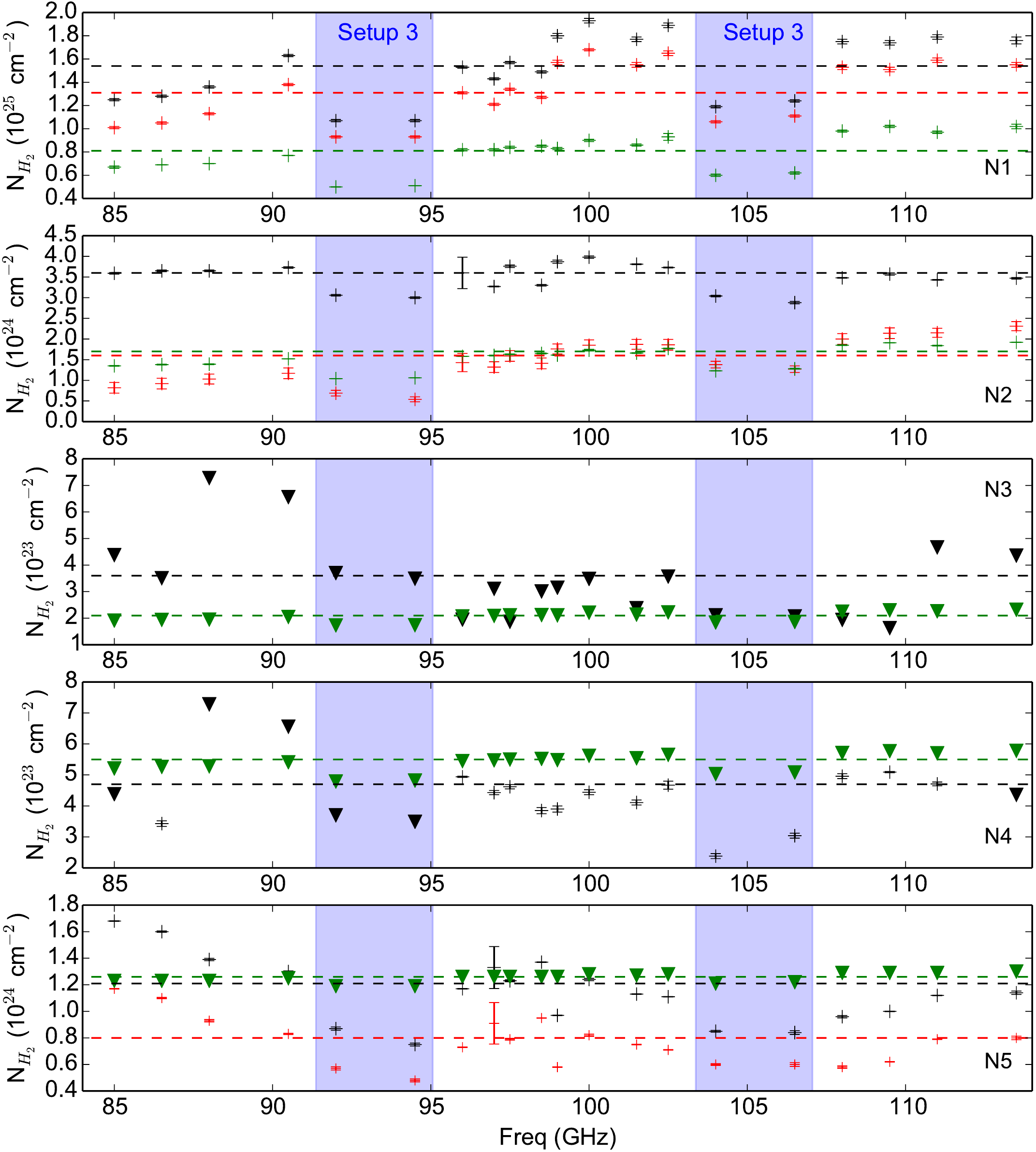} 
 \caption{\label{continuum-plot}  H$_2$ column densities as a function of frequency for the five hot cores embedded in Sgr~B2(N). The values obtained based on the ALMA data before and after correction for the free-free contribution are shown as black and red crosses respectively. The green crosses represent the results obtained based on the SMA map at 343~GHz smoothed to the angular resolution of the ALMA maps. Error bars are calculated from the error on $S^{\rm beam}_{\nu}$ given by the GAUSS-2D routine and take into account the uncertainty on the correction for the free-free emission. The triangles represent $5\sigma$ upper limits. The dashed line in each panel is the average H$_2$ column density (or upper limit) excluding setup 3.}
\end{center}
\end{figure}

From Eqs.~\ref{eq-radiative-transfer} and \ref{eq-tau2} we can now calculate the H$_2$ column density for each hot core using the following equation:
\begin{equation}
\label{eq-column-density}
N_{\rm H_2} = -\frac{1}{\mu_{\rm H_2}  m_{\rm H}  \kappa_{\nu}} \times \ln \left( 1 - \frac{S_{\nu}^{\rm beam}}{\Omega_{\rm beam}   B_{\nu}(T_d)}  \right).
\end{equation}
Figure~\ref{continuum-plot} shows the results of the H$_2$ column density calculations at $T_{\rm d}$ = 150~K, from both the SMA and ALMA data as a function of ALMA frequency. The figure shows that Sgr~B2(N4) is not systematically detected over the whole frequency range of the ALMA survey. All upper limits correspond to 5 times the noise level. All the results are also listed in Tables~\ref{tab-continuum-n1}-\ref{tab-continuum-n5}. One should keep in mind that the beam is slightly different as a function of frequency, especially in the frequency range covered by setup 3 which has the lowest angular resolution (HPBW $>$ 2$\arcsec$). We thus calculate the average peak H$_2$ column density of each hot core from both the SMA (smoothed to the ALMA resolution) and ALMA data excluding setup 3. Table~\ref{NH2calculations-bilan} summarizes the results obtained from the ALMA data before and after correction for the free-free contribution and for the SMA data after and before smoothing to the ALMA resolution. This table shows that the H$_2$ column densities of Sgr~B2(N2), Sgr~B2(N3), Sgr~B2(N4), and Sgr~B2(N5) are, respectively, 8, $>$36, 28, and 16 times lower than the one of the main hot core Sgr~B2(N1). Within the uncertainties, the column densities or upper limits obtained from the ALMA data for the faint hot cores Sgr~B2(N3), Sgr~B2(N4), and Sgr~B2(N5) are consistent with the upper limits derived from the SMA smoothed maps.

\begin{table}[!t]
\begin{center}
  \caption{\label{NH2calculations-bilan} H$_2$ column densities for a dust temperature of 150 K.} 
  \setlength{\tabcolsep}{1.0mm}
  \begin{tabular}{ccccc}
    \hline
Source  &  \multicolumn{4}{c}{$N_{\rm H_2}$ (10$^{24}$ cm$^{-2}$)}          \\
        & ALMA\tablefootmark{a}  & ALMA$_{\rm dust}$\tablefootmark{b}  & SMA$_{\rm (1.6\arcsec)}$ \tablefootmark{c} & SMA$_{\rm (0.3\arcsec)}$ \tablefootmark{d} \\    
    \hline
    \hline 
N1 &  15.4(2.2)      & 13.1(2.2)       & 8.1(0.9)     & \_ \tablefootmark{*}  \\
N2 &  3.6(0.2)       & 1.6(0.5)        & 1.7(0.2)      & 5.1(0.4) \\
N3 &  $<$0.36(0.16)  & $<$0.36(0.16)   & $<$0.21(0.01) & $<$0.5  \\
N4 &  0.47(0.05)     & 0.47(0.05)      & $<$0.55(0.02) & $<$1.0  \\
N5 &  1.21(0.21)     & 0.80(0.17)      & $<$1.26(0.02) & $<$2.1  \\
    \hline
\end{tabular}
\end{center}
\hspace{0cm}
\tablefoot{The uncertainties are given in parentheses and correspond to the standard deviations weighted by the error on $S_{\nu}^{\rm beam}$ and on the correction factor for the free-free emission.
\tablefoottext{a}{H$_2$ column densities calculated based on the ALMA data after correction for the primary beam attenuation, for a mean synthesized beam size of $\sim$1.6$\arcsec$.}
\tablefoottext{b}{ALMA$_{\rm dust}$ is in addition corrected for the free-free contribution.}
\tablefoottext{c}{H$_2$ column densities calculated based on the SMA map smoothed to the ALMA resolution ($\sim$1.6$\arcsec$).} 
\tablefoottext{d}{H$_2$ column densities calculated based on the SMA map at its original resolution ($\sim$0.3$\arcsec$)}. 
\tablefoottext{*}{The dust emission toward Sgr~B2(N1) being optically thick we cannot derive its H$_2$ column density.}}
\end{table}
\hspace{2cm}

        \subsection{Line identification}
                    \label{line-identification}

\begin{figure*}[!t]
\begin{center}
 \includegraphics[width=\hsize]{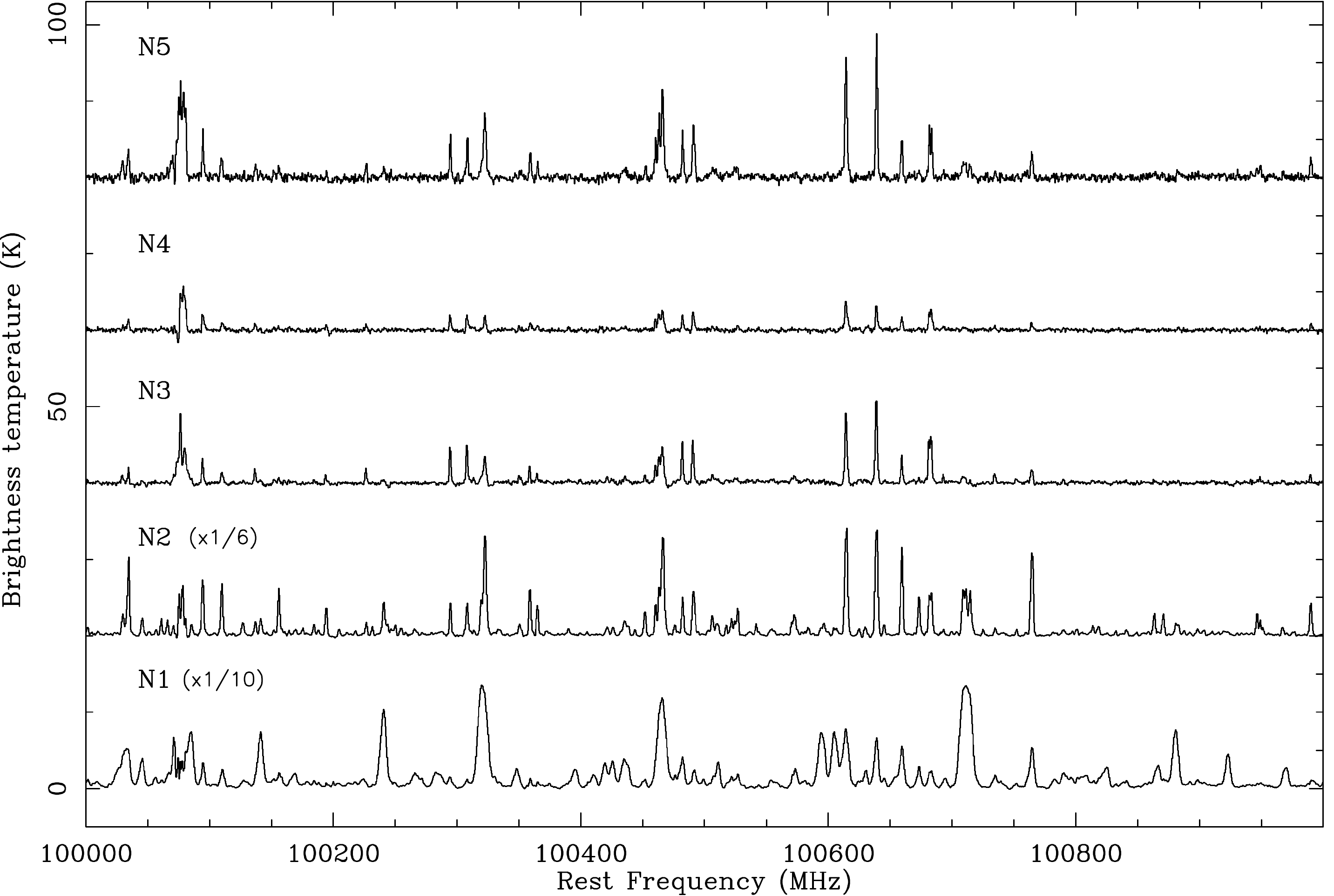} 
 \caption{\label{spectra-5-hot-cores}Part of the continuum-subtracted ALMA spectra observed toward the hot cores embedded in Sgr~B2(N). The spectra have been corrected for the primary beam attenuation, and shifted along the y axis for display purposes. The spectra of Sgr~B2(N1) and Sgr~B2(N2) have been divided by 10 and 6, respectively. The frequency axis corresponds to the systemic velocities derived in Sect.~\ref{kinematic-structure}.}
\end{center}
\end{figure*}

Table~\ref{result-fit-channel-map} clearly shows that the spectral line density is much lower toward Sgr~B2(N3), Sgr~B2(N4), and Sgr~B2(N5) than for Sgr~B2(N1) and Sgr~B2(N2), reducing considerably the occurence of line blending. This is also illustrated in Fig.~\ref{spectra-5-hot-cores} where a portion of the ALMA spectrum of each hot core is shown.
We use Weeds as described in Sect.~\ref{modelling} to perform the line identification and model the spectra observed toward the three new hot cores in order to derive their chemical composition. Table~\ref{tab-chemical-composition} shows the total number of emission lines detected above the 7$\sigma$ level, counted manually toward the peak position of each hot core, excluding setup 3. It also summarises the number of species identified so far and the fraction of remaining unidentified lines above 7$\sigma$ (U-lines hereafter). The line density derived here is somewhat lower than reported in Table~\ref{result-fit-channel-map}. In Table~\ref{result-fit-channel-map}, we divided the number of channels derived from Fig.~\ref{hot_cores_contour_map} by the typical FWHM of an emission line. This was a rough estimate because a faint line may have only a single channel emitting above the 7$\sigma$ threshold while a strong line will have more channels above 7$\sigma$ than the number of channels covered by its FWHM. This suggests that Sgr~B2(N1) and Sgr~B2(N2) probably contain less emission lines than indicated in Table~\ref{result-fit-channel-map}. 

Table~\ref{tab-chemical-composition} also shows that Sgr~B2(N4) has a low spectral line density compared to the other hot cores although roughly the same number of molecules were identified. This difference can be explain by the low number of less abundant isotopologs and vibrationally excited states detected in Sgr~B2(N4) compared to Sgr~B2(N3) and Sgr~B2(N5).

\begin{table}[!t]
\begin{center}
  \caption{\label{tab-chemical-composition} Statistics of the lines detected toward the three new hot cores.}
  \setlength{\tabcolsep}{1.0mm}
  \footnotesize
  \begin{tabular}{ccccccc}
    \hline
Source & $N_{\rm l}$\tablefootmark{a} & $n_{\rm l}$\tablefootmark{b}        & $N_{\rm species}$\tablefootmark{c} & $N_{\rm iso}$\tablefootmark{c} & $N_{\rm exc}$\tablefootmark{c}  & U-lines\tablefootmark{d} \\
       &  & (GHz$^{-1}$)                         &                          &                          &                    & (\%)    \\
    \hline
    \hline
N3  & 714  & 31  & 23  & 20  & 19  &  9  \\
N4  & 249  & 11  & 22  & 13  &  9  & 11  \\
N5  & 508  & 22  & 25  & 16  & 12  &  7  \\
    \hline
\end{tabular}
\end{center}
\tablefoot{\tablefoottext{a}{Total number of emission lines detected above the 7$\sigma$ level (rms~$\sim$3~mJy/beam) excluding setup 3.}
\tablefoottext{b}{Line density above 7$\sigma$, excluding setup 3.}
\tablefoottext{c}{Number of identified molecules, less abundant isotopologs, and vibrationally excited states.}
\tablefoottext{d}{Fraction of remaining unidentified lines above 7$\sigma$.}}
\end{table}

        \subsection{Spatial distribution of the molecules}
                   \label{spatial-distribution} 

\begin{figure*}[!t]
\begin{center}
 \includegraphics[width=\hsize]{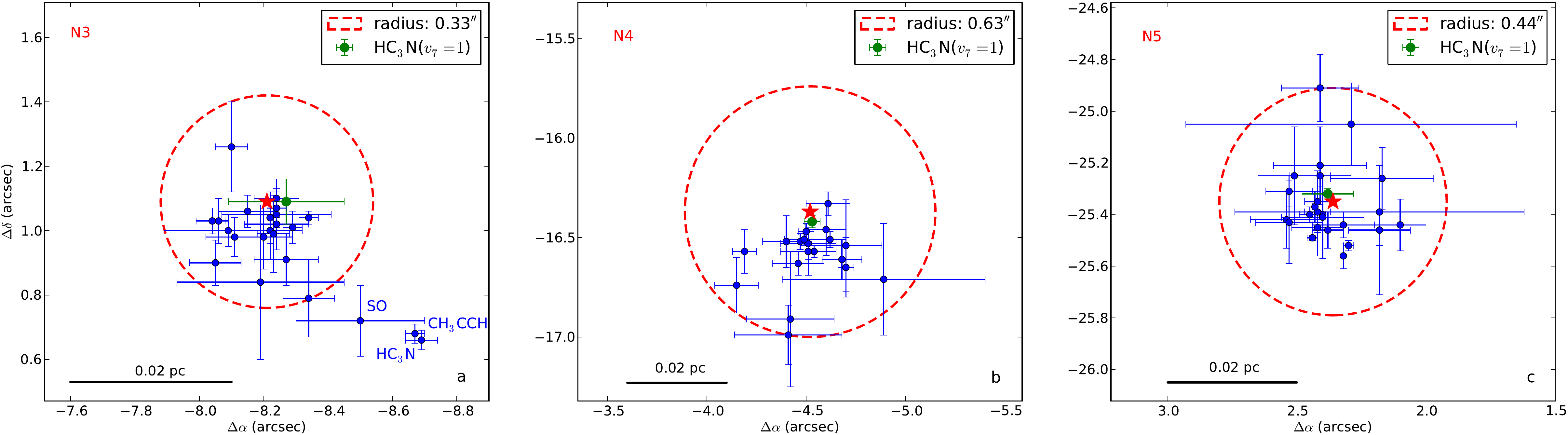} 
 \caption{\label{spatial-distribution-plot} Spatial distribution of the molecules identified around Sgr~B2(N3), Sgr~B2(N4), and Sgr~B2(N5). In each panel, the red star represents the position of the hot core derived from Fig.~\ref{result-fit-channel-map}. Blue crosses represent the mean peak position for each species. Error bars correspond to the standard deviation weighted by the uncertainties given by the GAUSS-2D routine. The radius of the dashed circle is given in the top right corner. The average peak position of HC$_3$N, $\varv_7=1$ is shown in green.}
\end{center}
\end{figure*}

In order to get information about the spatial structure of the three new hot cores, we investigate for each identified molecule the integrated intensity maps produced from its vibrationnal ground state transitions that are well detected and free of contamination from other species. The position of the emission peak of each spectral line is derived from a 2D-Gaussian fit to its map using the GAUSS-2D task of the GILDAS software. Figure~\ref{spatial-distribution-plot} shows the spatial distribution of the molecules identified so far toward Sgr~B2(N3), Sgr~B2(N4), and Sgr~B2(N5). Each cross represents the mean peak position of a given species. Figure~\ref{spatial-distribution-plot}a shows that almost all species identified toward Sgr~B2(N3) peak within a distance of $\sim$0.3$\arcsec$ from the reference position of the hot core. Only SO, \ce{CH3CCH}, and \ce{HC3N} peak beyond 0.5$\arcsec$ South-West of this position. In the case of Sgr~B2(N4) (Fig.~\ref{spatial-distribution-plot}b) and Sgr~B2(N5) (Fig.~\ref{spatial-distribution-plot}c), all molecules peak within distances of about 0.6$\arcsec$ and 0.4$\arcsec$ from the hot core position, respectively. For both Sgr~B2(N4) and Sgr~B2(N5), \ce{HC3N} in its vibrational ground state shows broad lines contaminated by other species. For this reason we used transitions in its first vibrationally excited state, $\varv_7=1$. They peak close to the reference position of each hot core, in particular toward Sgr~B2(N3) for which the emission peak of the vibrational ground state transitions is offset to the South-West by $\sim$0.6$\arcsec$.


        \subsection{Size of the hot cores}
                   \label{source-size}

For each hot core we select the transitions that are well reproduced by the model, are not severely contaminated by other species, and have a high signal-to-noise ratio (typically $\ge$~8 up to $\sim$99) to fit 2D Gaussians to their integrated intensity maps and derive their emission size. 

\begin{figure*}[!t]
\resizebox{\hsize}{!}
{\begin{tabular}{cccc}
\vspace{0.5cm}
   \includegraphics[width=20.5cm]{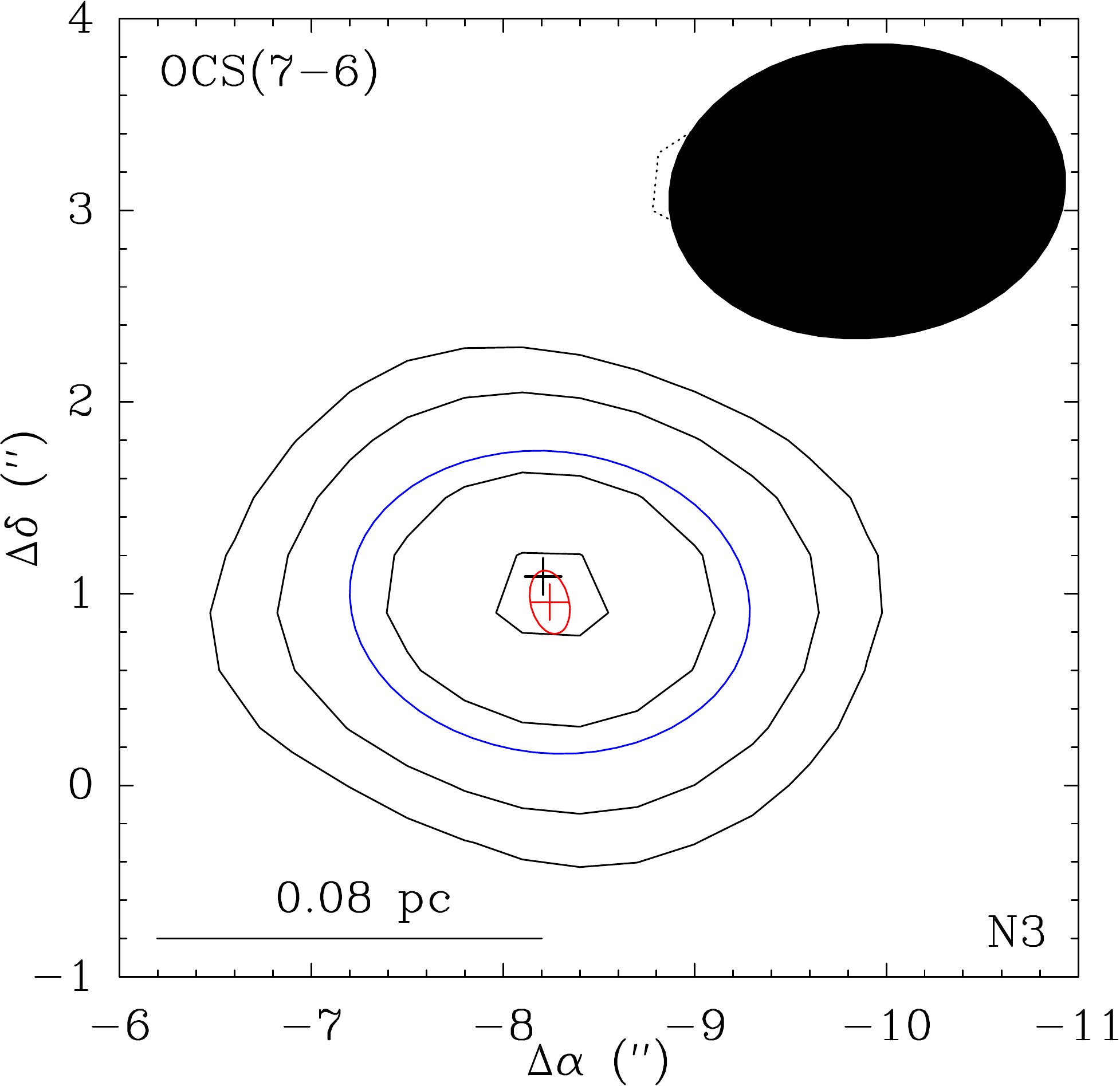} &
   \includegraphics[width=19cm]{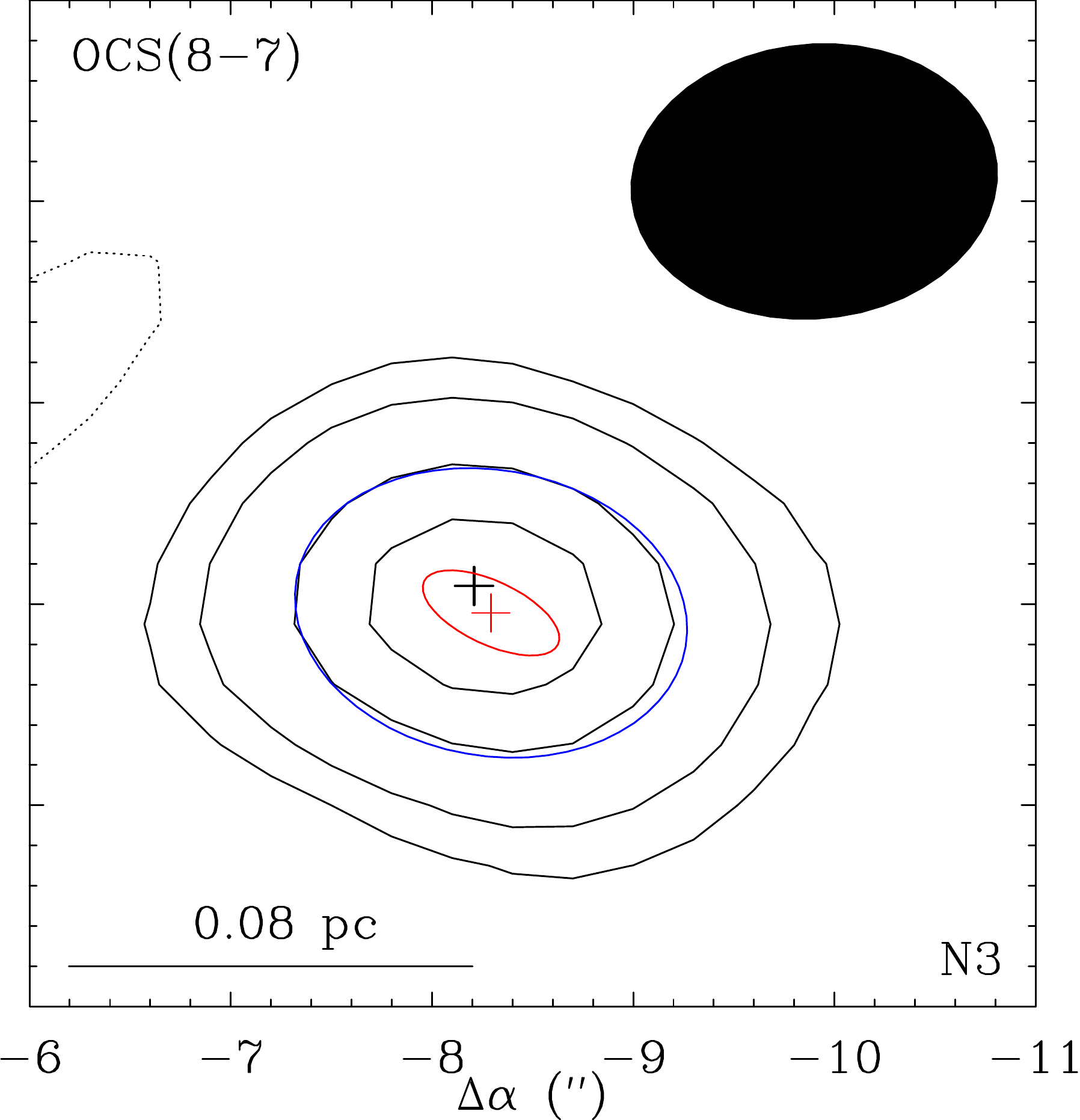} &
   \includegraphics[width=19cm]{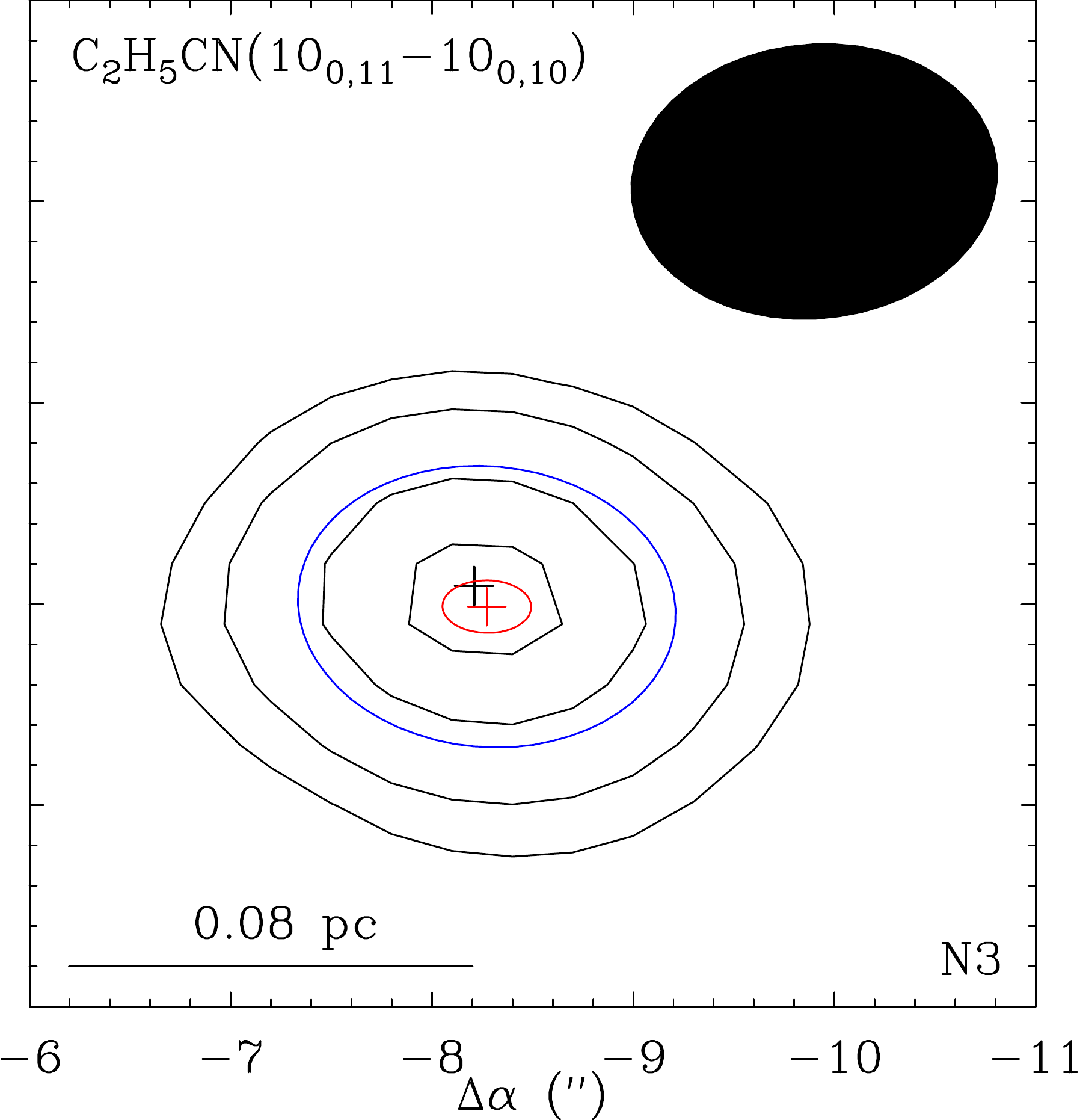} &
   \includegraphics[width=19cm]{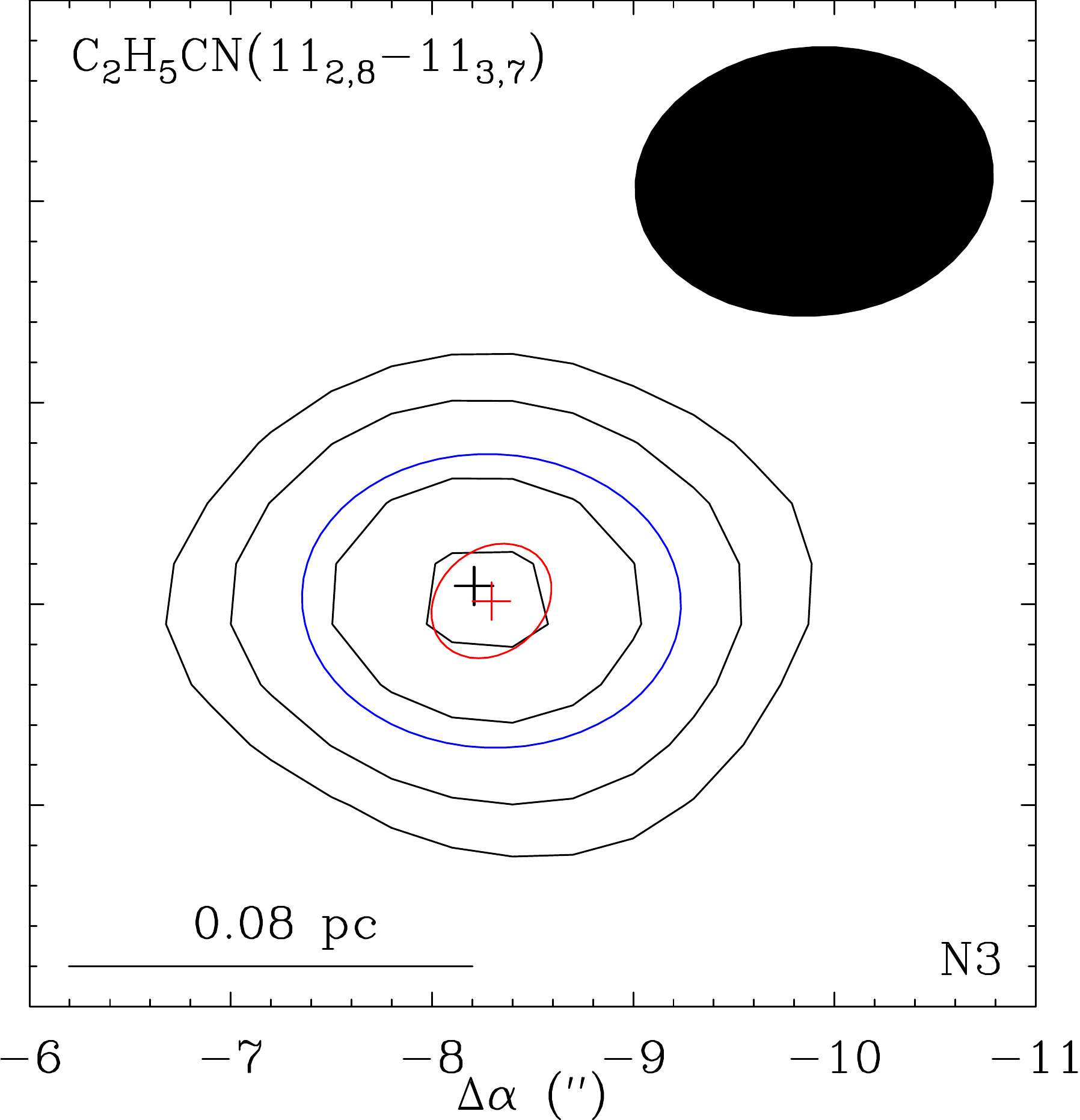} \\
\vspace{0.5cm}
   \includegraphics[width=20.5cm]{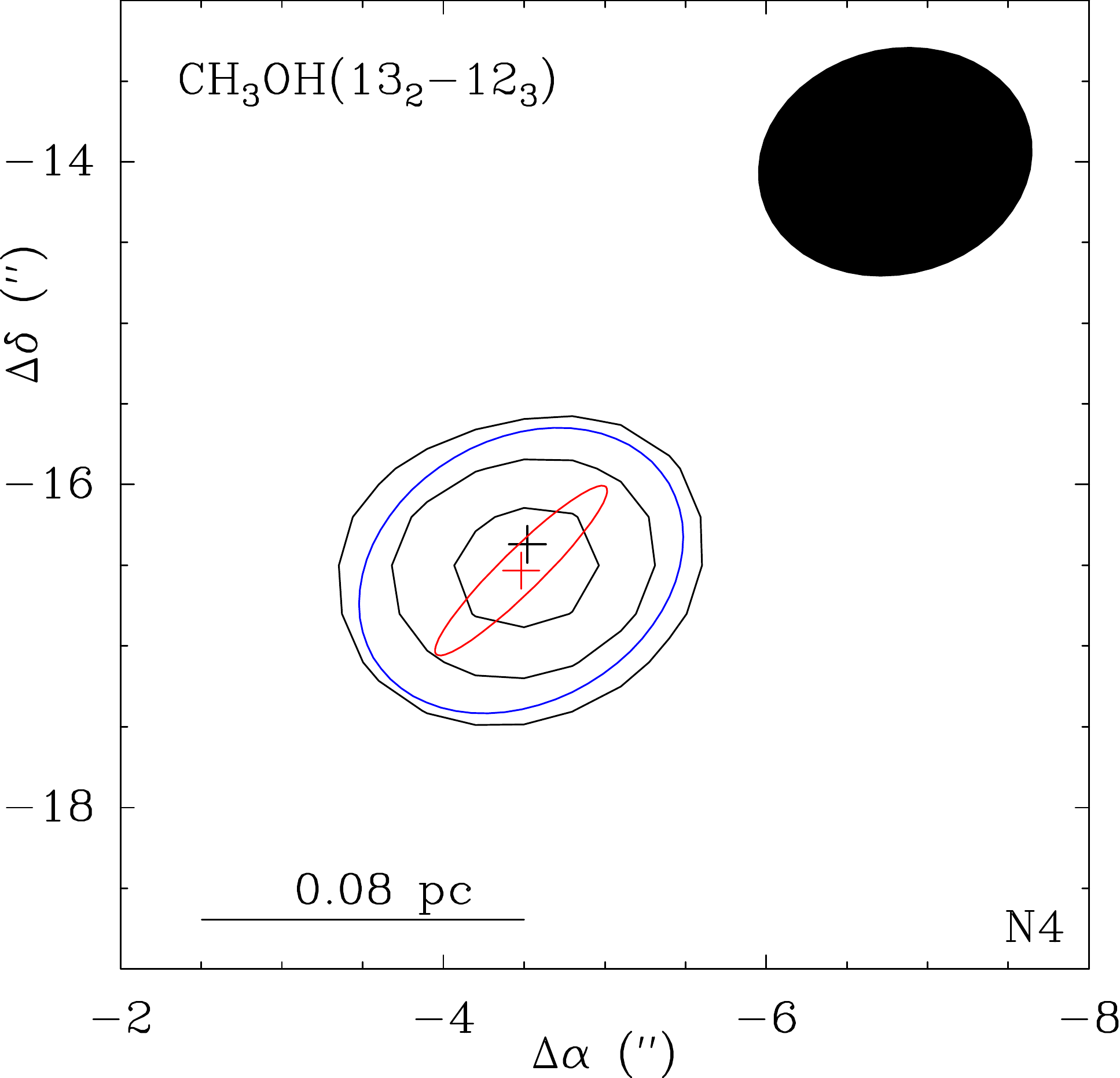} &
   \includegraphics[width=19cm]{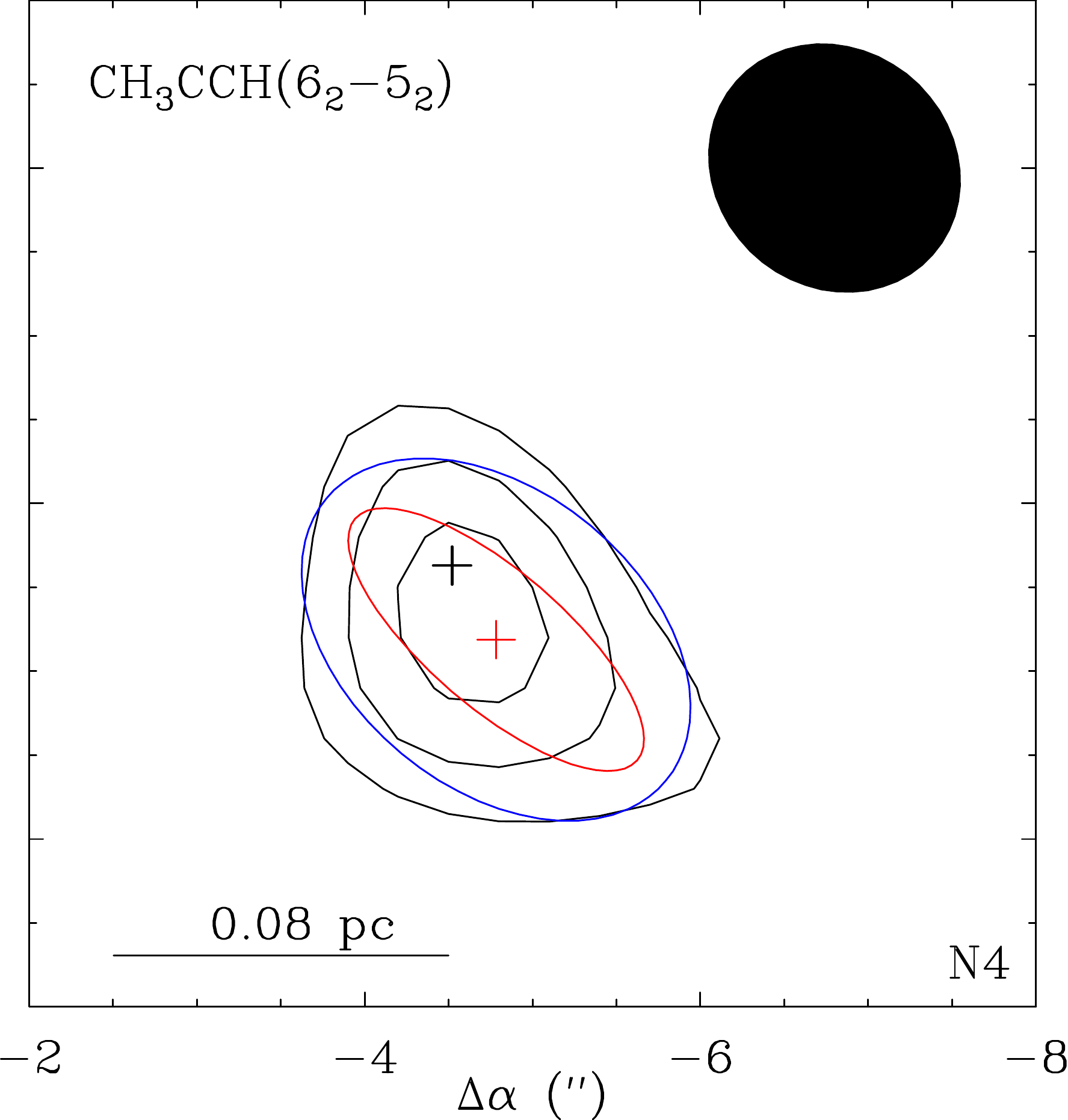} &
   \includegraphics[width=19cm]{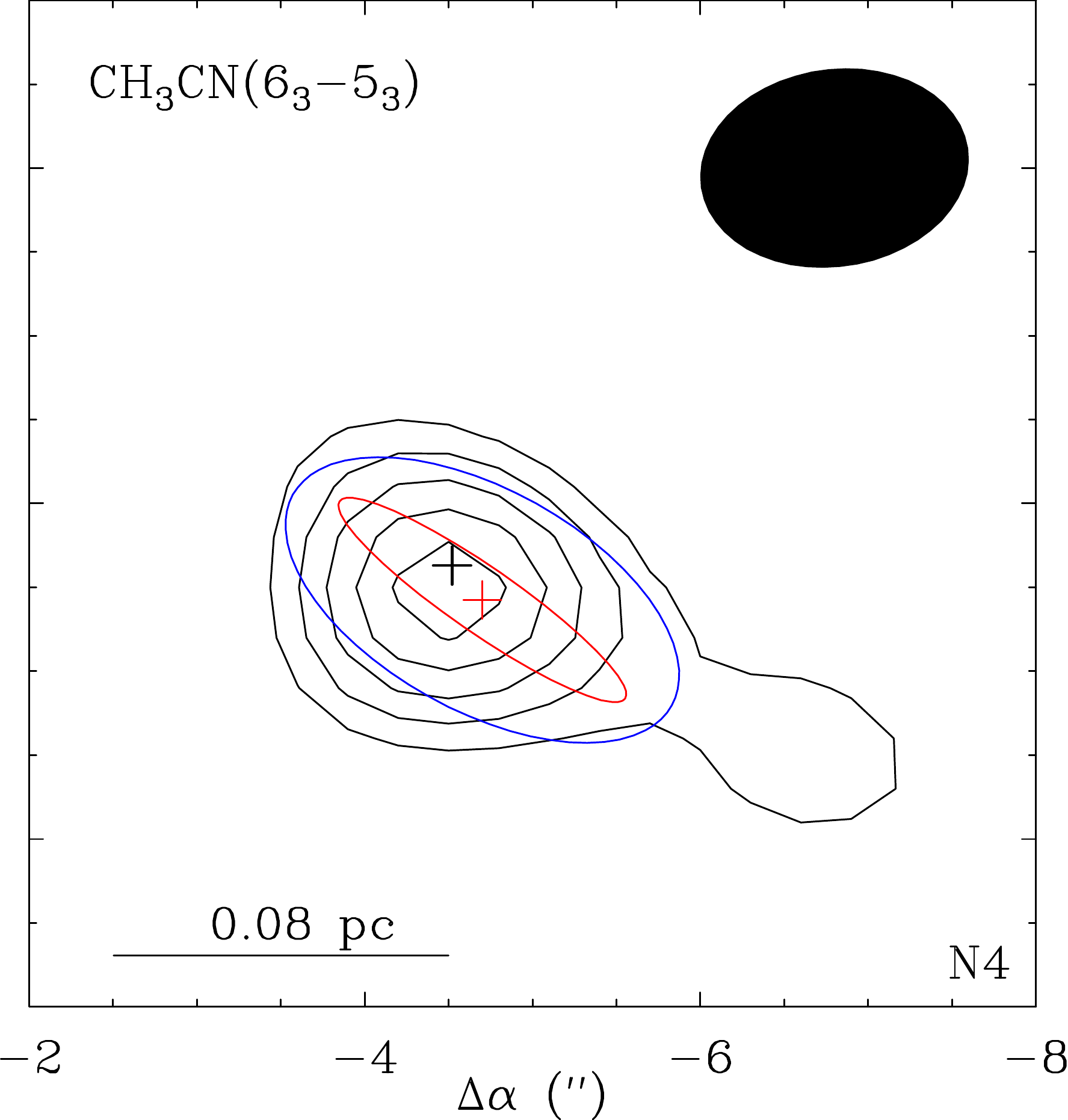} &
   \includegraphics[width=19cm]{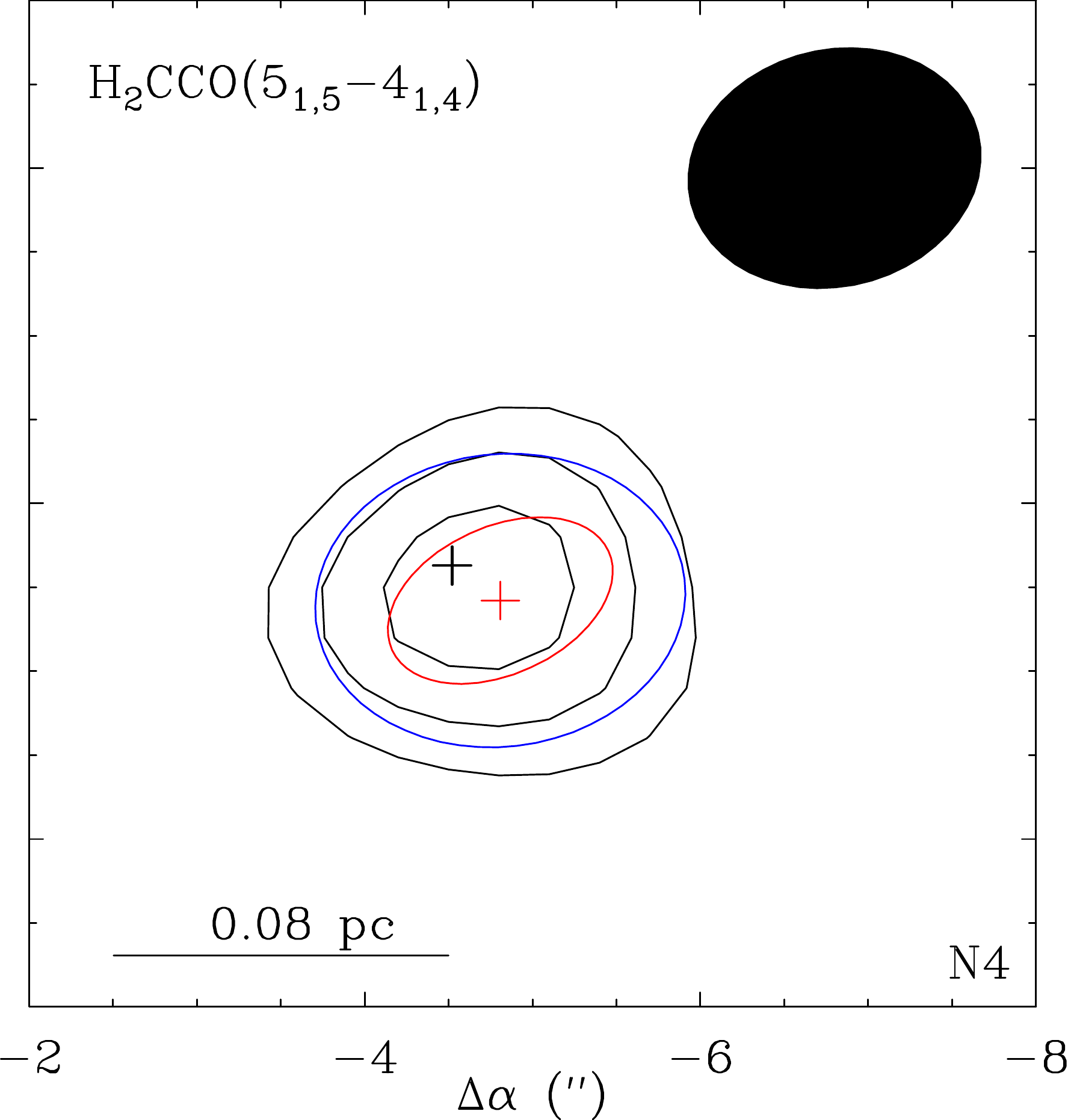} \\
   \includegraphics[width=20.cm]{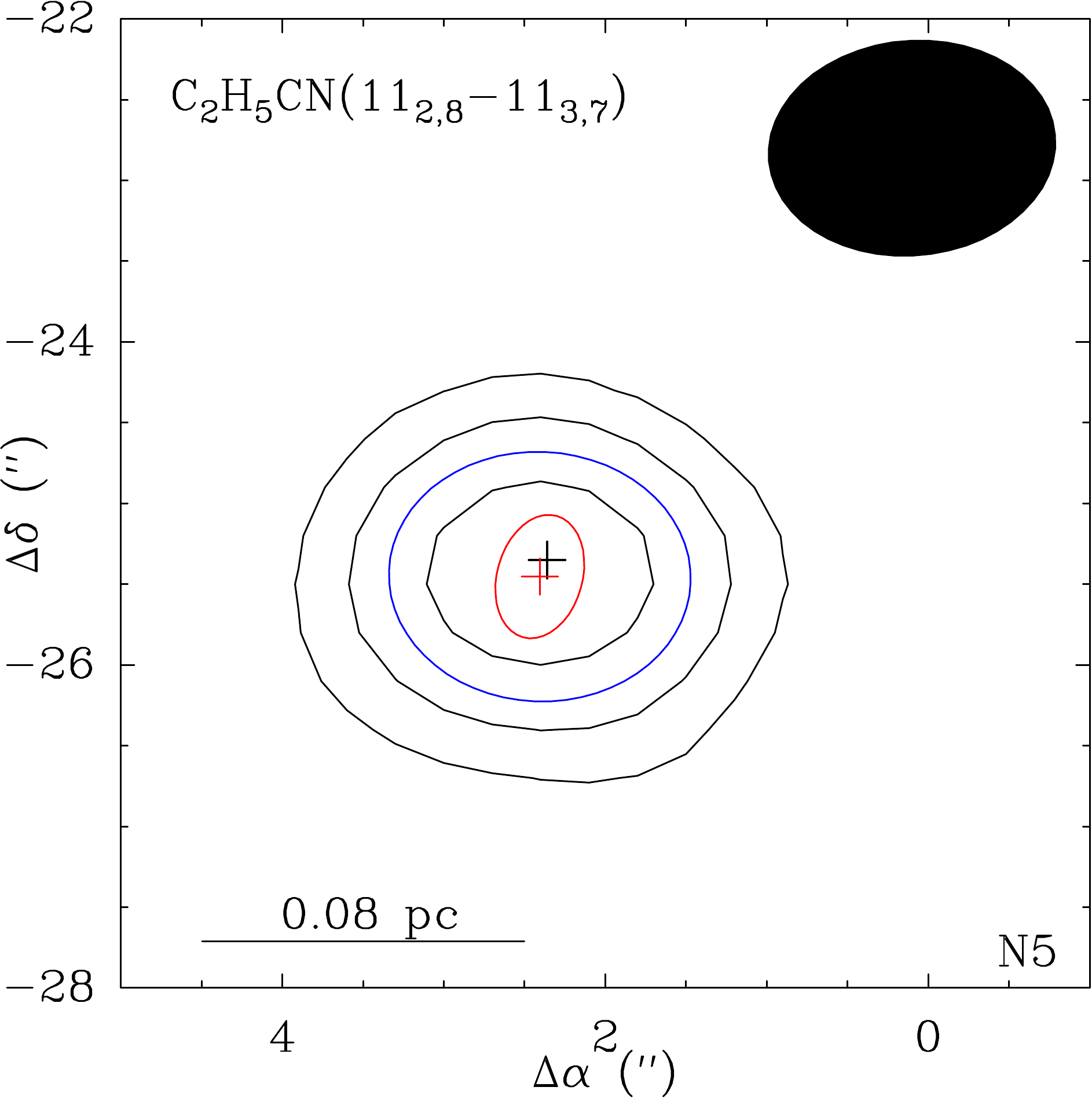} &
   \includegraphics[width=18.0cm]{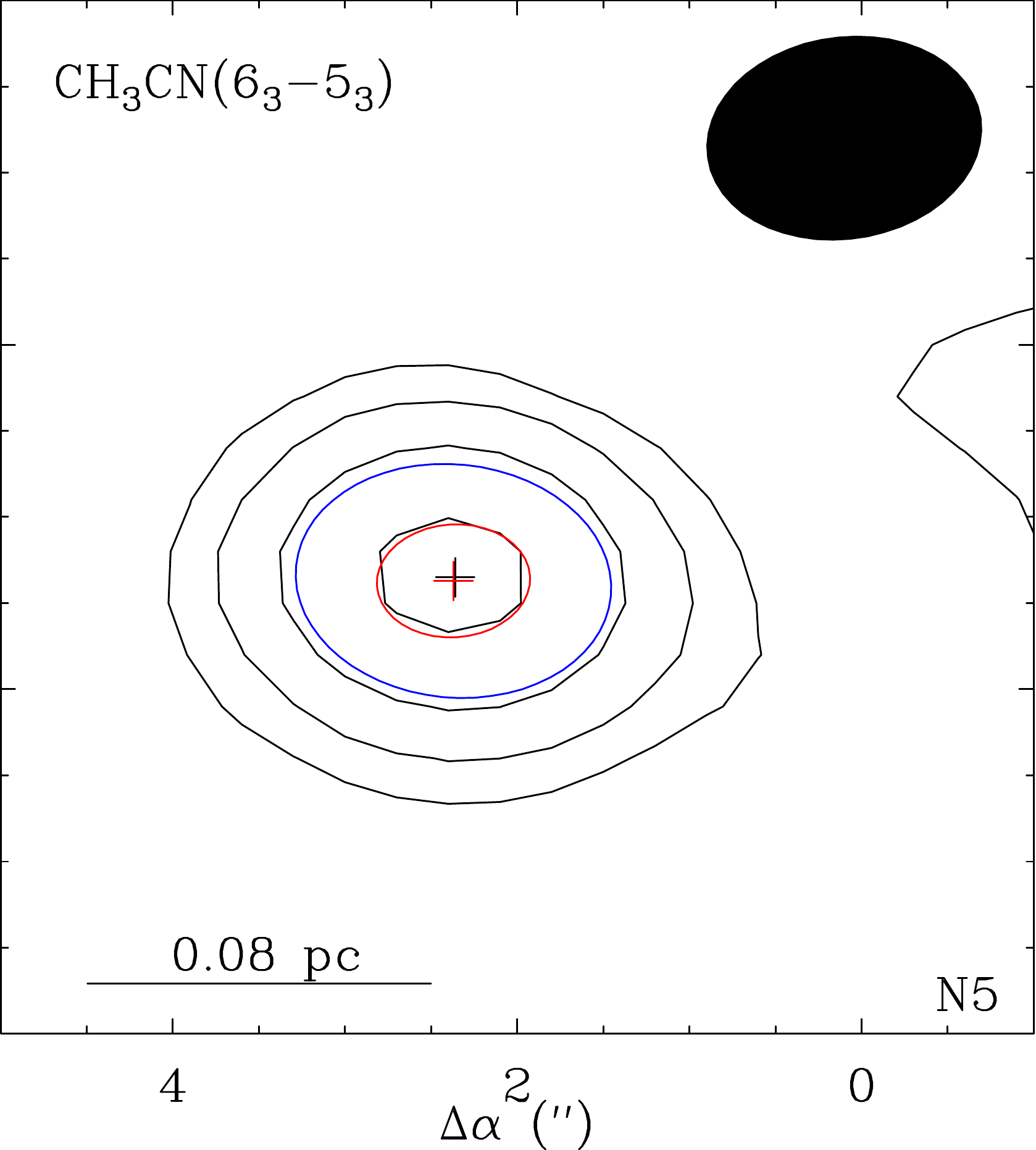} &
   \includegraphics[width=18.0cm]{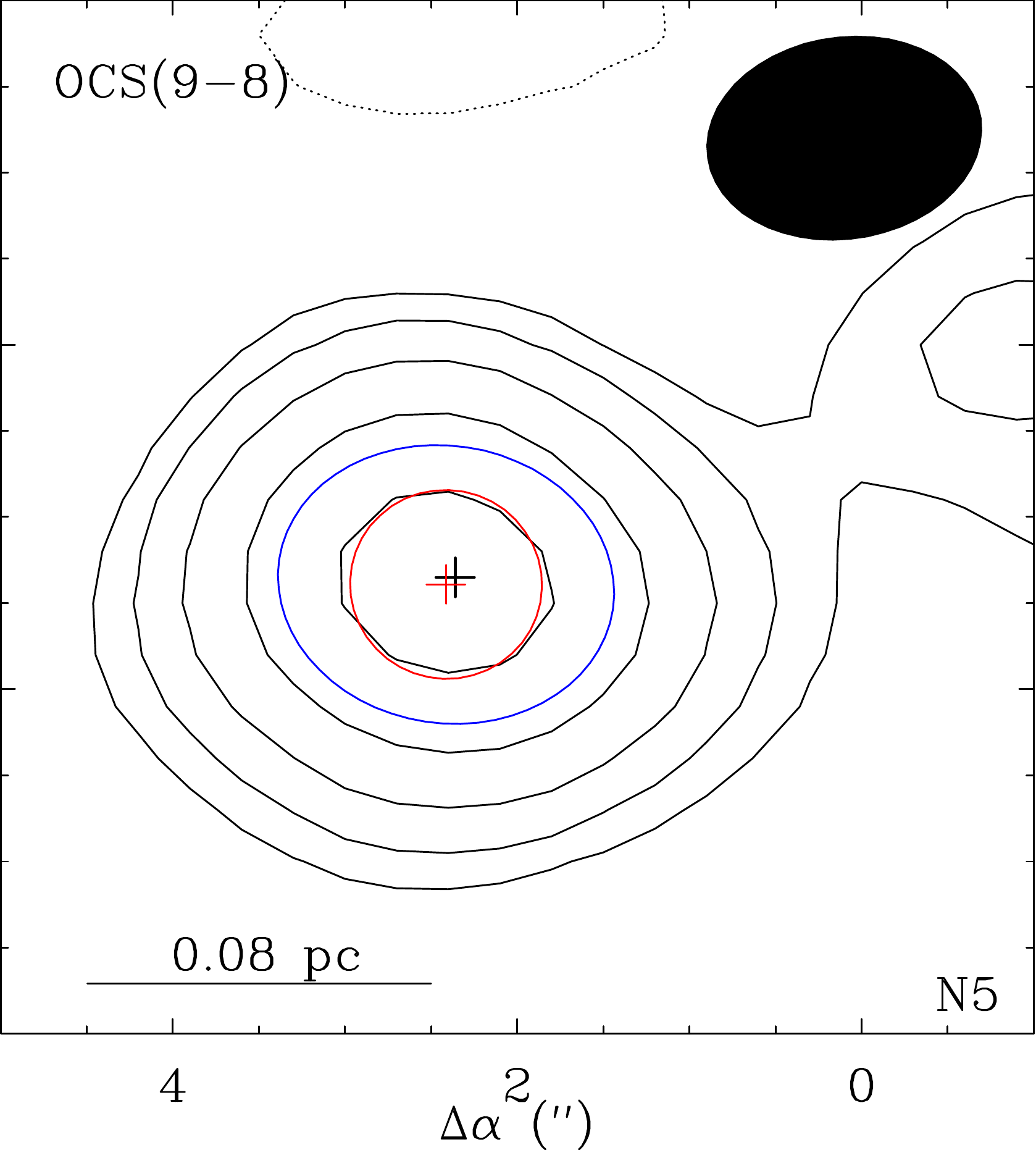} &
   \includegraphics[width=18cm]{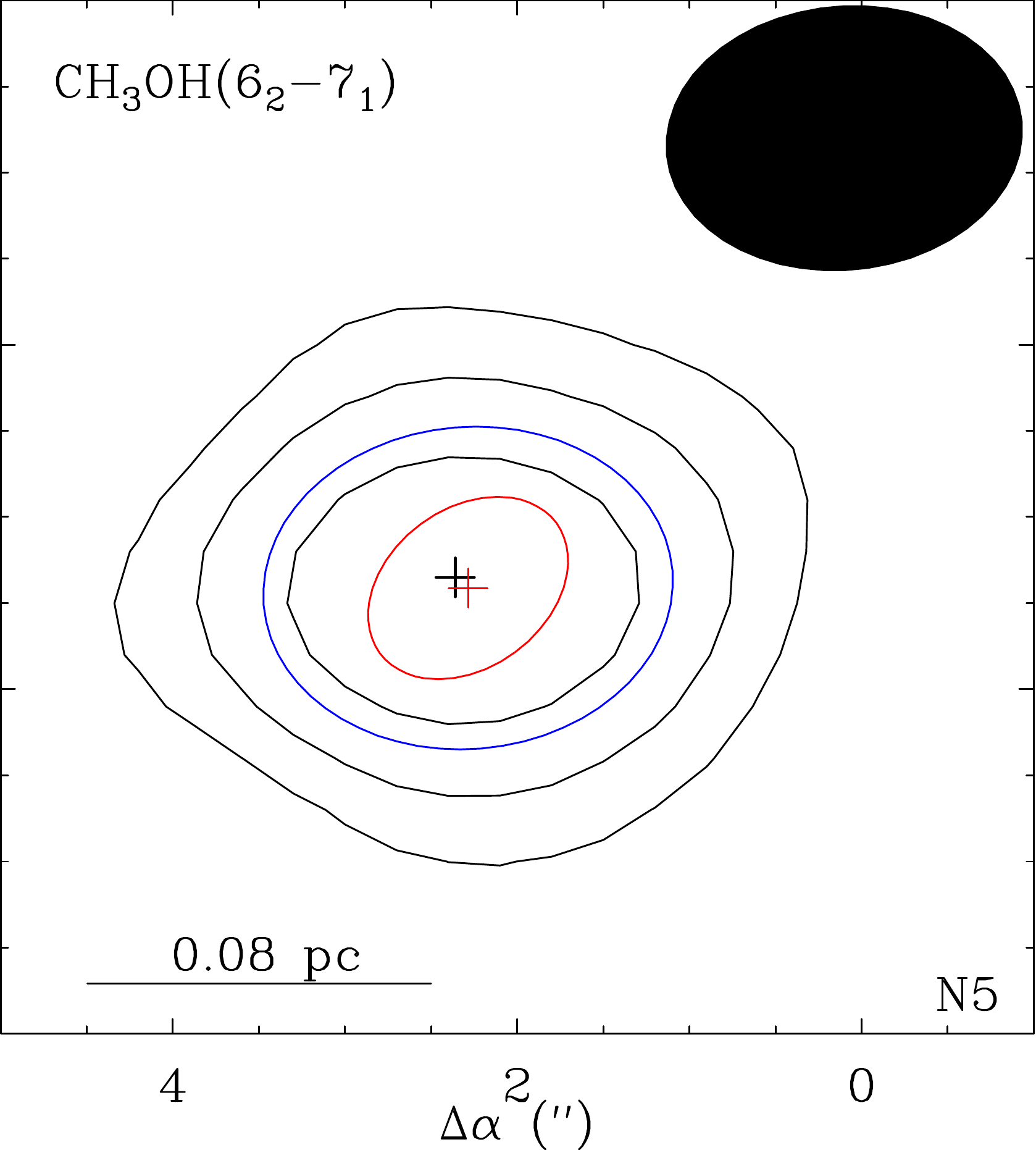} \\  
\end{tabular}}
\caption{\label{IIMap-N3-OCS} Integrated intensity maps of selected transitions toward Sgr~B2(N3) (top row), Sgr~B2(N4) (middle row), and Sgr~B2(N5) (bottom row). For each map the red cross shows the peak position of the emission and the black cross is the reference position of the hot core derived from Fig.~\ref{hot_cores_contour_map}. The blue ellipse represents the result of the Gaussian fit to the map while the red ellipse is the deconvolved emission size. The black filled ellipse represents the synthesized beam. The rms and contour levels are indicated in Table~\ref{contour-maps-size}.}   
\end{figure*}

It is difficult to constrain the size of Sgr~B2(N3) because the emission is unresolved for most species. Only four transitions of OCS and \ce{C2H5CN} show spatially resolved emission. The first row of Fig.~\ref{IIMap-N3-OCS} shows their integrated intensity maps. The result of the Gaussian fit is displayed in blue and the red ellipse shows the deconvolved emission size. The middle row of Fig.~\ref{IIMap-N3-OCS} shows the maps of four transitions for which the emission of Sgr~B2(N4) is resolved. In total for this source, 12 transitions from seven distinct species show resolved emission with a strong signal-to-noise ratio ($\ge$8 up to $\sim$16). Finally the bottom row of Fig.~\ref{IIMap-N3-OCS} shows four maps produced for Sgr~B2(N5). In total for this hot core, 22 transitions from six distinct species show resolved emission with a signal-to-noise ratio from $\sim$10 to 99. We decided to focus only on species showing compact emission, for this reason we did not take into account the transitions of \ce{CH3CCH} detected toward Sgr~B2(N5), also spatially resolved but showing extended emission, with a deconvolved size of $\sim$3.6$\arcsec$, around the position of the hot core.

Table~\ref{gaussian-fit} gives the results of the Gaussian fits to the integrated intensity maps of the lines discussed above that show resolved emission. These results are also plotted in Figs.~\ref{source-size-deconvolved-python-N3}, \ref{source-size-python-N4}, and \ref{source-size-python-N5} for Sgr~B2(N3), Sgr~B2(N4), and Sgr~B2(N5), respectively. For each transition, the deconvolved major and minor diameters of the emission ($\theta_{\rm maj}$ and $\theta_{\rm min}$), given in columns 10 and 11 of Table~\ref{gaussian-fit}, allow us to calculate the average deconvolved size of the emitting region ($\sqrt{ \theta_{maj} \times \theta_{min}}$). The results are given in column 13 of Table~\ref{gaussian-fit}. The mean deconvolved size of each hot core is derived from these values. We obtain sizes of 0.4$\pm$0.1$\arcsec$ for Sgr~B2(N3), 1.0$\pm$0.3$\arcsec$ for Sgr~B2(N4), and 1.0$\pm$0.4$\arcsec$ for Sgr~B2(N5).

\begin{table}[!t]
\begin{center}
  \caption{\label{contour-maps-size} Rms noise levels and contour levels used in Fig.~\ref{IIMap-N3-OCS}.}
  \setlength{\tabcolsep}{1.0mm} 
  \begin{tabular}{clcc}
    \hline
    Source & \multicolumn{1}{c}{Transition}    & $\sigma$ \tablefootmark{a}   &  Levels \tablefootmark{b} \\
    \hline
    \hline
    N3 & OCS(7-6)                           & 49.8 &  4, 8, 16, 22 \\
       & OCS(8-7)                           & 50.8 &               \\
       & \ce{C2H5CN}(10$_{0,11}$-10$_{0,10}$)  & 37.2 &               \\
       & \ce{C2H5CN}(11$_{2,8}$-11$_{3,7}$)   & 37.5 &               \\
    \hline
    N4 & \ce{CH3OH}(13$_2$-12$_3$) & 23.5 & 4, 6, 8, 10, 12  \\
       & \ce{CH3CN}(6$_3$-5$_3$)         & 30.2 &                  \\
       & \ce{H2CCO}(5$_{1,5}$-4$_{1,4}$)   & 17.1 &                  \\
       & \ce{CH3CCH}(6$_2$-5$_2$)        & 20.8 &                  \\
    \hline
    N5 & \ce{C2H5CN}(11$_{2,8}$-11$_{3,7}$) & 36.0 & 4, 8, 16, 32, 64 \\
       & \ce{CH3CN}(6$_3$-5$_3$)          & 47.4 &                  \\
       & OCS(9-8)                        & 20.0 &                  \\
       & \ce{CH3OH}(6$_2$-7$_1$)   & 28.0 &                   \\
        \hline
\end{tabular}
\end{center}
\tablefoot{\tablefoottext{a}{Rms noise level measured in the integrated intensity map in mJy beam$^{-1}$ km s$^{-1}$.}
\tablefoottext{b}{Contour levels in unit of $\sigma$.}}
\end{table}

\begin{sidewaystable*}
\begin{center}
  \caption{\label{gaussian-fit}  Results of elliptical 2D-Gaussian fits to the integrated intensity maps of the transitions with resolved emission.}
    \setlength{\tabcolsep}{1.0mm} 
    \footnotesize
  \begin{tabular}{cllrcclccccccccccccc}
    \hline
 Source & \multicolumn{1}{c}{Molecule}  & \multicolumn{1}{c}{Transition} & \multicolumn{1}{c}{Freq.}      &  $E_{\rm up}$ & $I_{\rm peak}$ \tablefootmark{a} & rms \tablefootmark{a}  &  \multicolumn{1}{c}{SNR} \tablefootmark{b} & Beam \tablefootmark{c} & PA \tablefootmark{c}  & maj. \tablefootmark{d} & min. \tablefootmark{d} &  PA \tablefootmark{d}   &  $\Delta \alpha$ \tablefootmark{e}  & $\Delta \beta$ \tablefootmark{e}  & $\theta _{\rm maj}$ \tablefootmark{f} & $\theta _{\rm min}$ \tablefootmark{f} &  PA \tablefootmark{f}       &  $D_l$ \tablefootmark{g}   \\
  &  &  & \multicolumn{1}{c}{(MHz)} & (K) & \multicolumn{2}{l}{(Jy beam$^{-1}$ km s$^{-1}$)} &  & ($\arcsec$ $\times$ $\arcsec$) & ($^{\rm o}$) & ($\arcsec$) & ($\arcsec$) & ($^{\rm o}$) & ($\arcsec$) & ($\arcsec$)  &  ($\arcsec$) & ($\arcsec$) & ($^{\rm o}$) & ($\arcsec$)   \\
    \hline
    \hline 
N3   & OCS           & 7-6                    & 85139.103 & 16.3  & 1.26(2)  & 0.050  & 25.2 & 2.08$\times$1.54  & -84.6 & 2.09(3) & 1.57(3) & +83.9(23)  & -8.24(2)  & 0.96(1)    & 0.34  & 0.20  & +13.7  & 0.26   \\
     &               & 8-7                    & 97301.208 & 21.0  & 1.49(4)  & 0.051  & 29.2 & 1.83$\times$1.37  & -84.5 & 1.96(3) & 1.42(5) & +81.2(29)  & -8.29(2)  & 0.96(2)    & 0.74  & 0.32  & +64.8  & 0.48   \\
     & \ce{C2H5CN}   & 10$_{0,11}$-10$_{0,10}$   & 96919.762 & 28.1  & 0.98(2)  & 0.037  & 26.5 & 1.83$\times$1.37  & -84.5 & 1.88(3) & 1.40(4) & +84.5(27)  & -8.27(2)  & 0.99(1)    & 0.44  & 0.26  & +89.3  & 0.34   \\
     &               & 11$_{2,8}$-11$_{3,7}$    & 98701.101 & 38.4  & 0.96(2)  & 0.038  & 25.3 & 1.79$\times$1.34  & -84.7 & 1.88(2) & 1.46(3) & +87.6(6)  & -8.30(1)  & 1.02(1)    & 0.64  & 0.52  & -51.4  & 0.58   \\
    \hline
 N4  &  OCS          & 7-6                    & 85139.103 & 16.3  & 0.62(1)  & 0.050  & 12.4 & 2.08$\times$1.54  & -84.6 & 2.38(4) & 1.86(5) & -97.8(34)  & -4.72(2)  & -16.68(2)  & 1.24  & 0.95  & -60.2  & 1.08   \\
     &               & 8-7                    & 97301.208 & 21.0  & 0.64(1)  & 0.055  & 11.6 & 1.83$\times$1.37  & -84.5 & 2.01(5) & 1.75(4) & -50.5(18)  & -4.73(2)  & -16.43(2)  & 1.34  & 0.23  & -22.0  & 0.55   \\
     & \ce{CH3OH}    & 13$_2$-12$_3$    & 100638.872 & 233.6 & 0.22(1)  & 0.024  & 9.2  & 1.72$\times$1.40  & -75.5 & 2.13(3) & 1.63(2) & -59.5(4)  & -4.48(1)  & -16.53(1)  & 1.48  & 0.26  & -45.4  & 0.62  \\
     & \ce{CH3OCH3}  & 4$_{1,4}$-3$_{0,3}$ \tablefootmark{*} & 99324.362 & 10.2  & 0.32(1)  & 0.020  & 16.0 & 1.77$\times$1.42  & -76.2 & 2.09(4) & 1.57(6) & -86.6(31)  & -4.45(2)  & -16.58(2)  & 1.20  & 0.51  & -70.5  & 0.78   \\
     &               & 7$_{0,7}$-6$_{1,6}$ \tablefootmark{*} & 111782.600 & 25.2  & 0.43(1)  & 0.053  & 8.1  & 1.58$\times$1.31  & -72.1 & 2.57(3) & 1.78(4) & +80.6(16)  & -4.51(2)  & -16.46(1)  & 2.03  & 1.19  & +83.0  & 1.56   \\
     & \ce{H2CCO}    & 5$_{1,5}$-4$_{1,4}$      & 100094.514 & 27.5  & 0.17(1)  & 0.017  & 10.0  & 1.77$\times$1.42  & -76.2 & 2.21(4) & 1.74(9) & -83.4(44)  & -4.81(2)  & -16.58(2)  & 1.42  & 0.88  & -65.4  & 1.12   \\ 
     &               & 5$_{1,4}$-4$_{1,3}$      & 101981.429 & 27.7  & 0.17(1)  & 0.018 & 9.4  & 1.72$\times$1.40  & -75.5 & 2.59(4) & 1.80(6) & -85.5(27)  & -4.64(3)  & -16.60(2)  & 1.97  & 1.06  & -79.1  & 1.44   \\
     & \ce{CH3CN}    & 6$_3$-5$_3$            & 110364.354 & 82.8  & 0.36(1)  & 0.030  & 12.0 & 1.61$\times$1.18  & -81.7 & 2.58(1) & 1.32(1) & +60.9(6)  & -4.70(2)  & -16.58(1)  & 2.07  & 0.39  & +55.3  & 0.90  \\
     &               & 6$_2$-5$_2$            & 110374.989 & 47.1  & 0.41(1)  & 0.031  & 13.2 & 1.61$\times$1.18  & -81.7 & 2.06(5) & 1.28(11) & +62.1(38)  & -4.55(4)  & -16.60(3)  & 1.37  & 0.20  & +49.9  & 0.52   \\
     & \ce{CH3CCH}   & 6$_3$-5$_3$            & 102530.348 & 82.3  & 0.18(1)  & 0.022  & 8.2  & 1.57$\times$1.42  & +48.6 & 3.39(1) & 1.63(4) & +38.7(2)  & -4.59(1)  & -16.57(1)  & 3.07  & 0.43  & +38.9  & 1.15   \\
     &               & 6$_2$-5$_2$            & 102540.145 & 46.1  & 0.19(1)  & 0.021  & 9.0  & 1.57$\times$1.42  & +48.6 & 2.63(6) & 1.77(3) & +50.3(24)  & -4.78(2)  & -16.81(2)  & 2.21  & 0.81  & +49.4  & 1.34   \\
     & \ce{CH3OCHO}  & 8$_{3,5}$-7$_{3,4}$      & 100308.179 & 27.4  & 0.14(1)  & 0.017  & 8.2  & 1.77$\times$1.42  & -76.2 & 2.40(3) & 1.70(6) & -80.7(31)  & -4.32(3)  & -16.57(2)  & 1.70  & 0.79  & -70.3  & 1.16   \\
    \hline
N5   &  \ce{C2H5CN}  & 10$_{0,11}$-10$_{0,10}$  & 96919.762  & 28.1  & 0.85(1)  & 0.032  & 26.9 & 1.83$\times$1.37  & -84.5 & 1.85(2) & 1.56(2) & -87.4(30)  & 2.39(1)   & -25.43(1)  & 0.78  & 0.10  & -19.3  & 0.28   \\   
     &               & 11$_{2,10}$-11$_{2,9}$   & 98177.574  & 32.8  & 0.86(1)  & 0.035  & 24.6 & 1.79$\times$1.34  & -84.7 & 1.85(2) & 1.51(2) & +87.1(1)  & 2.38(1)   & -25.47(1)  & 0.7   & 0.46  & -15.2  & 0.57   \\
     &               & 11$_{2,8}$-11$_{3,7}$    & 98701.101  & 38.4  & 0.85(1)  & 0.036  & 23.6 & 1.79$\times$1.34  & -84.7 & 1.87(2) & 1.55(2) & +87.7(29)  & 2.41(1)   & -25.45(1)  & 0.78  & 0.53  & -15.5  & 0.64   \\
     &               & 11$_{2,9}$-10$_{2,8}$    & 99681.461  & 33.0  & 0.95(1)  & 0.038  & 25.0 & 1.79$\times$1.34  & -84.7 & 1.84(2) & 1.57(2) & +86.3(1)  & 2.39(1)   & -25.43(1)  & 0.81  & 0.43  & -8.6   & 0.59   \\
     & \ce{CH3OCH3}  & 15$_{2,13}$-15$_{2,14}$ \tablefootmark{*}   & 88706.231  & 116.9 & 0.90(1)  & 0.062  & 14.5 & 1.91$\times$1.65  & +86.3 & 2.27(3) & 1.78(4) & +58.2(28)  & 2.19(2)   & -25.60(1)  & 1.39  & 0.18  & +44.4  & 0.50   \\
     & \ce{CH3CN}    & 5$_3$-4$_3$            & 91971.130  & 77.5  & 1.29(1)  & 0.067  & 19.3 & 2.88$\times$1.45  & +83.5 & 3.79(1) & 2.09(1) & +76.0(1)  & 2.38(3)   & -25.52(1)  & 2.73  & 0.90  & +57.3  & 1.57   \\  
     &               & 6$_3$-5$_3$            & 110364.354 & 82.8  & 1.81(3)  & 0.047  & 38.5 & 1.61$\times$1.18  & -81.7 & 1.84(2) & 1.35(2) & +84.4(15)  & 2.37(1)   & -25.37(1)  & 0.89  & 0.66  & -86.4  & 0.76   \\
     &               & 6$_2$-5$_2$            & 110374.989 & 47.1  & 1.65(2)  & 0.051  & 32.4 & 1.61$\times$1.18  & -81.7 & 2.19(2) & 1.45(3) & +80.6(17)  & 2.33(1)   & -25.37(1)  & 1.48  & 0.83  & +79.7  & 1.11   \\
     &  OCS          & 8-7                   & 91301.208  & 21.0  & 1.94(2)  & 0.026  & 74.6 & 1.82$\times$1.33  & -81.8 & 2.25(2) & 1.75(2) & +84.5(16)  & 2.41(1)   & -25.37(1)  & 1.32  & 1.13  & -86.4  & 1.22   \\   
     &               & 9-8                   & 109463.063 & 26.3  & 1.98(2)  & 0.020  & 99.0 & 1.61$\times$1.18  & -81.7 & 1.96(1) & 1.61(2) & +81.6(17)  & 2.41(1)   & -25.39(1)  & 1.12  & 1.10  & +79.1  & 1.11   \\
     & \ce{CH3OH}    & 6$_2$-7$_1$     & 85568.131  & 74.7  & 0.75(1)  & 0.028  & 26.8 & 2.08$\times$1.54  & -84.6 & 2.38(2) & 1.87(2) & -85.2(15)  & 2.29(1)   & -25.41(1)  & 1.27  & 0.92  & -53.7  & 1.08   \\    
     &               & 8$_0$-7$_1$     & 95169.391  & 83.5  & 1.00(2)  & 0.049  & 20.4 & 1.86$\times$1.36  & -82.1 & 2.26(2) & 1.45(5) & -84.8(16)  & 2.28(2)   & -25.31(1)  & 1.35  & 0.33  & -72.6  & 0.66   \\    
     &               & 2$_{1,0}$-1$_{1,0}$     & 96755.501  & 28.0  & 0.48(2)  & 0.026  & 18.5 & 1.86$\times$1.36  & -82.1 & 3.34(5) & 1.78(17) & -89.6(20)  & 2.20(5)   & -25.43(3)  & 2.78  & 1.12  & -87.6  & 1.77  \\  
     & \ce{CH3OCHO}  & 7$_{2,6}$-6$_{2,5}$(E)\tablefootmark{**} &  84449.169  & 19.0  & 0.14(1)  & 0.014  &  10.0 & 2.08$\times$1.54  & -84.6 & 3.42(5) & 2.03(8) & +82.6(15)  & 2.00(2)   & -25.47(1)  & 2.71  & 1.33  & +81.9  & 1.9  \\ 
     &               & 7$_{2,6}$-6$_{2,5}$(A) &  84454.754  & 19.0  & 0.14(1)  & 0.012  & 11.7 & 2.08$\times$1.54  & -84.6 & 2.85(3) & 2.02(3) & +88.8(14)  & 2.14(2)   & -25.33(1)  & 1.95  & 1.30  & -87.3  & 1.59  \\
     &               & 7$_{3,5}$-6$_{3,4}$(A) &  86265.796  & 22.5  & 0.14(1)  & 0.010  & 14.0 & 2.04$\times$1.51  & -83.3 & 2.41(1) & 1.75(1) & -70.7(1)   & 2.39(1)   & -25.35(1)  & 1.52  & 0.38  & -49.2  & 0.76  \\ 
     &               & 7$_{3,5}$-6$_{3,4}$(E)  & 86268.739  & 22.5  & 0.15(1)  & 0.010  & 15.0 & 1.86$\times$1.36  & -82.1 & 2.85(1) & 1.69(1) & -74.4(1)  & 2.53(1)   & -25.33(1)  & 2.08  & 0.45  & -65.0  & 0.97  \\
     &               & 7$_{1,6}$-6$_{1,5}$(A) &  88851.607  & 17.9  & 0.21(1)  & 0.019  & 11.1 & 1.91$\times$1.65  & +96.3 & 2.37(6) & 1.79(6) & +78.6(42)  & 2.05(3)   & -25.56(2)  & 1.44  & 0.62  & -70.7  & 0.94  \\
     &               & 7$_{2,5}$-6$_{2,4}$(E)  & 90145.723  & 19.7  & 0.18(1)  & 0.013  & 13.8 & 1.77$\times$1.58  & +51.5 & 2.11(5) & 1.83(5) & -89.3(107) & 2.14(3)   & -25.37(1)  & 1.30  & 0.74  & +73.5  & 0.98  \\
     &               & 7$_{2,5}$-6$_{2,4}$(A)  & 90156.473  & 19.7  & 0.16(1)  & 0.012  & 13.3 & 1.77$\times$1.58  & +51.5 & 2.65(10) & 2.00(13) & +87.7(62)  & 2.13(4)   & -25.42(1)  & 2.05  & 1.10  & +81.4  & 1.50  \\  
     &               & 8$_{5,3}$-7$_{5,2}$(E) &  98424.207  & 37.9  & 0.14(1)  & 0.012  & 11.7 & 1.79$\times$1.34  & -84.7 & 1.95(4) & 1.51(5) & -86.6(30)  & 2.38(2)   & -25.49(2)  & 0.88  & 0.56  & -52.7  & 0.70  \\ 
     &               & 8$_{1,7}$-7$_{1,6}$(E) & 100482.241  & 22.8  & 0.25(1)  & 0.016  & 15.6 & 1.77$\times$1.42  & -76.2 & 2.10(2) & 1.56(3) & +83.4(23)  & 2.50(2)   & -25.27(1)  & 1.15  & 0.60  & -88.2  & 0.83  \\
    
\hline
\end{tabular}
\end{center}
\tablefoot{Numbers in parentheses are uncertainties given by the fitting routine GAUSS-2D in units of the last digits.}
\tablefoottext{a}{Peak flux density and noise level measured in the integrated intensity map.}
\tablefoottext{b}{Signal-to-noise ratio.}
\tablefoottext{c}{Size of synthesized beam (HPBW) and position angle (East from North).}
\tablefoottext{d}{Size (FWHM) and position angle of the fitted Gaussian.}
\tablefoottext{e}{Equatorial offset with respect to phase center (see Sect.~\ref{obs-ALMA}).}
\tablefoottext{f}{Deconvolved major and minor diameters of the emission (FWHM) and position angle.}
\tablefoottext{g}{Average deconvolved size of the emitting region.}
\tablefoottext{*}{The line is a group of transitions from the same molecule blended together.}
\tablefoottext{**}{The A and E labels mark the two substates of the ground torsional level of methyl formate.}
\end{sidewaystable*}

         \subsection{Rotational temperature}
                    \label{Trot}


Population diagrams are plotted based on transitions that are well detected and not severely contaminated by lines from other species to derive rotational temperatures. We use the following equation \citep{snyder2005} to compute ordinate values (upper level column density divided by statistical weight, $N_u$/$g_u$), assuming optically thin emission in first approximation: 

\begin{equation}
\label{eq-rot-diagram}
\ln \left( \frac{N_u}{g_u} \right) = \ln \left( \frac{8\pi k_b \nu^2 W}{h c^3 A_{\rm ul} B g_u}  \right) = -\frac{E_u}{k_b T_{\rm rot}} + \ln \left( \frac{N_{\rm mol}}{Z}  \right),
\end{equation}
where $k_b$ is the Boltzmann constant, $\nu$ the frequency, $W$ the integrated intensity in brightness temperature scale, $h$ the Planck constant, $c$ the speed of light, $A_{\rm ul}$ the Einstein coefficient for spontaneous emission, $B = \frac{\text{source size} ^2}{\text{beam size} ^2 + \text{source size} ^2}$ the beam filling factor, $g_u$ the statistical weight of the upper level, $E_{\rm u}$ the upper level energy, $T_{\rm rot}$ the rotational temperature, $Z$ the partition function, and $N_{\rm mol}$ the molecular column density.

\begin{figure}[!t]
\begin{center}
 \includegraphics[width=\hsize]{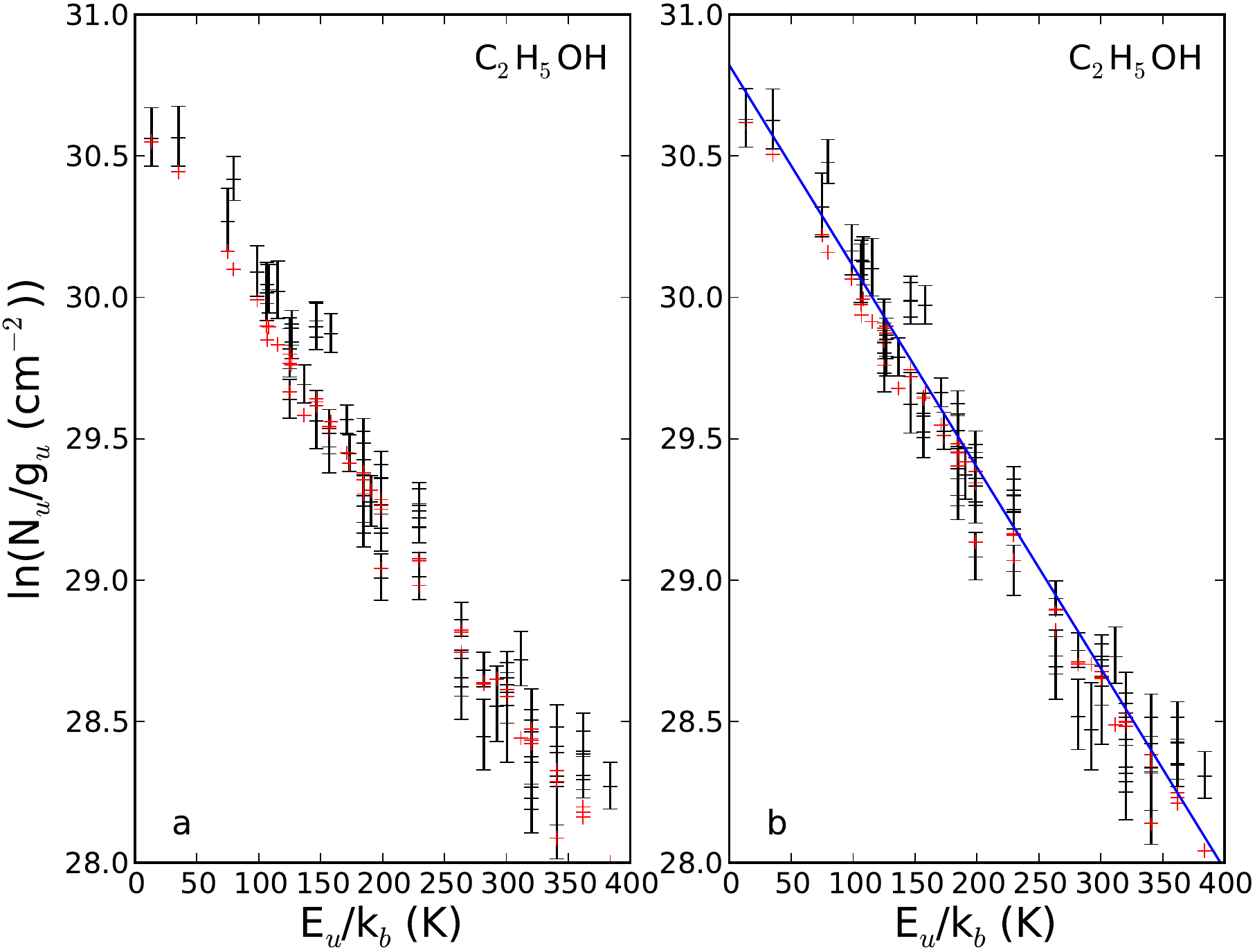}
 \includegraphics[width=\hsize]{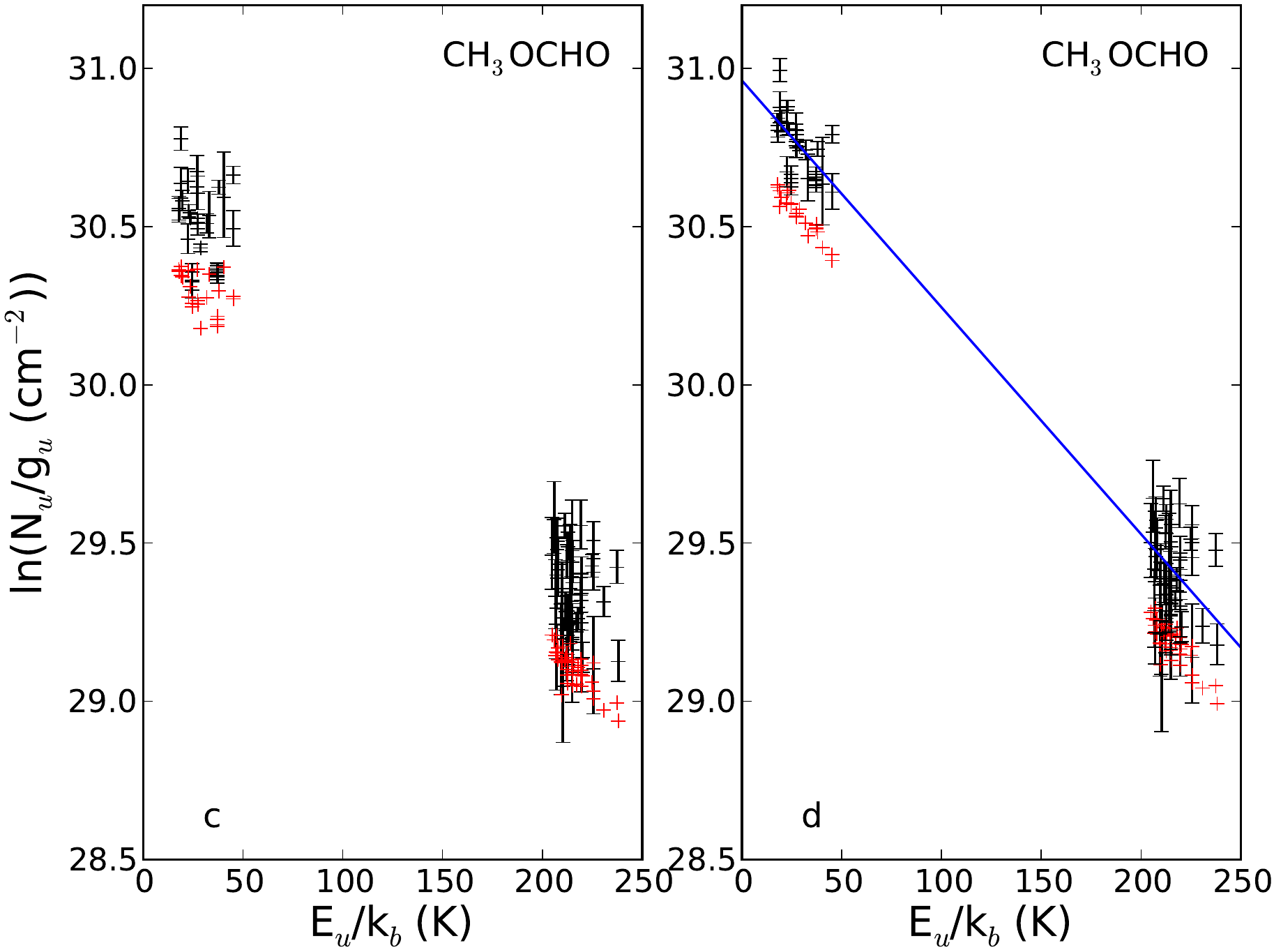} \\
 \caption{\label{popdiagram-N3} Population diagrams of \ce{C2H5OH} (panels a and b) and \ce{CH3OCHO} (panels c and d) for Sgr~B2(N3). The black points are computed using the integrated intensities of the observed spectrum while the red points are computed using the integrated intensities of our synthetic model. The error bars on the observed data points are 1$\sigma$ uncertainties on $N_u$/$g_u$. No correction is applied in panels a and c, while in panels b and d the optical depth correction has been applied to both the observed and synthetic populations and the contamination from all other species included in the full LTE model has been removed from the observed data points. The blue line is the weighted linear fit to the observed populations.}
\end{center}
\end{figure}

\begin{figure}[!t]
\begin{center}
 \includegraphics[width=\hsize]{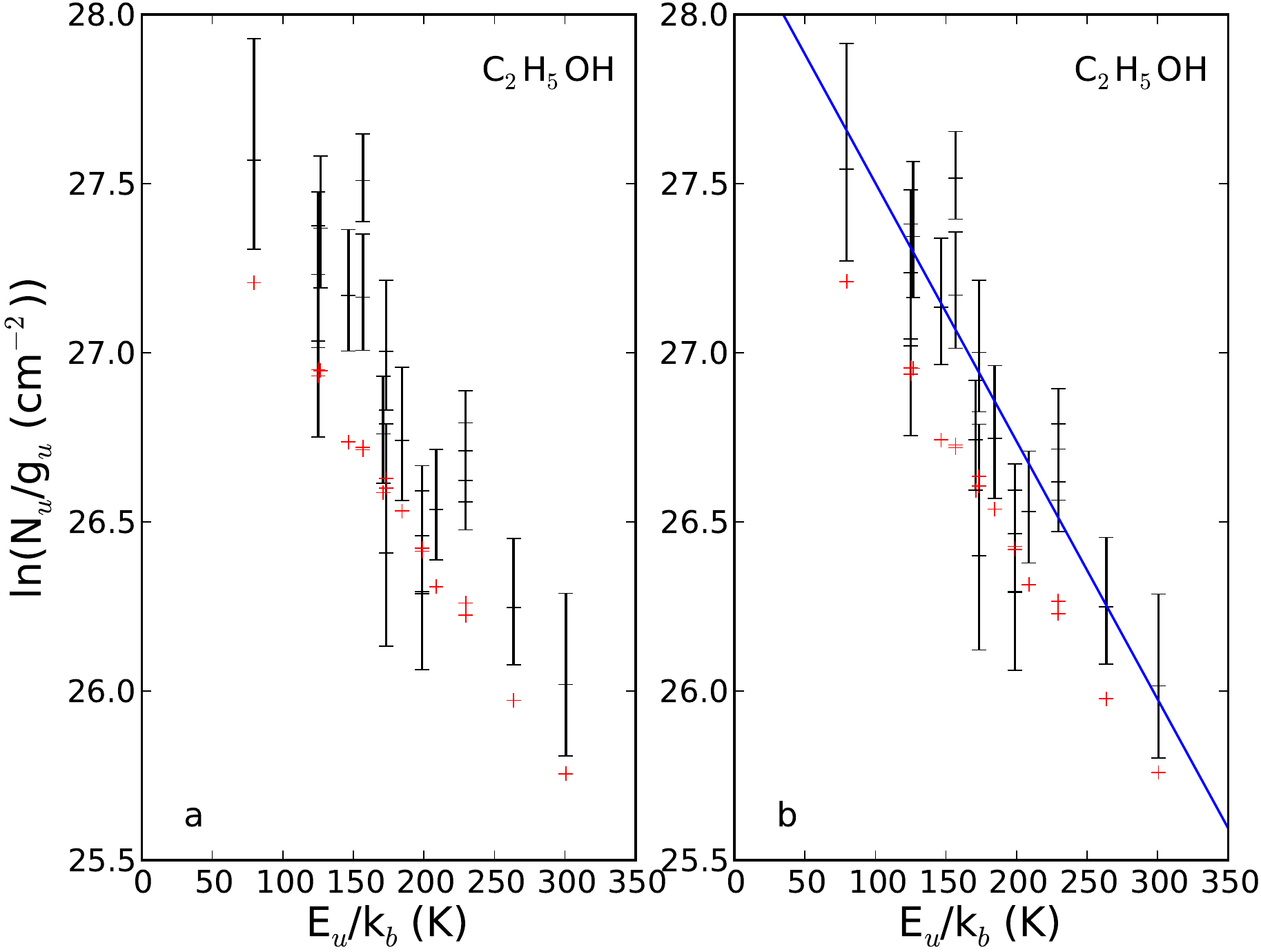}
 \includegraphics[width=\hsize]{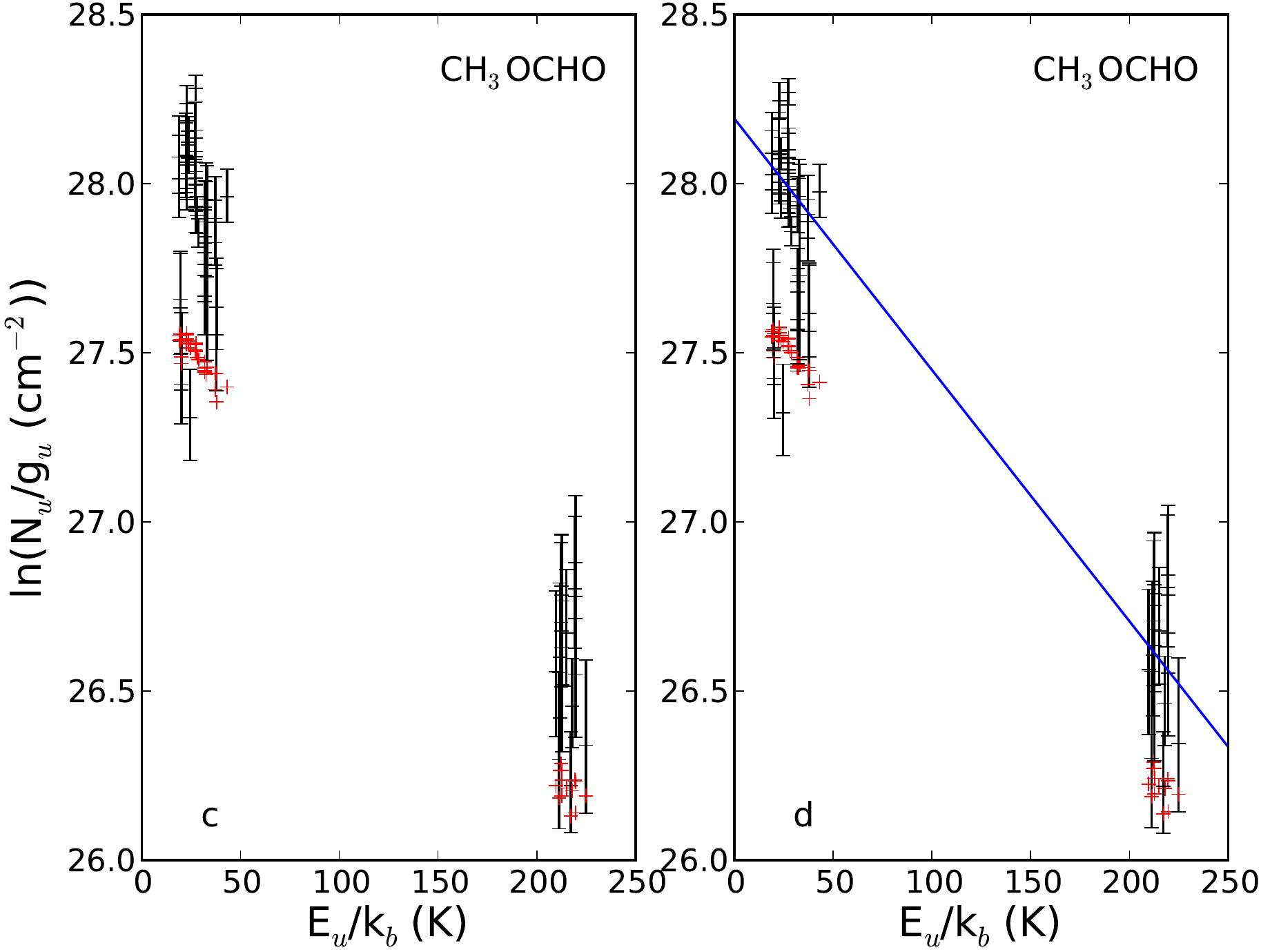} \\
 \caption{\label{popdiagram-N4} Same as Fig.~\ref{popdiagram-N3} but for Sgr~B2(N4).}
\end{center}
\end{figure}

\begin{figure}[!t]
\begin{center}
 \includegraphics[width=\hsize]{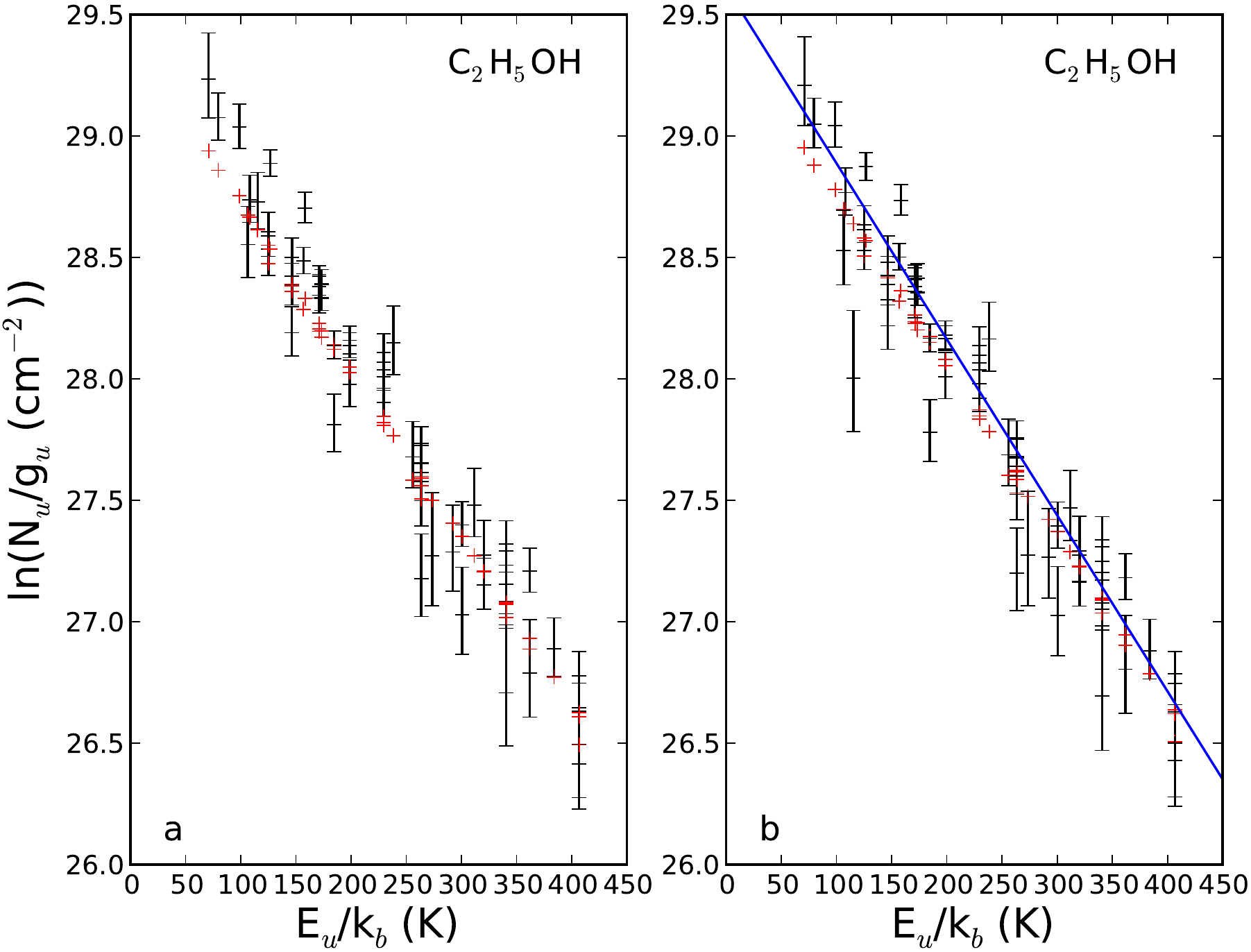}
 \includegraphics[width=\hsize]{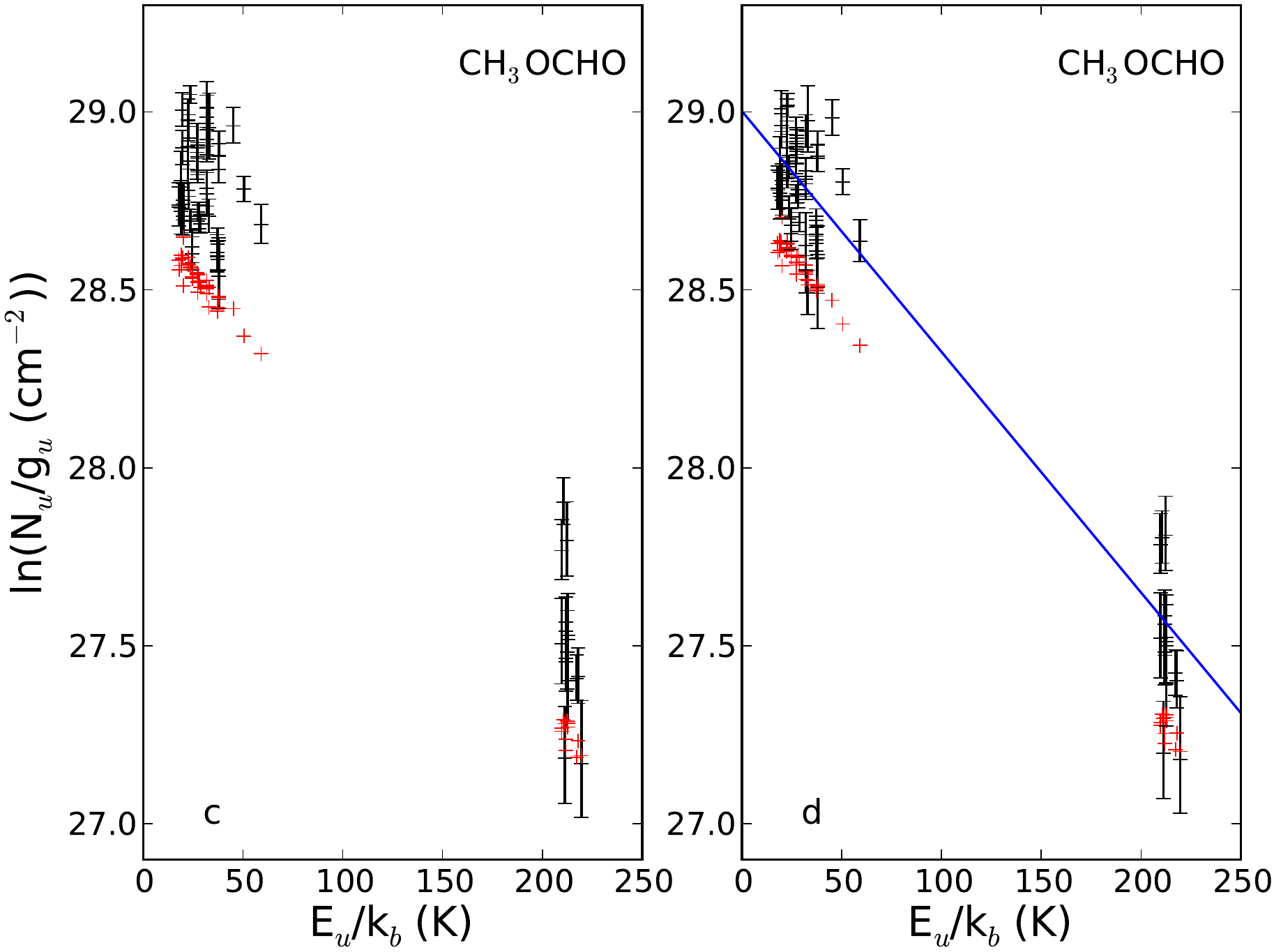} \\
 \caption{\label{popdiagram-N5} Same as Fig.~\ref{popdiagram-N3} but for Sgr~B2(N5).}
\end{center}
\end{figure}

We present here the results of this analysis for ethanol and methyl formate, two species that have many well detected lines spread over a large energy range (see Table~\ref{best-fit-parameters}). To derive the rotational temperature of \ce{CH3OCHO}, we use both its ground and first vibrationally excited states, modelled with the same parameters ($N_{\rm mol}$, $T_{\rm rot}$, source size, $v_{\rm off}$, $\Delta v$). Only the ground state transitions are used for ethanol. Figures~\ref{popdiagram-N3}a,c, \ref{popdiagram-N4}a,c, and \ref{popdiagram-N5}a,c show the population diagrams of both species for Sgr~B2(N3), Sgr~B2(N4), and Sgr~B2(N5), respectively. In Figs.~\ref{popdiagram-N3}b,d, \ref{popdiagram-N4}b,d, and \ref{popdiagram-N5}b,d we applied to both the observed and synthetic populations the opacity correction factor $C_{\tau} = \frac{\tau}{1-e^{- \tau}}$ \citep[see][]{goldsmith1999, snyder2005} using the opacities of our radiative transfer model. In all cases the opacity correction only slightly affects the population diagram (see values of $\tau_{\rm max}$ in column 5 of Table~\ref{tab-Trot}). Since a transition can be partially contaminated by other species, we also subtracted from the measured integrated intensities the contribution of contaminating molecules using our model that includes all species identified so far. The synthetic and observed points are closer to each other after this correction, however the synthetic data points are globally below the observed ones. This can be explained by our modelling procedure in which we try to not overestimate the observed peak flux density of each spectral line. The modelled spectrum can then well reproduce the observed spectrum in terms of peak flux density, but it will not necessarily exactly fit the whole line profile because we use a single linewidth to model all detected lines. In addition one has to keep in mind that even after removing the contamination from other species, the measured integrated intensities can still be affected by residual contamination from U-lines. 

The synthetic data points are not affected by contamination from other species and should be strictly aligned. The residual dispersion of the synthetic data points seen in Figs.~\ref{popdiagram-N3}b,d, \ref{popdiagram-N4}b,d, and \ref{popdiagram-N5}b,d can be explained by the frequency boundaries set to integrate the intensity which are a compromise between covering the line as much as possible and limiting the contamination from other species emitting at nearby frequencies in the observed spectrum. 

After correction for the opacity and contamination from other species, the observed data points are roughly aligned and can be fitted by a single straight line, meaning one single temperature component. The blue line in Figs.~\ref{popdiagram-N3}b,d, \ref{popdiagram-N4}b,d, and \ref{popdiagram-N5}b,d shows the weighted linear fit to the observed data points. According to Eq.~\ref{eq-rot-diagram} and considering the LTE approximation, the inverse of the slope gives us the kinetic temperature of the hot cores. The results are presented in Table~\ref{tab-Trot} and show that all hot cores have temperatures around 140~K, also consistent with the kinetic temperature of Sgr~B2(N2) \citep{muller2016,belloche2016}. 

For all the species emitting numerous lines spread over a broad energy range,  we use the same method to plot population diagrams and derive rotational temperatures. For the rest of the molecules, we set the temperature to 145~K in our LTE model which allows in most cases to reproduce well the observed spectra of the three new hot cores.

\begin{table}[!t]
\begin{center}
  \caption{\label{tab-Trot} Rotational temperatures derived from population diagrams.}
  \setlength{\tabcolsep}{1.5mm} 
  \begin{tabular}{cllccc}
    \hline
    Source    & \multicolumn{1}{c}{Species}    & \multicolumn{1}{c}{States \tablefootmark{a}} & $N_{\rm l}$\tablefootmark{b} & $\tau_{\rm max}$ \tablefootmark{c} & $T_{\rm rot}$\tablefootmark{d} \\   
              &            &                           &    &                & (K)         \\
    \hline
    \hline
     N3     & \ce{C2H5OH}  & $\varv$ = 0                & 57   & 0.24 & 140.9(4.6)   \\
            & \ce{CH3OCHO} & $\varv$ = 0, $\varv_{\rm t}$ = 1 & 70   & 0.80 & 140.8(2.6)   \\
     N4     & \ce{C2H5OH}  & $\varv$ = 0                & 18   & 0.01 & 131.1(21.5)  \\
            & \ce{CH3OCHO} & $\varv$ = 0, $\varv_{\rm t}$ = 1 & 47   & 0.04 & 134.6(10.5)  \\ 
     N5     & \ce{C2H5OH}  & $\varv$ = 0                & 50   & 0.07 & 138.1(5.6)  \\
            & \ce{CH3OCHO} & $\varv$ = 0, $\varv_{\rm t}$ = 1 & 58   & 0.15 & 148.6(8.0)  \\
    \hline
     N2\tablefootmark{*}   & \ce{C2H5OH}  & $\varv$ = 0 & 156  & 0.87 & 139.6(1.6)   \\
            & \ce{CH3OCHO} & $\varv$ = 0, $\varv_{\rm t}$ = 1 & 106  & 0.70 & 142.4(4.4)   \\
    \hline
\end{tabular}
\end{center}
\tablefoot{\tablefoottext{a}{Vibrational or torsional states taken into account to fit the population diagrams.}
\tablefoottext{b}{Number of lines plotted in the population diagram.}
\tablefoottext{c}{Maximum opacity.}
\tablefoottext{d}{Rotational temperature derived from the fit. The standard deviation is indicated in parentheses.}
\tablefoottext{*}{Sgr~B2(N2)'s parameters are from \citet{muller2016} and \citet{belloche2016}}.}
\end{table}

           \subsection{Kinematic structure of the hot cores}
                       \label{kinematic-structure}

We derived the systemic velocity of the faint hot cores embedded in Sgr~B2(N) by fitting 1D Gaussians to numerous well detected transitions over the whole spectrum. We obtained a velocity of 74~km~s$^{-1}$ for Sgr~B2(N3), 64~km~s$^{-1}$ for Sgr~B2(N4), and 60~km~s$^{-1}$ for Sgr~B2(N5). Sgr~B2(N3) and Sgr~B2(N4) have the same velocity as Sgr~B2(N2) and  Sgr~B2(N1), respectively.  
     
These fits also give information on the typical linewidth of each species. The three hot cores have a median linewidth of about 5.0~km~s$^{-1}$, ranging from 4.0 to 7.7~km s$^{-1}$ for Sgr~B2(N3), from 3.5 to 5.5~km~s$^{-1}$ for Sgr~B2(N4), and from 3.5 to 6.8~km~s$^{-1}$ for Sgr~B2(N5) (see Table~\ref{best-fit-parameters}).  

\begin{figure*}[!t]
\resizebox{\hsize}{!}
{\begin{tabular}{cccc}
   \includegraphics[width=\hsize]{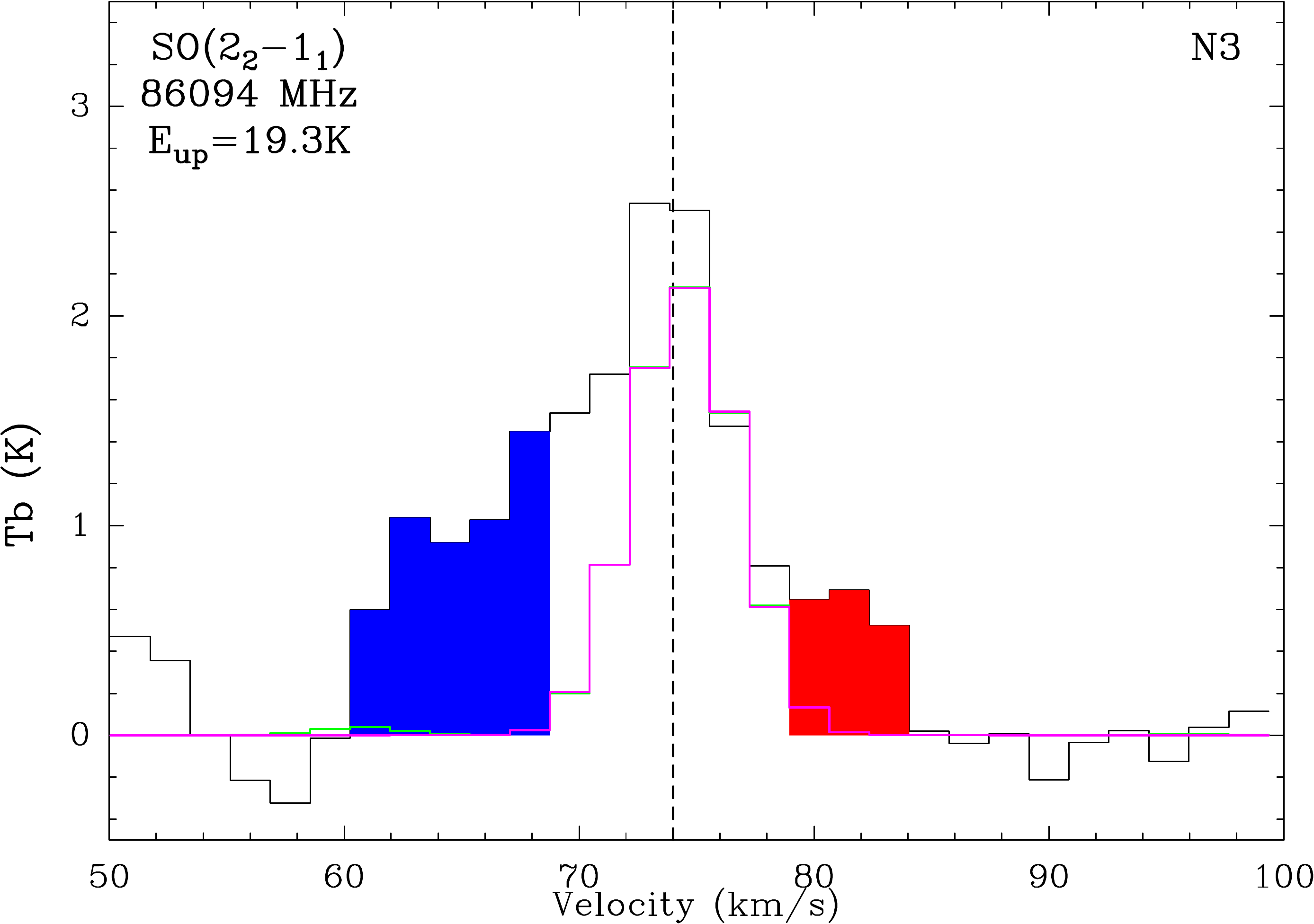} &
   \includegraphics[width=\hsize]{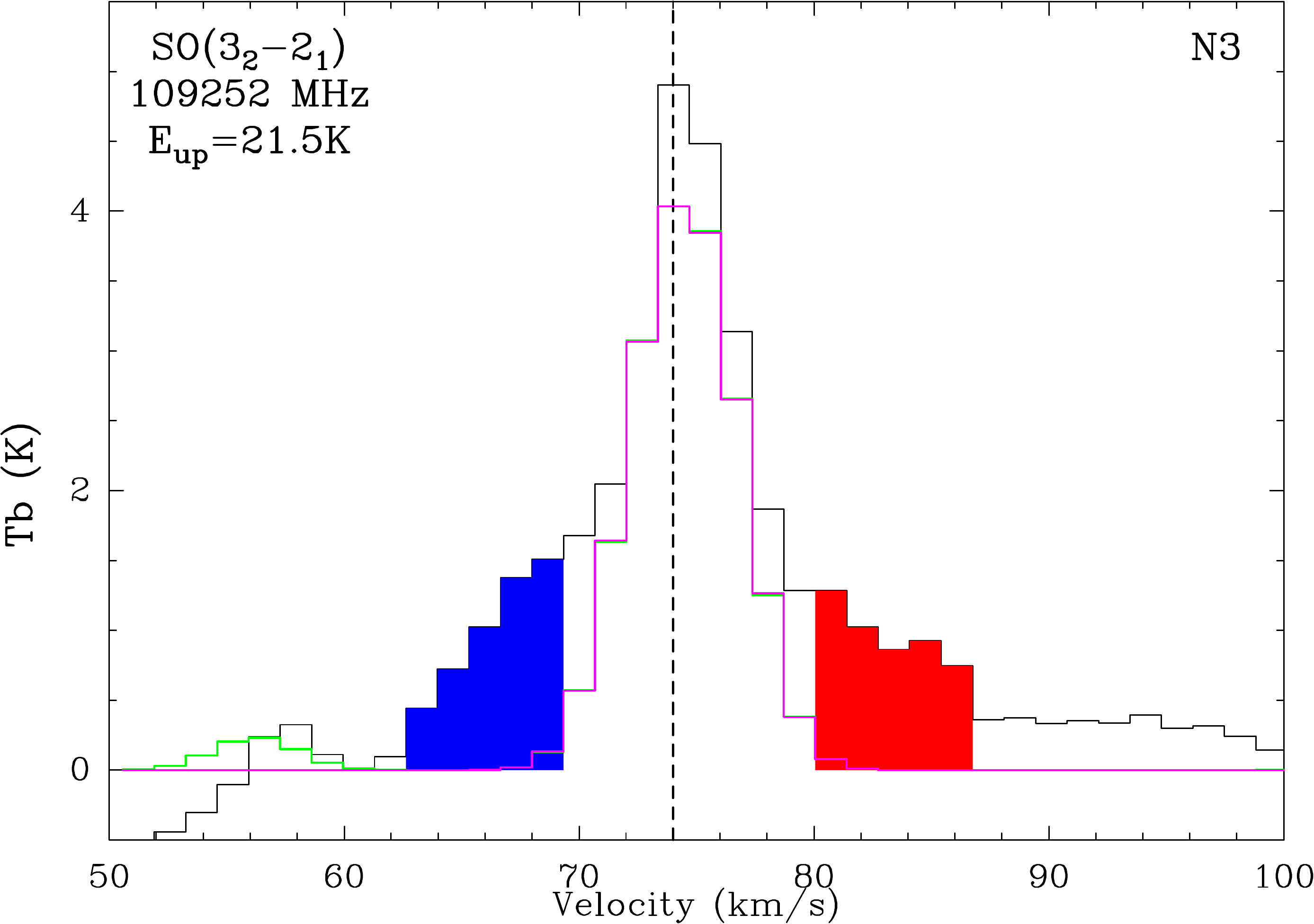} &
   \includegraphics[width=\hsize]{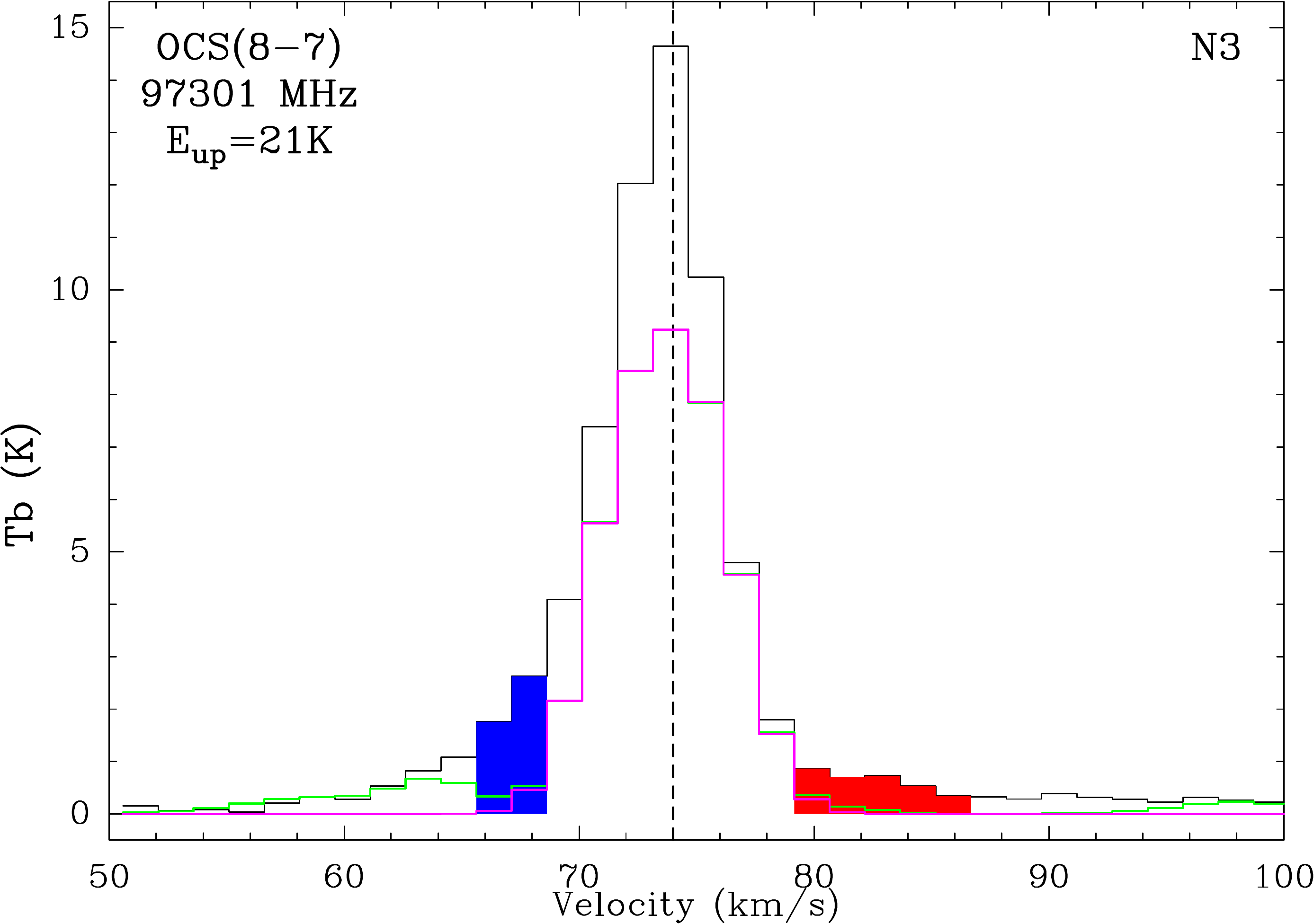} \\
   \includegraphics[width=\hsize]{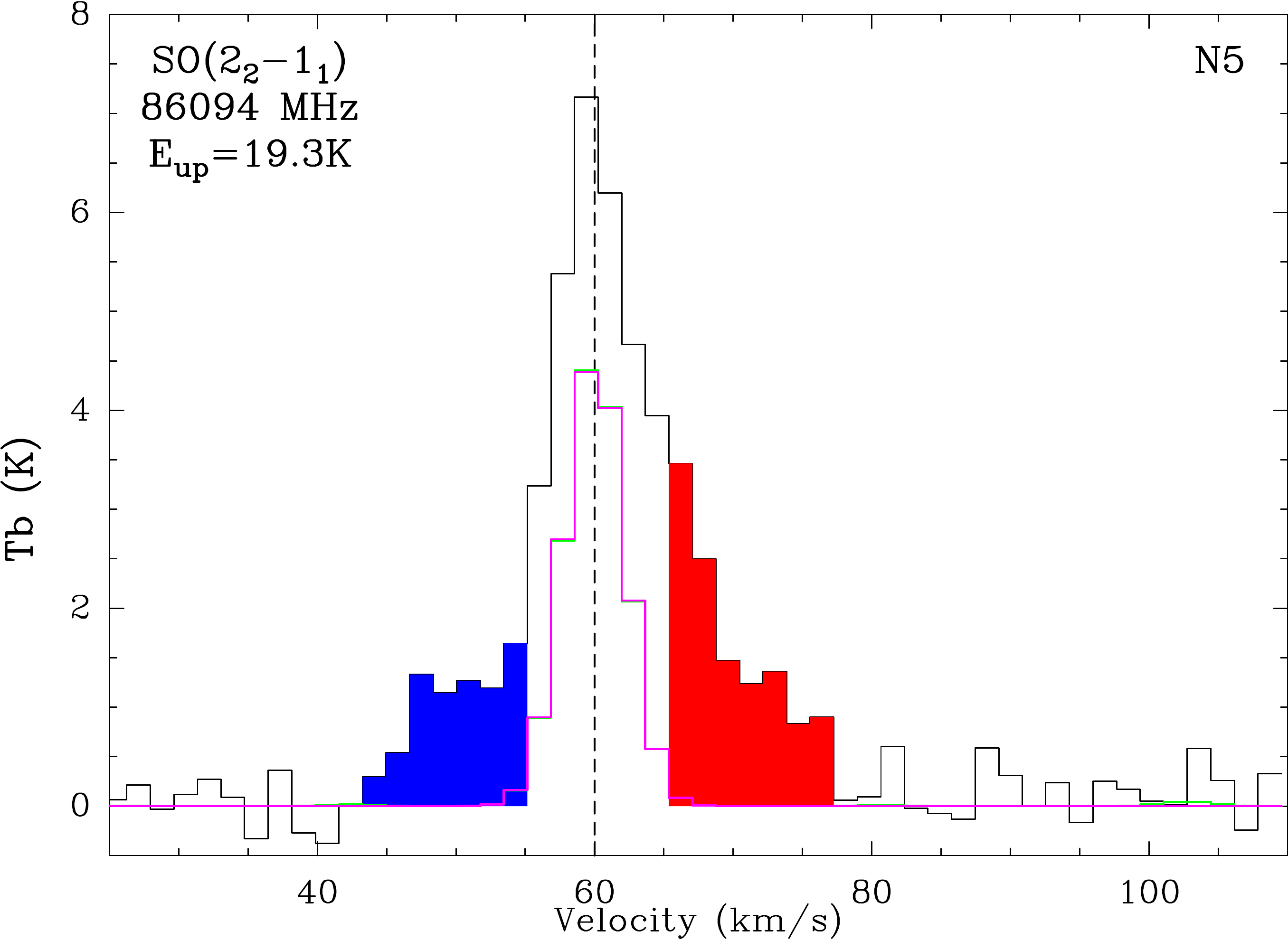} &
   \includegraphics[width=\hsize]{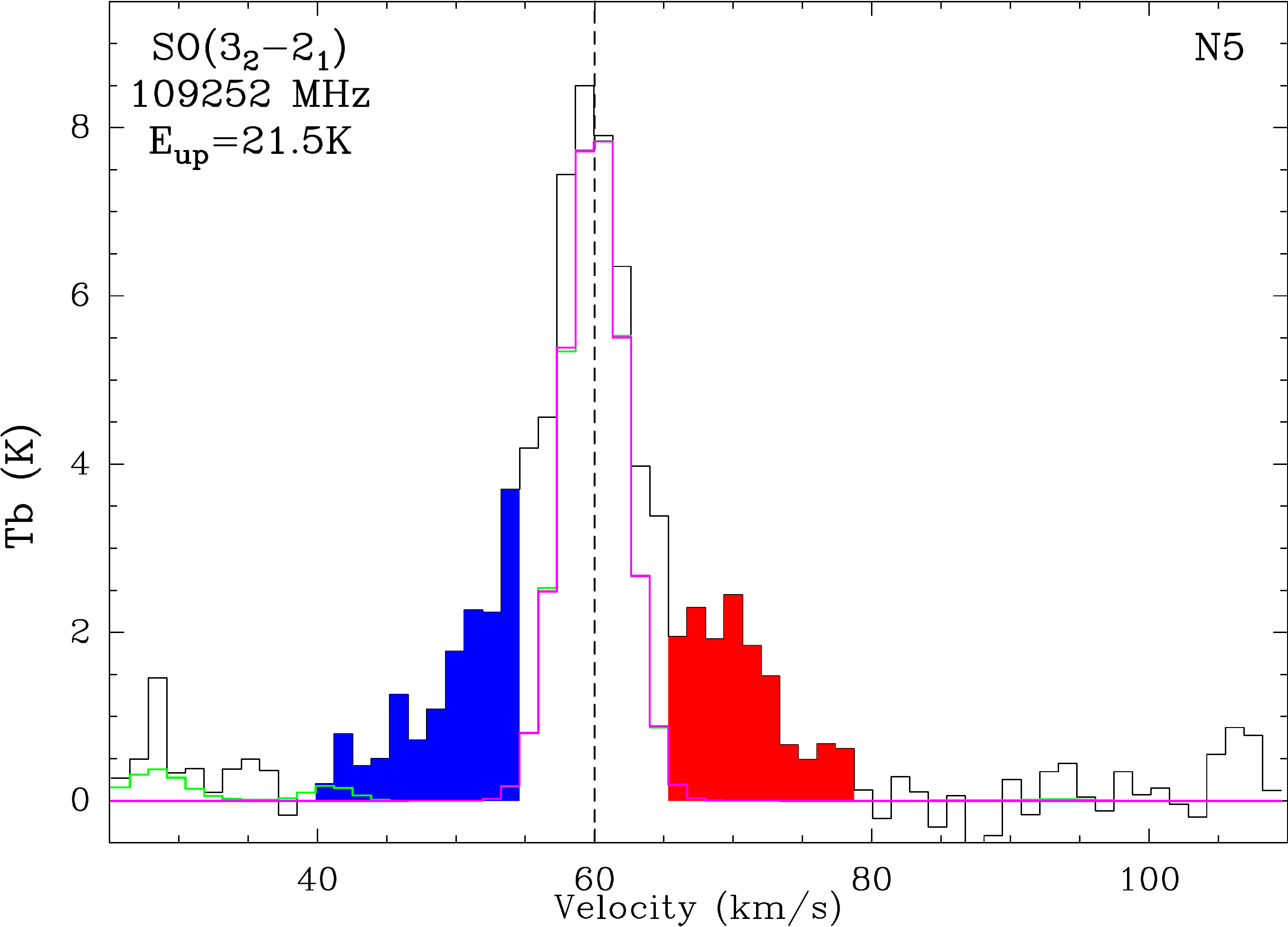} &
   \includegraphics[width=\hsize]{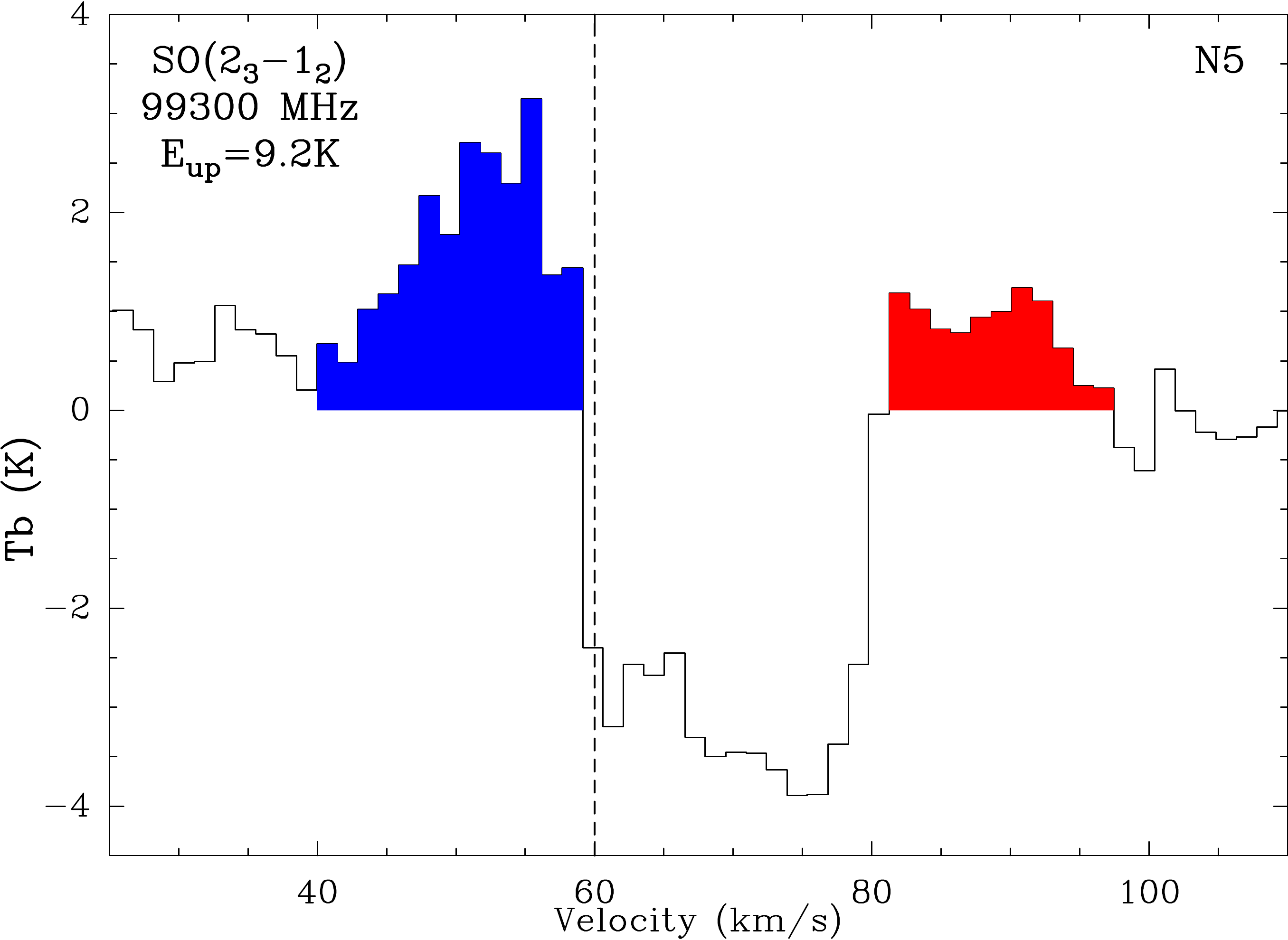} &
   \includegraphics[width=\hsize]{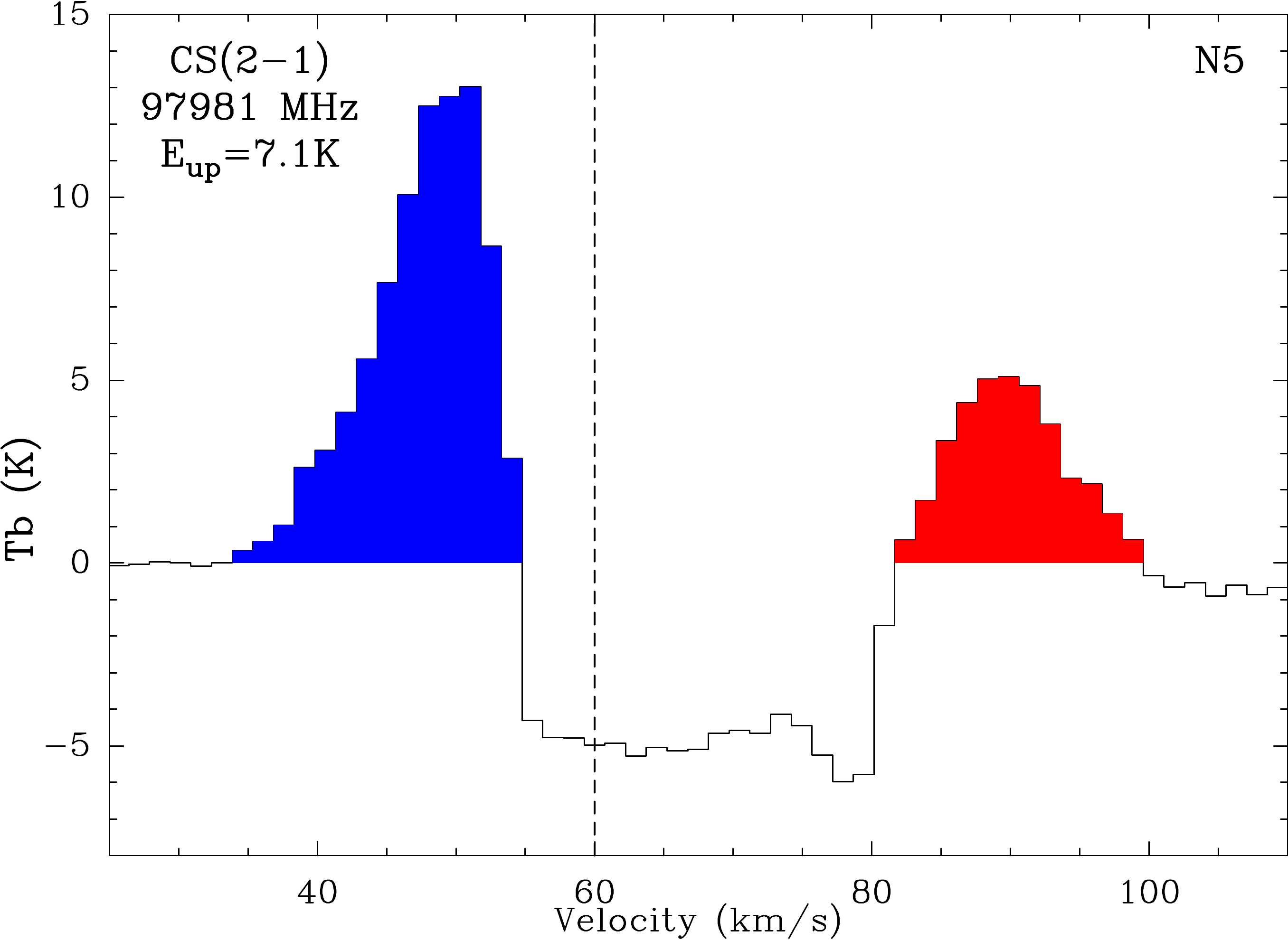} \\    
\end{tabular}}
\caption{\label{kinematic-structure-spectra} Spectra of the lines investigated to search for outflows toward Sgr~B2(N3) (top row) and Sgr~B2(N5) (bottom row). The dashed vertical line marks the systemic velocity of the source and the high velocity wings are highlighted in blue and red. The magenta spectrum represents our best-fit model while the green spectrum shows the model that contains all the identified species. The rest frequency and upper level energy (in temperature unit) of each transition are indicated in each panel.}   
\end{figure*}

The spectra observed toward the three new hot cores show some broader lines with wing emission at high velocities which could suggest the presence of outflows. Investigating these molecular outflows can provide information on the evolutionary stage during the star formation process. To assess the presence of outflows and characterize their properties, we investigate spectral lines from molecules known as typical tracers of outflows. Figure~\ref{kinematic-structure-spectra} presents the spectra of some of these lines detected toward Sgr~B2(N3) and Sgr~B2(N5). All of them show broad wing emission at blue- and red-shifted velocities compared to the systemic velocity of the source. The top row of Fig.~\ref{kinematic-structure-spectra} shows two transitions of SO and the OCS(8-7) transition observed toward Sgr~B2(N3). The blue wing of the latter appears narrower because its boundaries have been defined in order to avoid contamination from another species (see green spectrum in Fig.~\ref{kinematic-structure-spectra}). The bottom row of Fig.~\ref{kinematic-structure-spectra} presents the CS(2-1) transition and three transitions of SO observed toward Sgr~B2(N5). The SO transition with the lowest upper energy level, SO(2$_3$-1$_2$), shows a deep absorption profile and strong broad wings in emission, like the CS(2-1) transition. For all lines, the wing boundaries are chosen to avoid contamination from the line core emission predicted by the LTE model (magenta spectrum in Fig.~\ref{kinematic-structure-spectra}) and contamination from other species (green spectrum in Fig.~\ref{kinematic-structure-spectra}). The velocity ranges used to integrate the blue- and red-shifted emission are summarized in columns 6 and 7 of Table~\ref{kinematic-structure-results}. 

\begin{figure*}[!t]
\resizebox{\hsize}{!}
{\begin{tabular}{cccc}
       \vspace{1.0cm}
   \includegraphics[width=20.5cm]{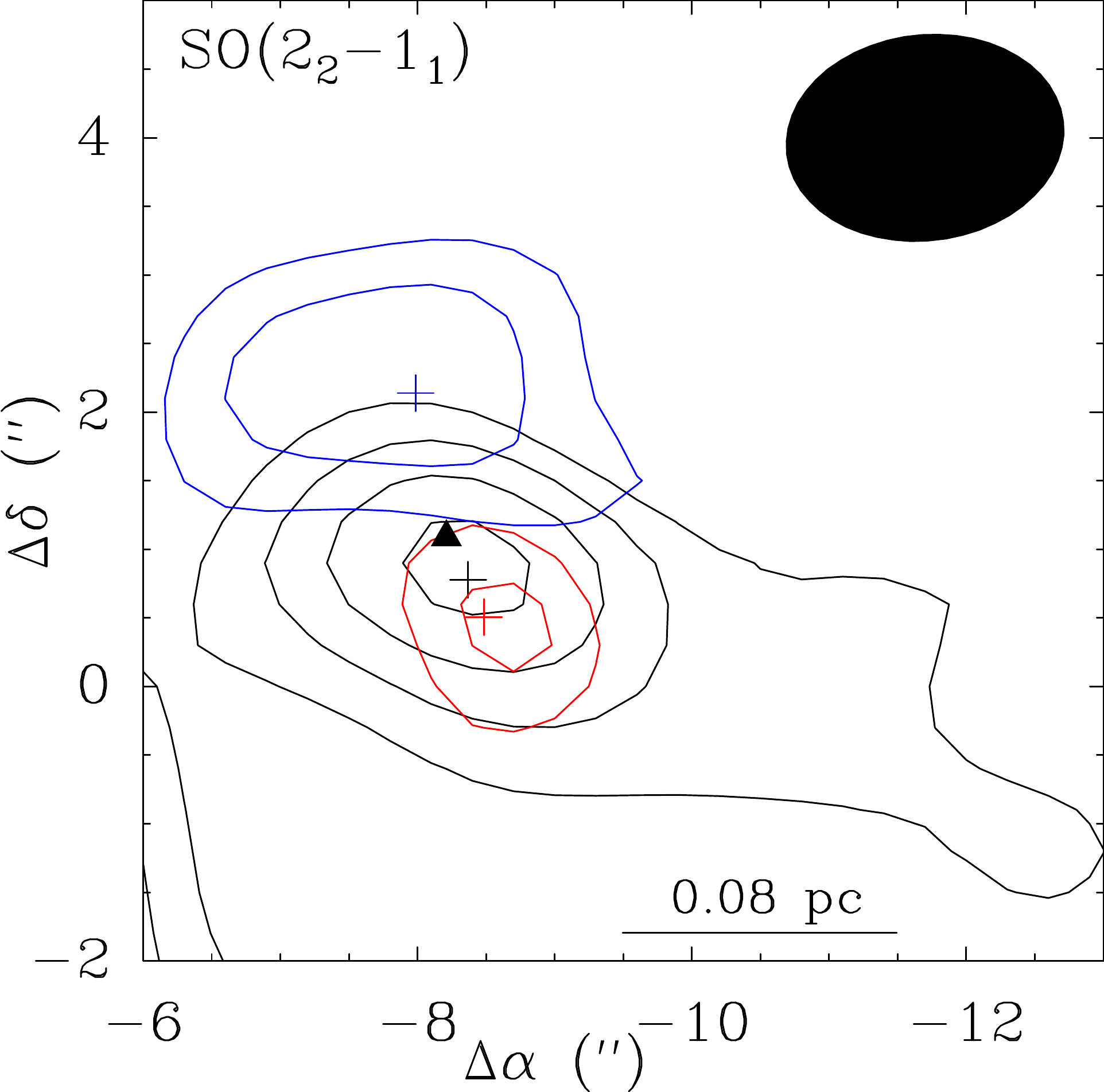} &
   \includegraphics[width=\hsize]{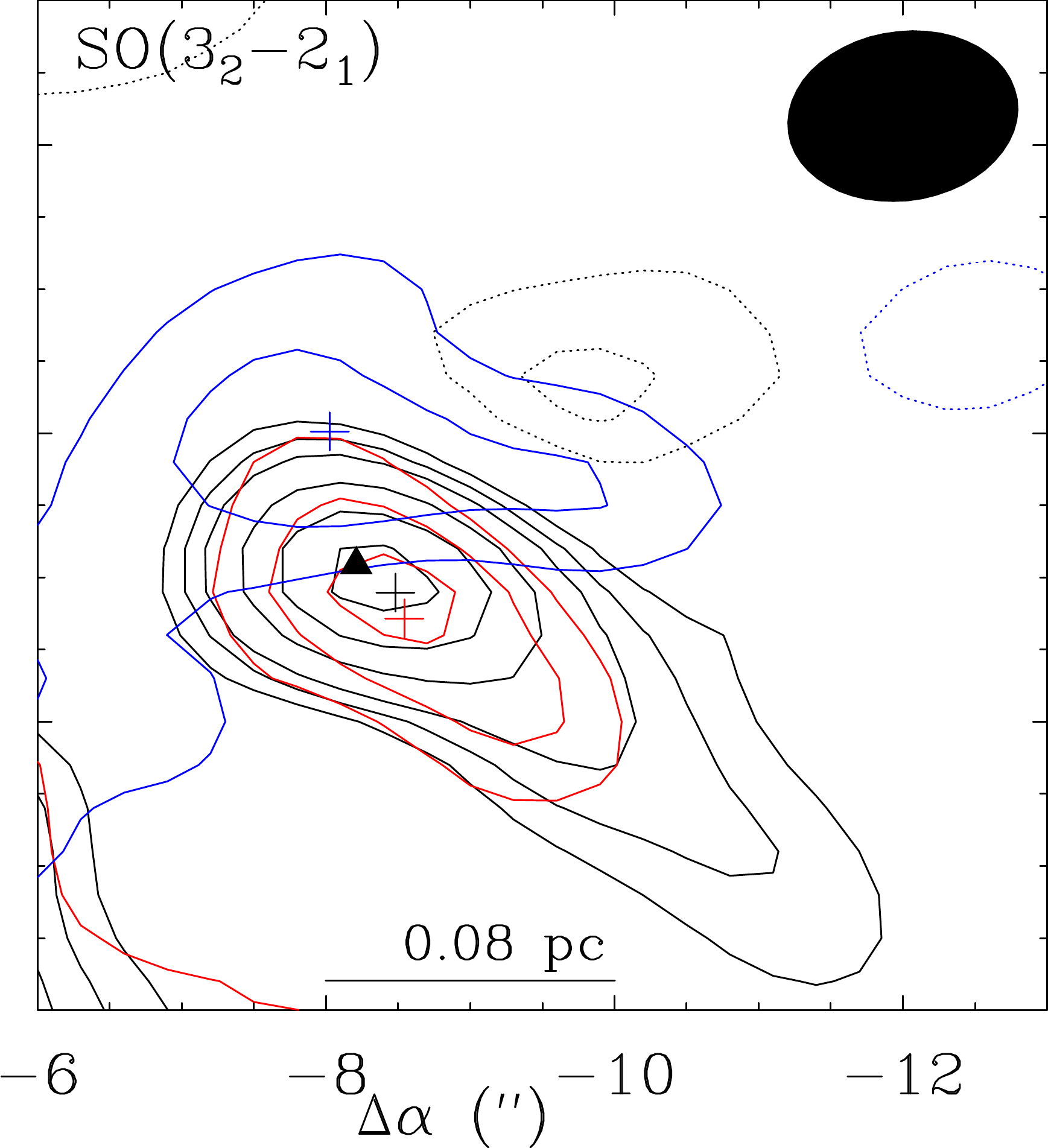} &
   \includegraphics[width=\hsize]{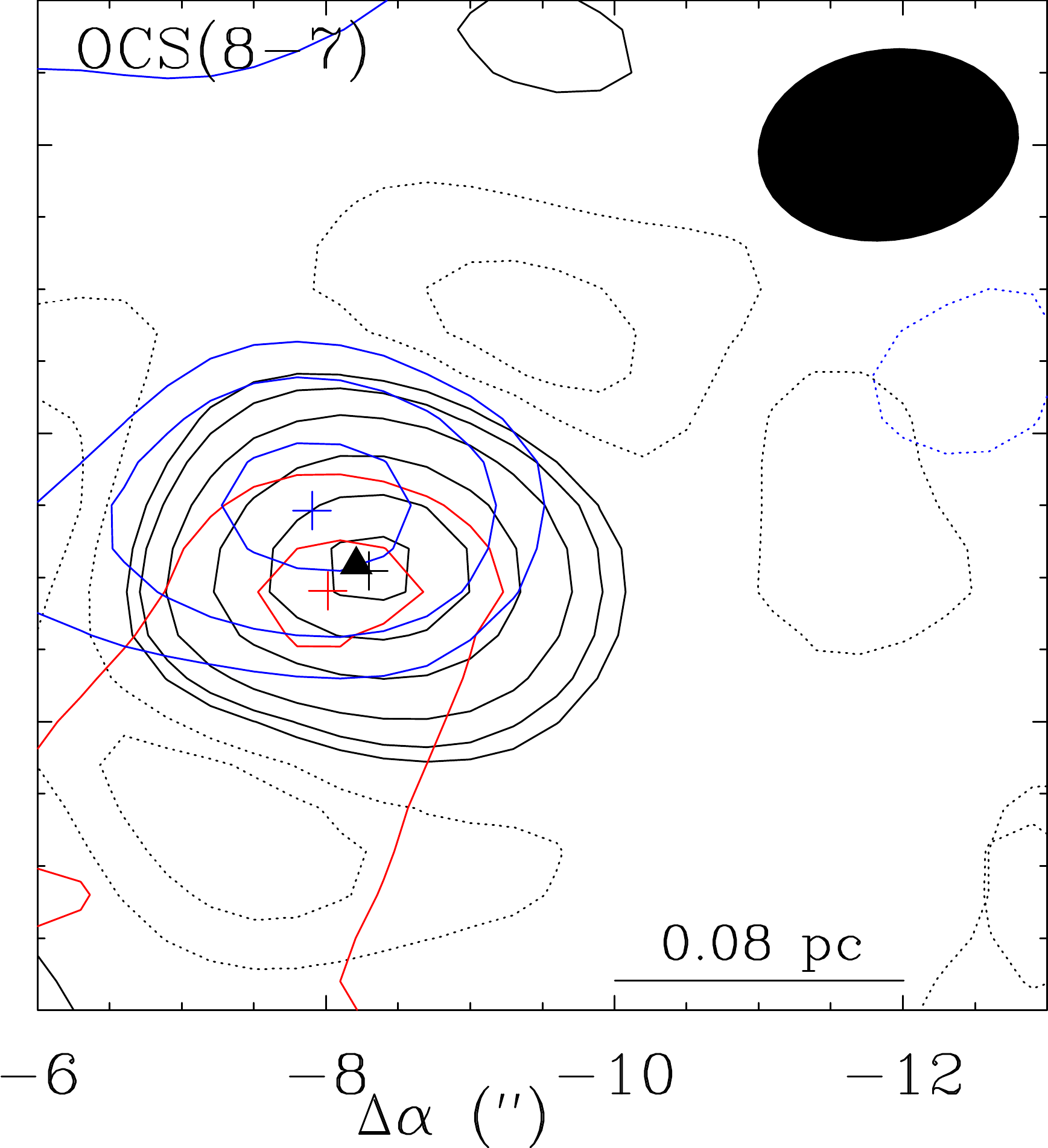} \\
   \includegraphics[width=22cm]{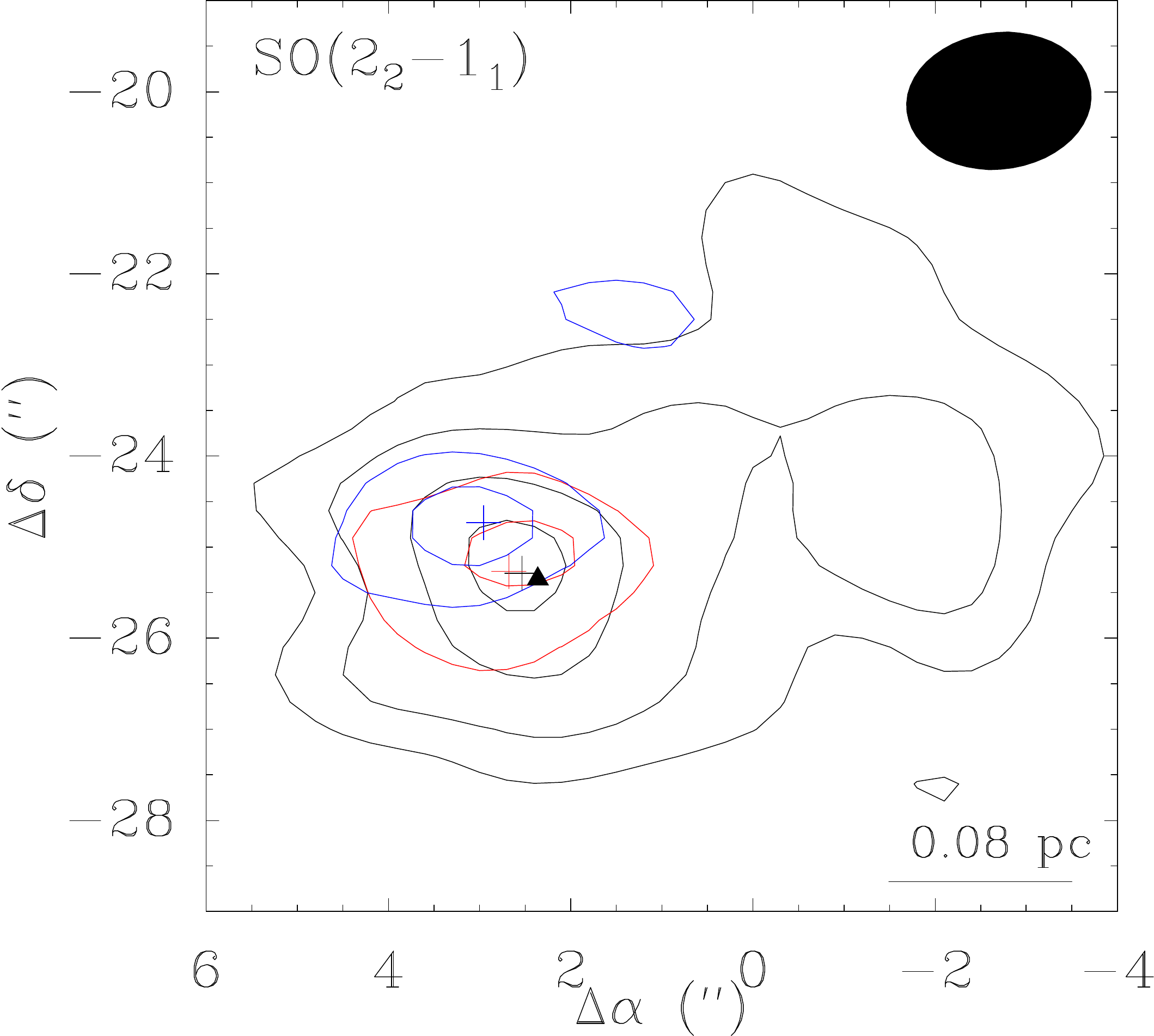} &
   \includegraphics[width=\hsize]{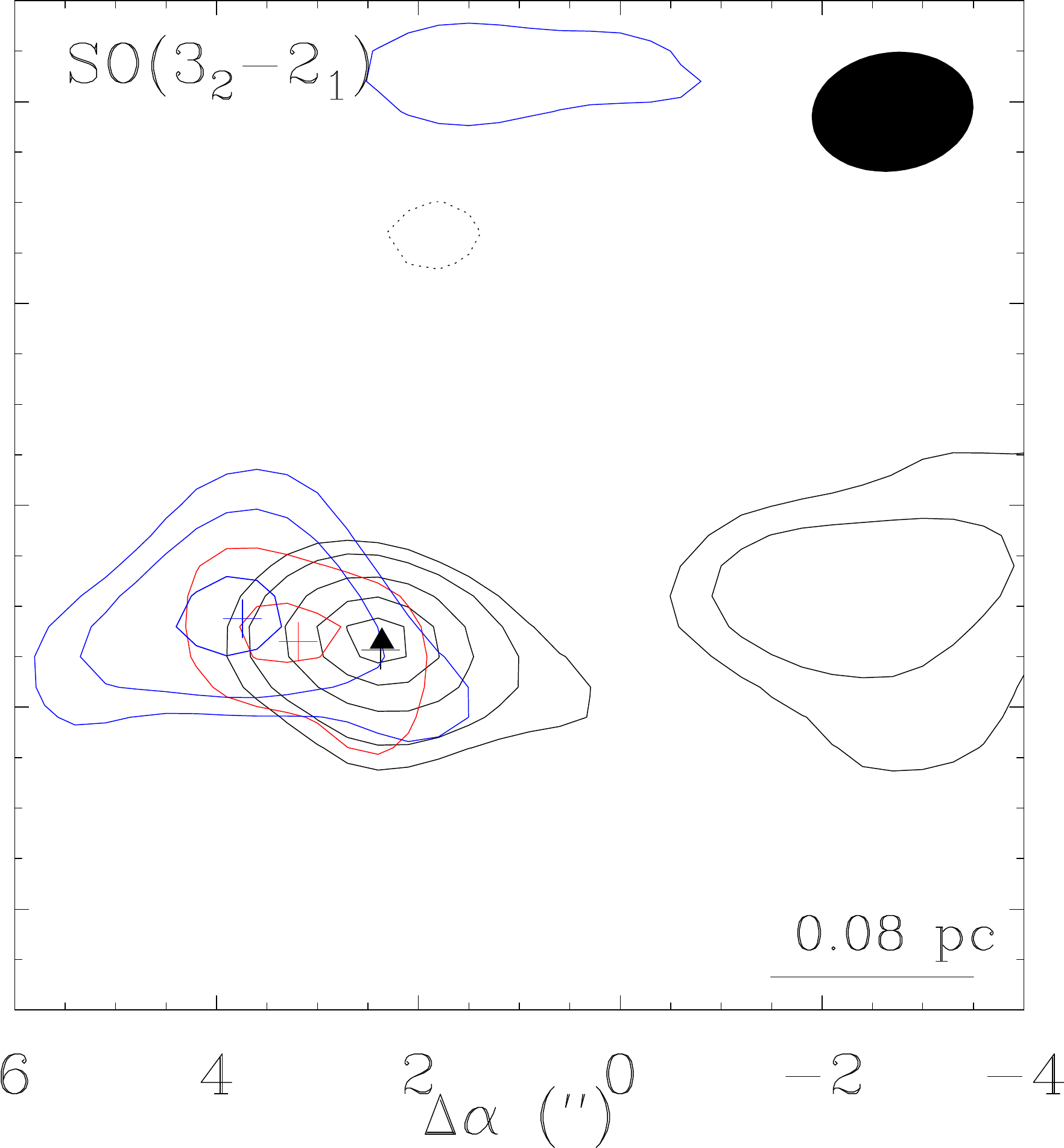} &
   \includegraphics[width=20.8cm]{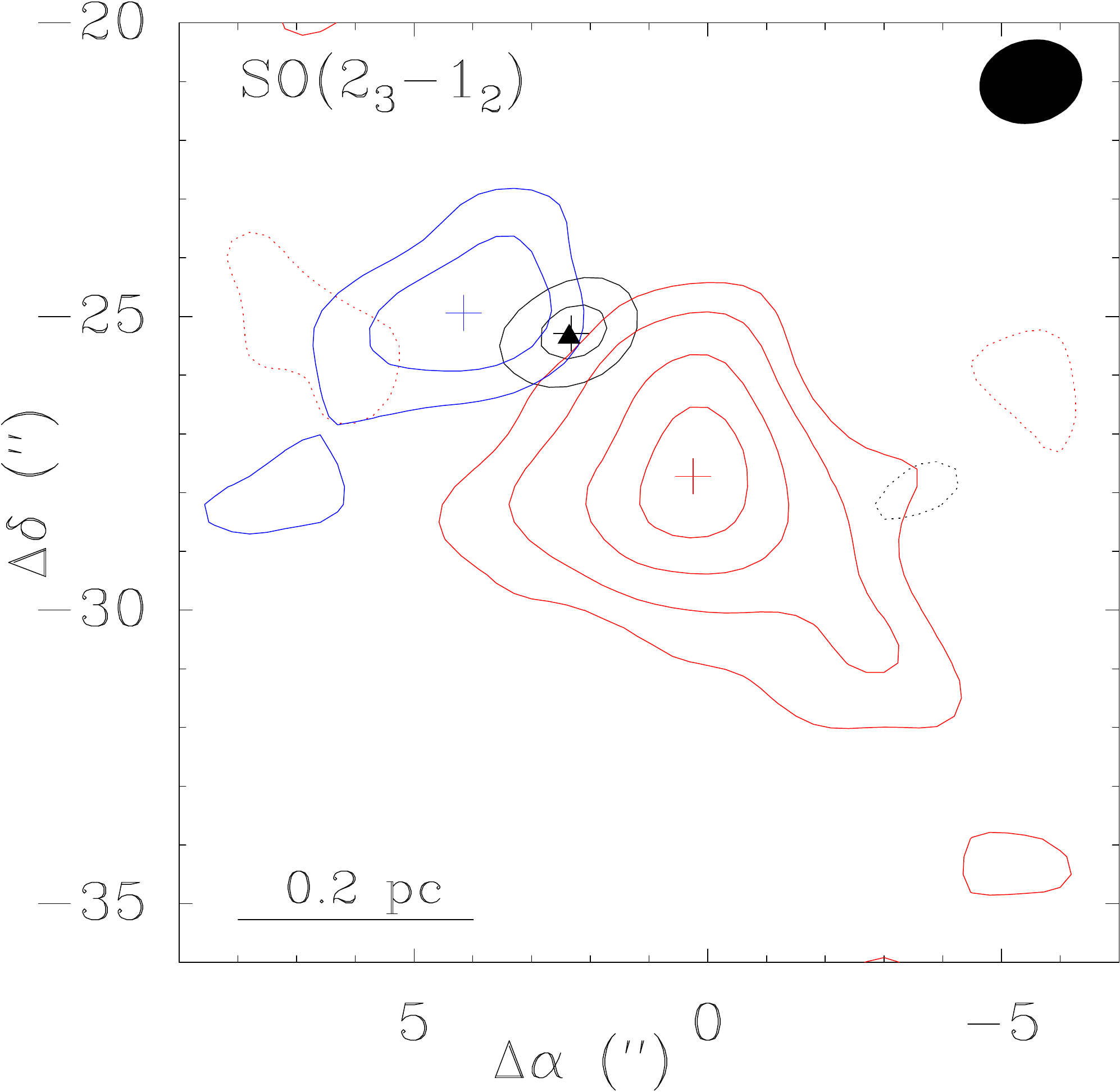} & \hspace{1cm}
   \includegraphics[width=17.6cm]{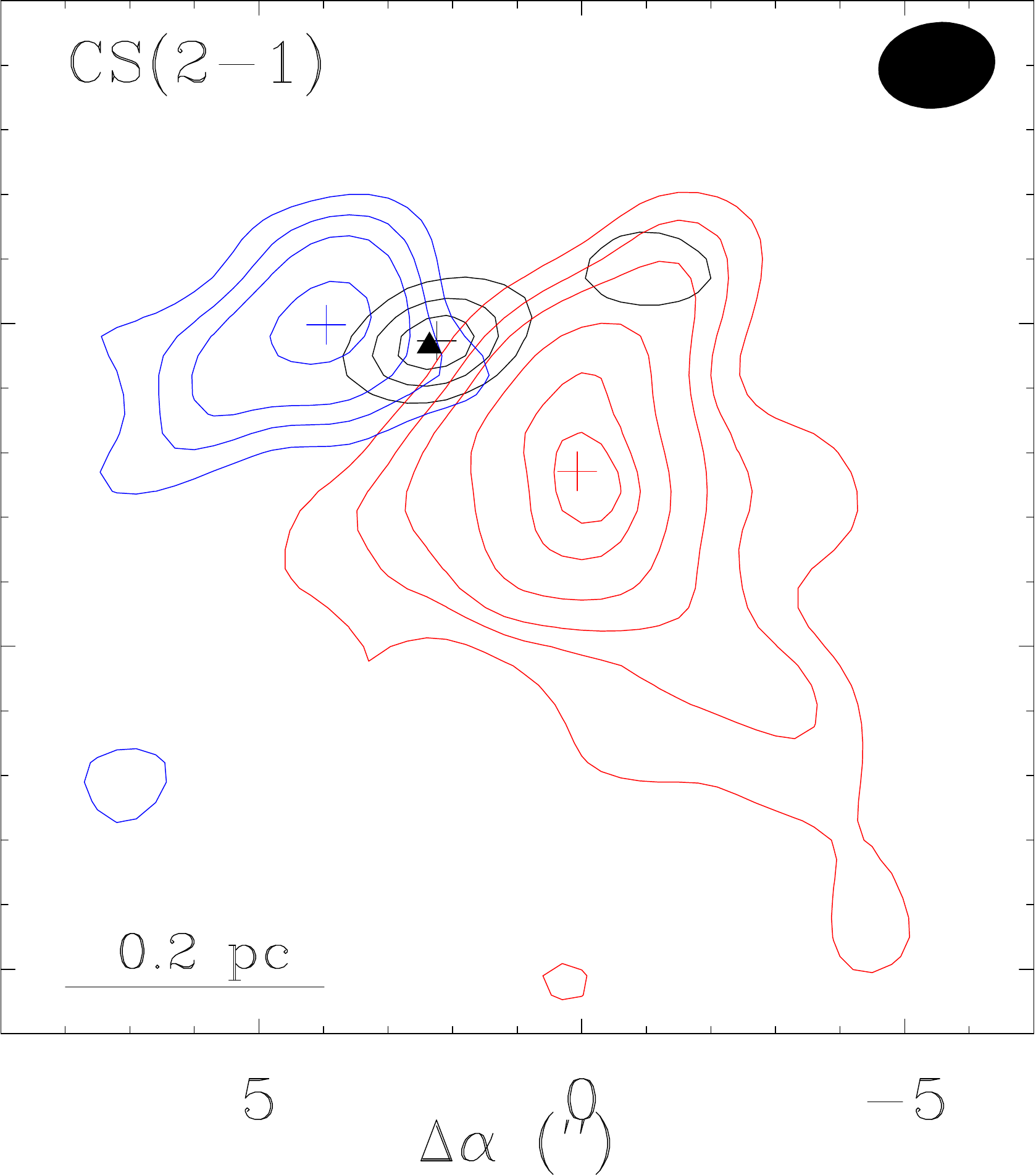} \\
\end{tabular}}
\caption{\label{kinematic-structure-maps} Integrated intensity maps of the lines shown in Fig.~\ref{kinematic-structure-spectra} toward Sgr~B2(N3) (top row) and Sgr~B2(N5) (bottom row). In each panel, maps of the blue- and red-shifted wings are presented in blue and red contours, respectively, overlaid on the integrated emission of the line core (black contours). For the line cores affected by absorption, the black contours represent the continuum emission. Rms noise levels and contour levels used for each map are listed in Table~\ref{contour-levels-map-kinematic}. Each cross corresponds to the peak position of the emission. The black triangle marks the position of the hot core (Sgr~B2(N3) or Sgr~B2(N5)) derived from Fig.~\ref{hot_cores_contour_map}.}  
\end{figure*}

The integrated intensity maps of blue- and red-shifted emission are presented in Fig.~\ref{kinematic-structure-maps} for each line along with the map of the line core, or the continuum emission when the line core is in absorption. All maps show a bipolar morphology for both Sgr~B2(N3) and Sgr~B2(N5) with distinct blue and red lobes, shifted compared to the line core, which could suggest the presence of an outflow. The top row of Fig.~\ref{kinematic-structure-maps} shows the integrated intensity maps of the SO lines observed toward Sgr~B2(N3). Both maps are similar, showing the peak position of the blue wing clearly shifted North-East of the line core, while the red one is slightly shifted to the South. The map produced for the OCS(8-7) transition confirms the North-South velocity gradient observed in the SO maps. The bottom row of Fig.~\ref{kinematic-structure-maps} presents the maps of the line wings observed toward Sgr~B2(N5). The maps of the SO(2$_3$-1$_2$) and CS(2-1) transitions, with the line core seen in absorption, show a clear bipolar structure oriented NE-SW. The other two transitions of SO exhibit a different morphology, with the red lobe shifted East of the continuum emission. The reason for this behaviour is unclear but we note that the velocities over which the red-shifted wing emission is integrated are smaller than the ones used for SO(2$_3$-1$_2$) and CS(2-1) (see Table~\ref{kinematic-structure-results}).

\begin{table*}[!t]
\begin{center}
  \caption{\label{kinematic-structure-results} Properties of molecular outflows toward Sgr B2(N3) and Sgr B2(N5).}
  \setlength{\tabcolsep}{1.0mm} 
  \begin{tabular}{cccccccccc}
    \hline
    Source & Transition & $V_{\rm LSR}$\tablefootmark{a} & $r_{\rm blue}$\tablefootmark{b} & $r_{\rm red}$\tablefootmark{b} & $\Delta V_{\rm blue}$\tablefootmark{c} & $\Delta V_{\rm red}$\tablefootmark{c} & $V^{\rm max}_{\rm blue}$\tablefootmark{d}  & $V^{\rm max}_{\rm red}$\tablefootmark{d}  & $t_{\rm dyn}$ \tablefootmark{e}   \\

             &      & (km s$^{-1}$) &  ($\arcsec$) & ($\arcsec$) & (km s$^{-1}$) & (km s$^{-1}$)  &   (km s$^{-1}$) &   (km s$^{-1}$)  & (10$^{3}$yr) \\
    \hline
    \hline
        N3     & SO(2$_2$-1$_1$)  & 74.0 & 1.07 (NE) & 0.64 (SW) &  [61.1 ; 67.9] & [79.8 ; 83.2]  & 12.9 & 9.2   & 3.0 \\  
               & SO(3$_2$-2$_1$)  & 74.0 & 1.04 (NE) & 0.53 (SW) &  [63.2 ; 68.5] &  [80.7 ; 86.1] & 10.8 & 12.1  & 2.8  \\
               & OCS(7-8)         & 74.0 & 0.48 (NE) & 0.21 (SE) & [66.4 ; 68.0] &  [79.8 ; 85.8]  &  7.6 & 11.8  & 1.6  \\    
        \hline
        N5   & SO(2$_2$-1$_1$)    & 60.0 & 0.87 (NE) & 0.35 (E)  &  [44.0 ; 54.3] &  [66.1 ; 76.4] & 16.0 & 16.4  & 1.5   \\   
             & SO(3$_2$-2$_1$)    & 60.0 & 1.37 (E) & 0.83 (E)  &  [40.4 ; 53.9] &  [65.9 ; 78.0] & 19.6 & 18.0  & 2.3 \\   
             & SO(2$_3$-1$_2$)    & 60.0 & 1.86 (E) & 3.17 (SW) &  [40.6 ; 58.4] &  [81.9 ; 96.7] & 19.4 & 36.7  & 3.6 \\    
             & CS(2-1)           & 60.0 & 1.86 (E) & 3.00 (SW) &  [34.6 ; 53.9] & [82.4 ; 98.8]  & 25.4 & 38.8  & 3.0  \\       
        \hline
\end{tabular}
\end{center}
\tablefoot{\tablefoottext{a}{Systemic velocity of the source.}
\tablefoottext{b}{Distance of the emission peak of the blue/red-shifted lobe compared to the reference position of the hot core. The direction is indicated in parentheses.}
\tablefoottext{c}{Velocity range adopted to integrate the emission from the blue/red-shifted wing.}
\tablefoottext{d}{Maximum outflow velocity of the blue/red-shifted wing calculated as the difference between the high end of the velocity range and $V_{\rm LSR}$.}
\tablefoottext{e}{Average dynamical time of the outflow, assuming an inclination of $45^\circ$ ($r/V_{\rm max}$)}.}
\end{table*}

\begin{table}[!t]
\begin{center}
  \caption{\label{contour-levels-map-kinematic} Rms noise levels and contour levels used in Fig.~\ref{kinematic-structure-maps}.}
  \setlength{\tabcolsep}{1.5mm} 
  \begin{tabular}{ccccc}
    \hline
    Source & Transition    & Range\tablefootmark{a}    & rms\tablefootmark{b}   & Levels\tablefootmark{c} \\
    \hline
    \hline
         &                & blue  & 40.52  &      \\
   N3    & SO(2$_2$-1$_1$) & core  & 36.59 & 3,5,7,9 \\
         &                & red   & 15.45  &       \\
    \cline{2-5}          
         &                 & blue  & 23.75 &      \\
         & SO(3$_2$-2$_1$)  & core  & 22.96 & 4,6,8,12,16,20 \\
         &                 & red   & 15.00  &      \\
    \cline{2-5}          
         &                 & blue & 16.86  &      \\
         & OCS(8-7)        & core & 29.95  & 3,6,12,24,36,45 \\
         &                 & red  & 13.53  &       \\
    \hline       
    \hline     
         &                 & blue  & 45.30  &      \\
   N5    & SO(2$_2$-1$_1$) & core  & 44.90   & 4,6,10,14 \\
         &                 & red   & 44.60  &       \\
    \cline{2-5}
         &                 & blue  & 28.04  &      \\
         & SO(3$_2$-2$_1$)  & core  & 28.27  & 4,6,10,14,16 \\
         &                 & red   & 34.18  &       \\
    \cline{2-5}
         &                 & blue  & 83.53  &      \\
         & SO(2$_3$-1$_2$) & core  & 87.83   & 4,8,15,23 \\
         &                 & red   & 56.16  &       \\
    \cline{2-5}
         &                 & blue & 98.32  &       \\
         & CS(2-1)         & core & 63.52  & 8,13,18,28,38,43  \\
         &                 & red  & 102.02 &       \\
    \hline

        \hline
\end{tabular}
\end{center}
\tablefoot{
\tablefoottext{a}{The velocity ranges are shown in Fig.~\ref{kinematic-structure-spectra}.}
\tablefoottext{b}{Rms noise level, $\sigma$, in mJy beam$^{-1}$ km s$^{-1}$ measured in the integrated intensity map.}
\tablefoottext{c}{Contour levels in unit of $\sigma$.}}
\end{table}

For each transtion the distance $r$ between the peak positions of the blue- and red-shifted outflow lobes and the hot core reference positions are given in Table~\ref{kinematic-structure-results}. In columns 8 and 9 we calculated the maximum outflow velocities, $V_{\rm max}$, for the blue and red lobes as the difference between the high end of the velocity range set to integrate the wing emission and the systemic velocity of the source. From these values we calculate dynamical timescale of each outflow lobe as $t_{\rm dyn} = \frac{r}{V_{\rm max}}$ assuming that the inclination of the outflow axis with respect to the line of sight is about 45$^{\rm o}$ as the maps show two distinct blue and red lobes. The average dynamical times obtained from the blue- and red-shifted lobes are on the order of a few thousand years (Table~\ref{kinematic-structure-results}).

Finally, we found no evidence for a bipolar structure around Sgr~B2(N4) and Sgr~B2(N2). In the case of Sgr~B2(N4), the same lines as those investigated in Sgr~B2(N3) and Sgr~B2(N5) are too weak to show wing emission (see SO lines in Fig.~\ref{appendix-kinematic-n4-raies1}) or they show atypical shapes with blue-shifted wings but without red-shifted wings (see OCS lines in Fig.~\ref{appendix-kinematic-n4-raies1}). The same shape is also observed in the lines of other species not considered as typical tracers of outflows (see Fig.~\ref{appendix-kinematic-n4-raies2}). It is thus difficult to conclude whether this broad blue-shifted emission in Sgr~B2(N4) is due to an outflow or to a fainter component emitting at lower velocity.

          \subsection{Chemical composition}  
                     \label{chemical-composition}

\begin{figure*}[!t]
\resizebox{\hsize}{!}
{\begin{tabular}{ccc}
 \includegraphics[width=\hsize]{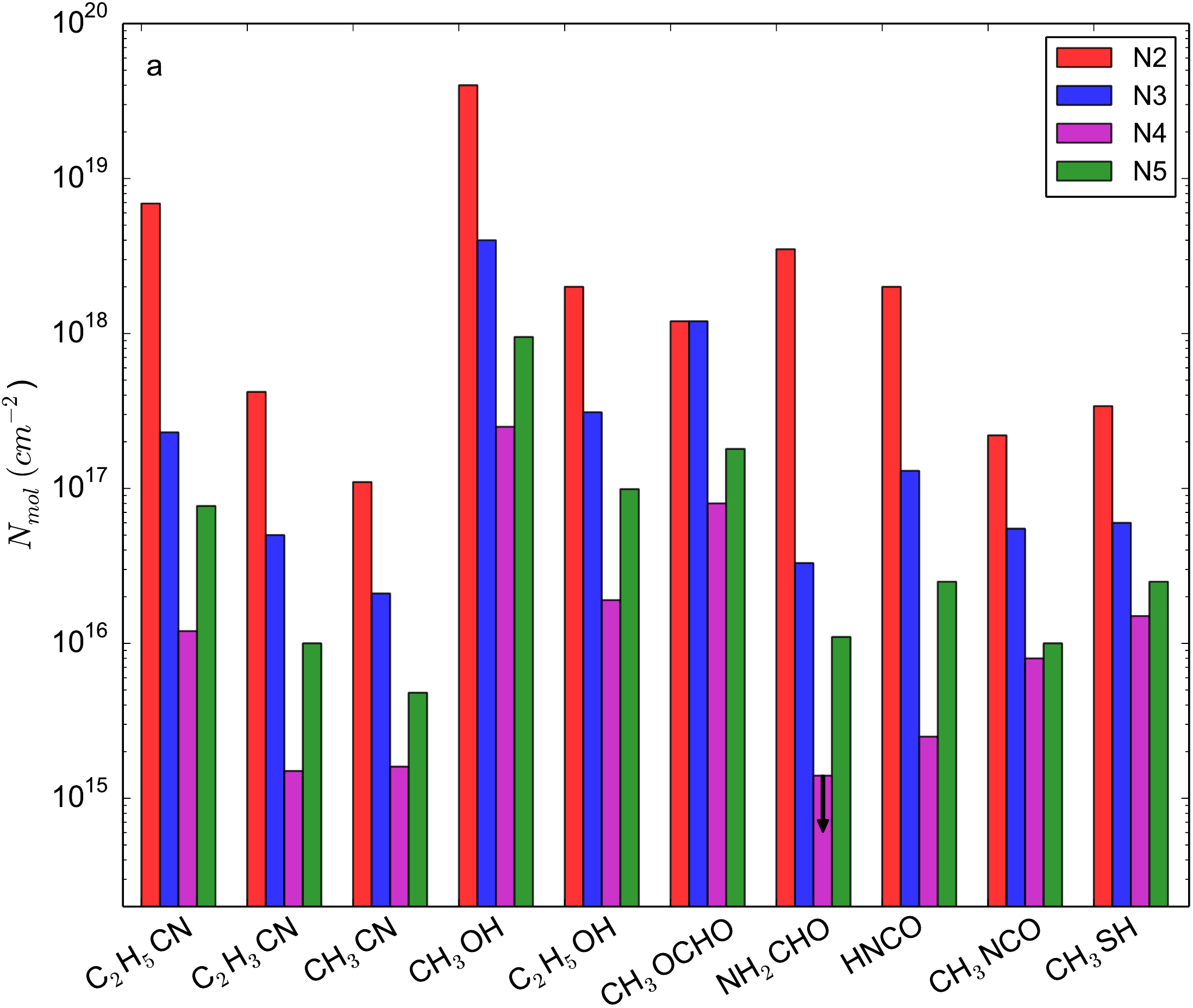} &
 \includegraphics[width=\hsize]{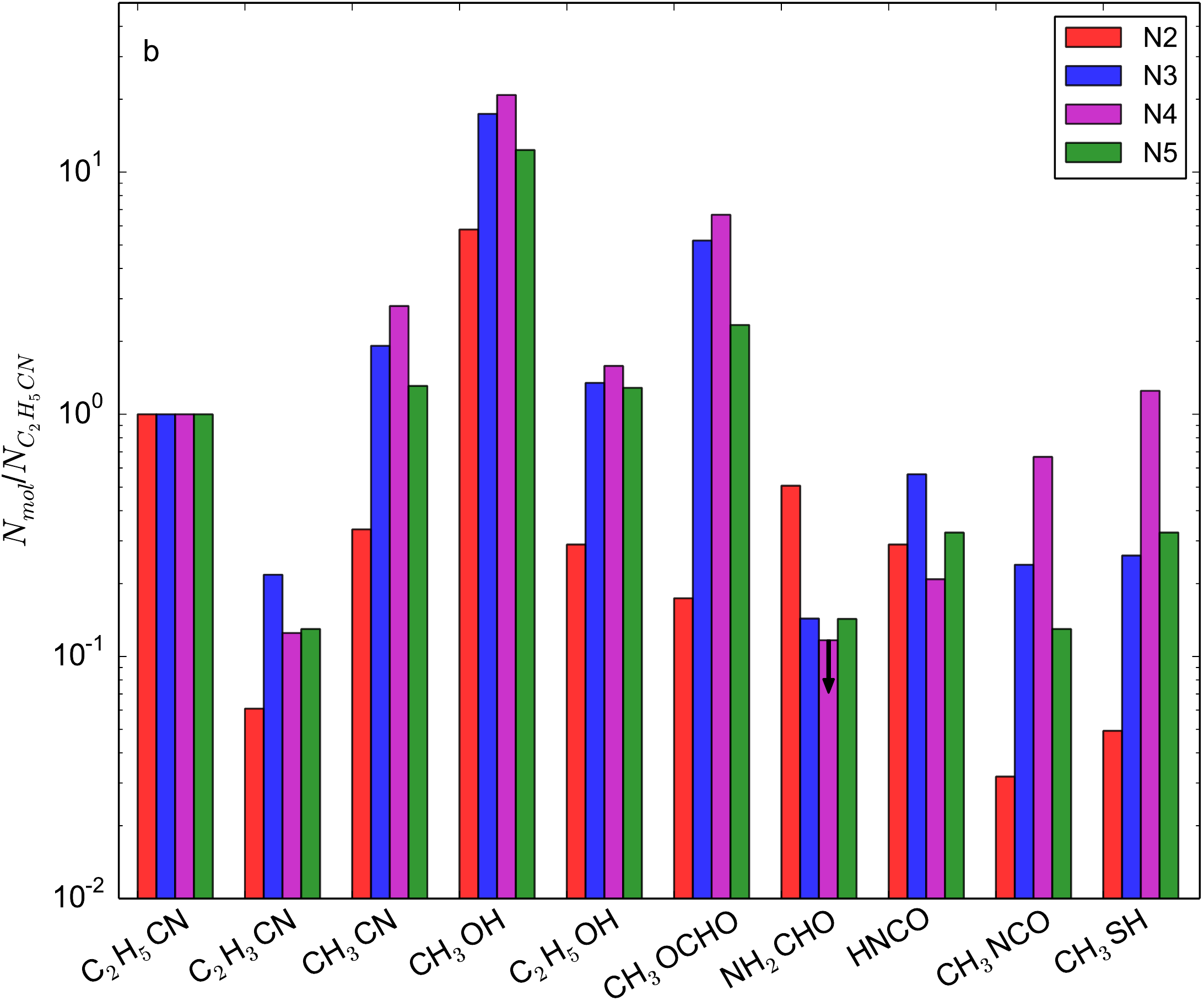} & 
 \includegraphics[width=\hsize]{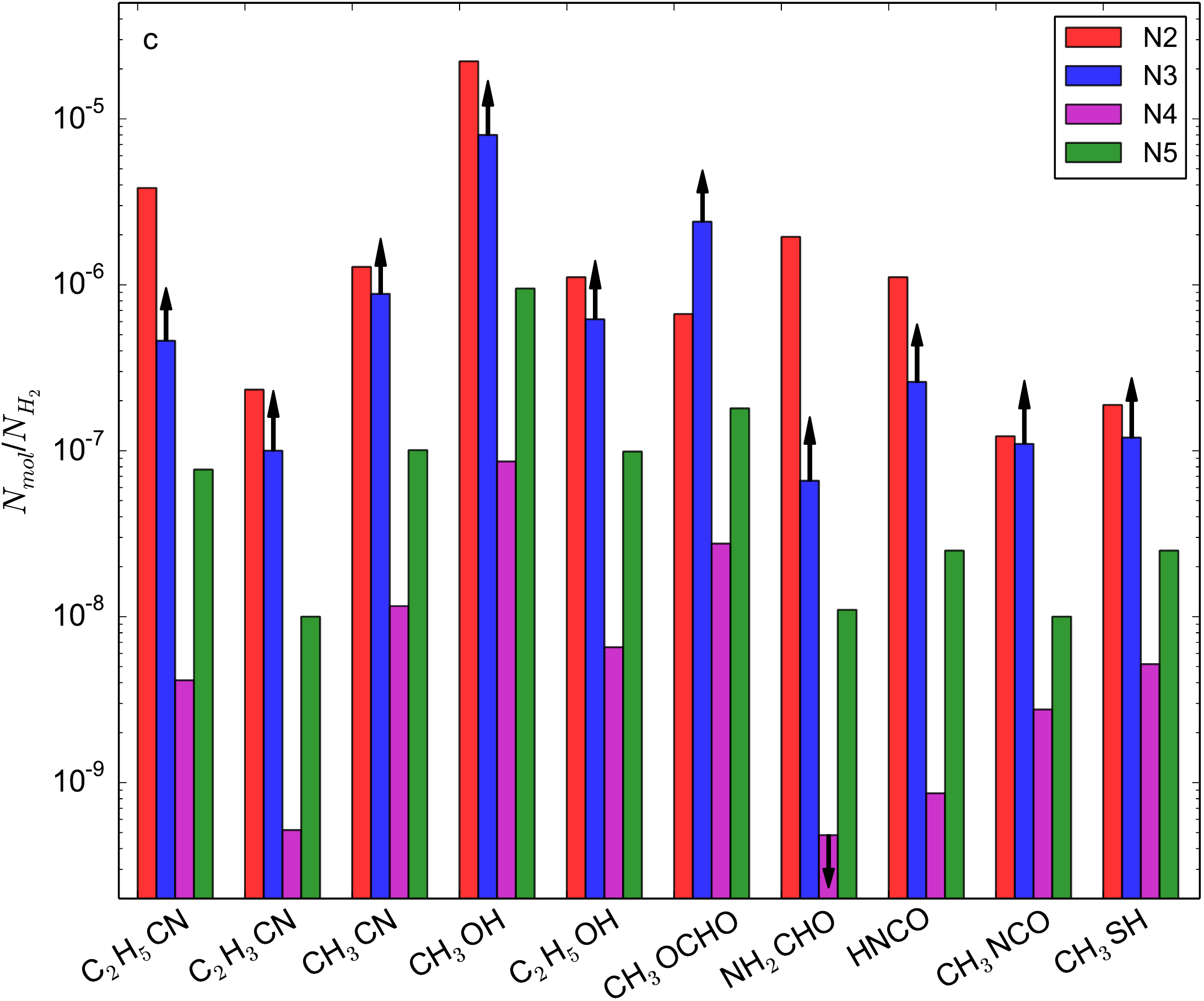} \\
\end{tabular}}
 \caption{\label{chemical_composition_plot} \textbf{a} Column densities of various molecules detected toward four of the five hot cores embedded in Sgr~B2(N) (see Table~\ref{best-fit-parameters}). \textbf{b} Column densities normalized to the column density of \ce{C2H5CN}. \textbf{c} Chemical abundances with respect to H$_2$. In each panel, lower and upper limits are indicated with arrows.}
\end{figure*}

We use Weeds as described in Sect.~\ref{modelling} to model the emission lines of ten (complex) organic molecules detected toward the three new hot cores. For each molecule, the source size, rotational temperature, and velocity and linewidth are derived as described in Sects.~\ref{source-size}, \ref{Trot}, and \ref{kinematic-structure}, respectively. All these parameters are listed in Table~\ref{best-fit-parameters} along with the molecular column densities obtained from our best-fit LTE models toward the three hot cores. Column densities derived for Sgr~B2(N2) are also shown for comparison \citep{muller2016, belloche2016, belloche2017}. We investigated the isotopologs $^{13}$\ce{CH3CN} and CH$_3^{13}$CN instead of \ce{CH3CN} because the vibrational ground state transitions of the latter are optically thick. We assume the isotopic ratio [\ce{CH3CN}]/[$^{13}$\ce{CH3CN}]~= [\ce{CH3CN}]/[CH$_3^{13}$CN]~=~21 derived by \citet{belloche2016} to obtain the column density of \ce{CH3CN}. The resulting chemical composition of the four hot cores is displayed in Fig.~\ref{chemical_composition_plot}a. Figure~\ref{chemical_composition_plot}b shows the column densties normalized to the column density of \ce{C2H5CN}.

\begin{table*}[!t]
\begin{center}
  \caption{\label{best-fit-parameters} Parameters of our best-fit LTE model.}
  \begin{tabular}{clrccccrc}
    \hline
Source &   \multicolumn{1}{c}{Species} & \multicolumn{1}{c}{$N_l$} \tablefootmark{a} & $N_{\rm mol}$ \tablefootmark{b} & $C_{\rm vib}$ \tablefootmark{c} &$T_{\rm rot}$ \tablefootmark{d}  & $D$ \tablefootmark{e} & \multicolumn{1}{c}{$v_{\rm off}$ \tablefootmark{f}} & $\Delta v$ \tablefootmark{g} \\
               &               &                                    & (cm$^{-2}$)                                   &  &  (K)                                        &   ($\arcsec$)                                      &  (km s$^{-1}$)                           &  (km s$^{-1}$)    \\
    \hline
    \hline
N3         & \ce{C2H5CN}, $\varv=0$   &  48  & 2.3$\times$10$^{17}$ & 1.53 & 170  &  0.4 & 0.0  & 5.5  \\
           & \ce{C2H3CN}, $\varv=0$   &  24  & 5.0$\times$10$^{16}$ & 1.00  & 150  &  0.4 & 0.0  & 7.0  \\
           & $^{13}$\ce{CH3CN}, $\varv=0$ & 8 & 2.1$\times$10$^{16}$ & 1.06  & 145  &  0.4 & 0.2  & 5.3 \\
           & \ce{CH3}$^{13}$CN, $\varv=0$ & 6 & 2.1$\times$10$^{16}$ & 1.06  & 145  &  0.4 & 0.2  & 5.3 \\
           & \ce{CH3OH}, $\varv=0$    &  31  & 4.0$\times$10$^{18}$ & 1.00  & 170  &  0.4 & 0.0  & 5.0  \\
           & \ce{C2H5OH}, $\varv=0$   &  71  & 3.1$\times$10$^{17}$ & 1.24  & 145  &  0.4 & 1.0  & 4.0 \\
           & \ce{CH3OCHO}, $\varv=0$  &  76  & 1.2$\times$10$^{18}$ & 1.23  & 145  &  0.4 & 0.6  & 4.1  \\
           & \ce{NH2CHO}, $\varv=0$   &  10  & 3.3$\times$10$^{16}$ & 1.09  & 145  & 0.4  & 0.5  & 5.6 \\
           & \ce{HNCO}, $\varv=0$     &   5  & 1.3$\times$10$^{17}$ & 1.005  & 145  & 0.4  & 0.6  & 7.7  \\
           & \ce{CH3NCO}, $\varv=0$  &  29  & 5.5$\times$10$^{16}$ & 1.00  & 145  & 0.4  & 0.5  & 4.0 \\
           & \ce{CH3SH}, $\varv=0$    &  5   & 6.0$\times$10$^{16}$ & 1.00  & 145  &  0.4 & 0.0  & 4.0 \\
           &  &  &  &  &  &  &  &  \\          
N4         & \ce{C2H5CN}, $\varv=0$   &  25  & 1.2$\times$10$^{16}$ & 1.38  & 150  &  1.0 & -0.5 & 5.5  \\
           & \ce{C2H3CN}, $\varv=0$   & 11   & 1.5$\times$10$^{15}$ & 1.00  & 150  &  1.0 & -0.6 & 4.5  \\
           & $^{13}$\ce{CH3CN}, $\varv=0$ & 6 & 1.6$\times$10$^{15}$ & 1.06  & 145  &  1.0 & 0.0  & 5.3 \\
           & \ce{CH3}$^{13}$CN, $\varv=0$ & 7 & 1.6$\times$10$^{15}$ & 1.06  & 145  &  1.0 & 0.0  & 5.3 \\
           & \ce{CH3OH}, $\varv=0$    &  13  & 2.5$\times$10$^{17}$ & 1.00  & 190  &  1.0 & -0.6 & 5.0  \\
           & \ce{C2H5OH}, $\varv=0$   &  18  & 1.9$\times$10$^{16}$ & 1.24  & 150  &  1.0 & -0.3 & 3.5  \\ 
           & \ce{CH3OCHO}, $\varv=0$  &  25  & 8.0$\times$10$^{16}$ & 1.23  & 150  &  1.0 & 0.0  & 4.0  \\
           & \ce{NH2CHO}, $\varv=0$   &  \_  & $<$1.4$\times$10$^{15}$ & 1.09  & 145  &  0.4  & 0.5  & 5.0 \\       
           & \ce{HNCO}, $\varv=0$     &   3  & 2.5$\times$10$^{15}$ & 1.006  & 150  & 1.0  & -0.9 & 4.1  \\
           & \ce{CH3NCO}, $\varv=0$  &  16  & 8.0$\times$10$^{15}$ & 1.00 & 150  & 1.0  & 0.0  & 4.0 \\
           & \ce{CH3SH}, $\varv=0$    & 3    & 1.5$\times$10$^{16}$ & 1.00  & 150  &  1.0 & 0.0  & 4.5 \\
           &  &  &  &  &  &  &  &  \\      
N5         & \ce{C2H5CN}, $\varv=0$   &  37  & 7.7$\times$10$^{16}$ & 1.53  & 170  &  1.0  & -0.5  & 5.5  \\    
           & \ce{C2H3CN}, $\varv=0$   &  14  & 1.0$\times$10$^{16}$ & 1.00  & 145  &  1.0  & -0.6  & 6.0  \\
           & $^{13}$\ce{CH3CN}, $\varv=0$ & 7 & 4.8$\times$10$^{15}$ & 1.06  & 145  &  1.0 & -0.6  & 4.6 \\
           & \ce{CH3}$^{13}$CN, $\varv=0$ & 5 & 4.8$\times$10$^{15}$ & 1.06  & 145  &  1.0 & -0.6  & 4.6 \\
           & \ce{CH3OH}, $\varv=0$    &  13  & 9.5$\times$10$^{17}$ & 1.00  & 180  &  1.0  & -0.7  & 4.5  \\
           & \ce{C2H5OH}, $\varv=0$   &  71  & 9.9$\times$10$^{16}$ & 1.24  & 145  &  1.0  & -0.8  & 3.5 \\
           & \ce{CH3OCHO}, $\varv=0$  &  48  & 1.8$\times$10$^{17}$ & 1.23  & 145  &  1.0  & -0.5  & 3.5  \\
           & \ce{NH2CHO}, $\varv=0$   &  7   & 1.1$\times$10$^{16}$ & 1.09  & 145  & 1.0   & -0.5  & 6.0 \\
           & \ce{HNCO}, $\varv=0$     &  4   & 2.5$\times$10$^{16}$ & 1.005  & 145  & 1.0   & -0.8  & 6.8  \\
           & \ce{CH3NCO}, $\varv=0$  &  8   & 1.0$\times$10$^{16}$ & 1.00 & 145  & 1.0   & -0.5  & 3.5 \\
           & \ce{CH3SH}, $\varv=0$    &  3   & 2.5$\times$10$^{16}$ & 1.00 & 145  &  1.0  & 0.0   & 4.5 \\
           &  &  &  &  &  &  &  &  \\
           \hline 
N2  \tablefootmark{*}  & \ce{C2H5CN}, $\varv=0$ & 154 & 6.9$\times$10$^{18}$ & 1.38 & 150  &  1.2  & -0.8  & 5.0  \\
           & \ce{C2H3CN}, $\varv=0$   &  44   & 4.2$\times$10$^{17}$ & 1.00  & 200  & 1.1  & -0.6  & 6.0  \\
           & $^{13}$\ce{CH3CN}, $\varv=0$ & 8  & 1.1$\times$10$^{17}$ & 1.08  & 170  & 1.4  & -0.5  & 5.4 \\
           & \ce{CH3}$^{13}$CN, $\varv=0$ & 7  & 1.1$\times$10$^{17}$ & 1.08  & 170  & 1.4  & -0.5  & 5.4 \\
           & \ce{CH3OH}, $\varv=0$    & 41    & 4.0$\times$10$^{19}$ & 1.00  & 160  & 1.4  & -0.5  & 5.4  \\
           & \ce{C2H5OH}, $\varv=0$   & 168   & 2.0$\times$10$^{18}$ & 1.24  & 150  & 1.5  & -0.4  & 4.7 \\
           & \ce{CH3OCHO}, $\varv=0$  & 90    & 1.2$\times$10$^{18}$ & 1.23  & 150  & 1.5  & -0.4  & 4.7  \\
           &  \ce{NH2CHO}, $\varv=0$  &  30   & 3.5$\times$10$^{18}$ & 1.17  & 200  & 0.8  & 0.2   & 5.5  \\
           & \ce{HNCO}, $\varv=0$     & 12    & 2.0$\times$10$^{18}$ & 1.06  & 240  &  0.9 &  0.0  & 5.5 \\
           & \ce{CH3NCO}, $\varv=0$  &  60   & 2.2$\times$10$^{17}$ & 1.00  & 150  &  0.9 & -0.6  & 5.0  \\
           & \ce{CH3SH}, $\varv=0$    & 12    & 3.4$\times$10$^{17}$  & 1.00  & 180  & 1.4  & -0.5  & 5.4 \\

    \hline
\end{tabular}
\end{center}
\tablefoot{\tablefoottext{a}{Number of lines detected above 3$\sigma$. One line may mean a group of transitions of the same molecule blended together.}
\tablefoottext{b}{Total column density of the molecule.}
\tablefoottext{c}{Correction factor applied to the column density to account for the contribution of vibrationally or torsionally excited states not included in the partition function.}
\tablefoottext{d}{Rotational temperature (see Sect.~\ref{Trot}).}
\tablefoottext{e}{Source diameter (FWHM) (see Sect.~\ref{source-size}).}
\tablefoottext{f}{Velocity offset with respect to the assumed systemic velocity of the source: 74~km s$^{-1}$ for Sgr~B2(N2) and Sgr~B2(N3), 64~km s$^{-1}$ for Sgr~B2(N4), and 60~km s$^{-1}$, for  Sgr~B2(N5) (see Sect.~\ref{kinematic-structure}).}
\tablefoottext{g}{Linewidth (FWHM) (see Sect.~\ref{kinematic-structure}).}
\tablefoottext{*}{Sgr~B2(N2)'s parameters are taken from \citet{belloche2016, muller2016, belloche2017}.}}
\end{table*}

We compute chemical abundances relative to H$_2$ using the H$_2$ column densities derived in Sect.~\ref{continuum}. From the emission lines detected toward Sgr~B2(N3), we derived a deconvolved source size of 0.4$\arcsec$ (see Sect.~\ref{source-size}), therefore we use  the upper limit on the H$_2$ column density derived from the SMA map at its original resolution of 0.3$\arcsec$ (Table~\ref{NH2calculations-bilan}). This gives us lower limits to the chemical abundances of the molecules detected toward Sgr~B2(N3). In the case of Sgr~B2(N2) and Sgr~B2(N5) we derived deconvolved source sizes $>$2.0$\arcsec$ from the ALMA continuum maps (see Table \ref{continuum-pics-table}), larger than the ALMA angular resolution of 1.6$\arcsec$. For both hot cores, we thus extrapolate the H$_2$ column density obtained at the ALMA angular resolution (Table~\ref{NH2calculations-bilan}) to the more compact region where the molecular emission comes from ($\sim$1.2$\arcsec$ and 1.0$\arcsec$ for Sgr~B2(N2) and Sgr~B2(N5) respectively, see Sect. \ref{source-size}). We assume spherical symmetry and a density profile proportional to $r^{-1.5}$, which implies a column density scaling with $r^{-0.5}$. Finally, the ALMA continuum maps yield an average deconvolved source size of 0.7$\arcsec$ for Sgr~B2(N4), smaller than the ALMA resolution. Therefore, we correct the H$_2$ column density obtained for Sgr~B2(N4) in Sect.~\ref{h2-column-density} for the beam dilution, that is we multiply it by $\frac{HPBW^2 + \theta_s^2}{\theta_s^2}$ = $\frac{1.6^2 + 0.7^2}{0.7^2}$. The resulting H$_2$ column densities are listed in Table~\ref{continuum-for-chemical-abundances} and used in Fig.~\ref{chemical_composition_plot}c to derive chemical abundances.

Figure~\ref{chemical_composition_plot} shows that the three new hot cores have similar compositions but differ from Sgr~B2(N2). Among the new hot cores, Sgr~B2(N3) appears to be closer to Sgr~B2(N5) than Sgr~B2(N4) in terms of chemical content. Figure~\ref{chemical_composition_plot}b shows indeed that Sgr~B2(N3) and Sgr~B2(N5) have, within a factor of 2, the same abundances relative to \ce{C2H5CN}. It is also the case for Sgr~B2(N4), except for \ce{CH3NCO} and \ce{CH3SH}, which are more abundant  in this source. \ce{NH2CHO} is not detected toward Sgr~B2(N4). Relative to H$_2$, Sgr~B2(N4) has COM abundances roughly one order of magnitude below that of Sgr~B2(N5), while Sgr~B2(N2) and Sgr~B2(N3) lie roughly one order of magnitude above (Fig.~\ref{chemical_composition_plot}c). \ce{C2H5CN} and \ce{NH2CHO} are in particular much more abundant in Sgr~B2(N2).

\begin{table}[!t]
\begin{center}
  \caption{\label{continuum-for-chemical-abundances} H$_2$ column densities used to derive chemical abundances.} 
  \setlength{\tabcolsep}{1.5mm}
  \begin{tabular}{cccc}
    \hline
Source &  $\theta_s$\tablefootmark{a} & $N_{\rm H_2}$\tablefootmark{b} & $n$\tablefootmark{c}   \\  
       & ($\arcsec$)&   (10$^{24}$ cm$^{-2}$) & (10$^7$ cm$^{-3}$)\\
    \hline
    \hline 
N2 & 1.2 & 1.8    &  1.4 \\
N3 & 0.3 & $<$0.5 &  $<$1.6 \\
N4 & 0.7 & 2.9    &  3.9 \\
N5 & 1.0 & 1.0    &  0.9 \\
    \hline
\end{tabular}
\end{center} 
\tablefoot{\tablefoottext{a}{Sizes for which the new H$_2$ column densities are calculated as described in Sect.~\ref{chemical-composition}.}
\tablefoottext{b}{H$_2$ column density calculated for the size indicated in the previous column. This value is used in Fig.~\ref{chemical_composition_plot} to plot chemical abundances relative to H$_2$.}
\tablefoottext{c}{Mean particule density calculated as $n = \frac{N_{\rm H_2}}{\theta_s} \times \frac{\mu_{\rm H_2}}{\mu}$, with $\mu$ = 2.37 the mean molecular weight per free particule \citep{kauffmann2008}.}}
\end{table}

\section{Discussion}

         \subsection{Discovery of three new hot cores in Sgr~B2(N)}
\label{newhotcores}

In Sect.~\ref{search-new-hot-cores} we presented the detection of three new hot cores in Sgr~B2(N), which we called Sgr~B2(N3), Sgr~B2(N4), and Sgr~B2(N5), based on the detection of high spectral line density regions in Fig.~\ref{hot_cores_contour_map}. While Sgr~B2(N3) is not detected in the ALMA continuum maps, Sgr~B2(N4), and Sgr~B2(N5) show a faint continuum level with H$_2$ column densities about 28 and 16 times lower than the one calculated for Sgr~B2(N1). The two main hot cores Sgr~B2(N1) and Sgr~B2(N2) have high H$_2$ column densities of 1.3$\times$10$^{25}$ cm$^{-2}$ and 1.6$\times$10$^{24}$ cm$^{-2}$ at a resolution of $\sim$1.6$\arcsec$. In a previous analysis of Sgr~B2(N) based on the same ALMA survey, \cite{bellochescience} calculated a peak H$_2$ column density at 98.8~GHz of 4.2$\times$10$^{24}$~cm$^{-2}$ for Sgr~B2(N2), for a beam size of $1.8\arcsec \times 1.3\arcsec$, which is about 3 times higher than our result. This difference results from the dust mass opacity coefficient they assumed ($\kappa_{\rm 98.8 GHz}$ = 4.3$\times$10$^{-3}$~cm$^2$~g$^{-1}$), $\sim$1.3 times smaller than the value we derive for this frequency based on our combined analysis of the ALMA and SMA data sets. In addition, they assumed that all the continuum emission is due to dust while we took into account the contribution of the free-free emission. This results in an additional factor $\sim$2 on the H$_2$ column density. From the flux density measured toward Sgr~B2(N2) in the SMA map at its original resolution we derived a H$_2$ column density of 5.1$\times$10$^{24}$ cm$^{-2}$. Based on the same map, \citet{qin2011} calculated an H$_2$ column density of 1.5$\times$10$^{25}$~cm$^{-2}$ toward Sgr~B2(N2). They used the same dust temperature, but they assumed optically thin emission and a dust mass opacity coefficient $\kappa_{\rm 343 GHz}$ = 6.8$\times$10$^{-3}$~cm$^2$~g$^{-1}$ that is 3.8 times smaller than the value we obtain at this frequency from our combined ALMA/SMA analysis. In addition, we derived a dust opacity of 0.61 for Sgr~B2(N2) in the original SMA map (see Table~\ref{tau-calculations}), which implies that they underestimated the column density by a factor $\sim$1.3. The two effects explain the factor $\sim$3 difference between the two studies.\\

Sgr~B2(N3), Sgr~B2(N4), and Sgr~B2(N5) are associated with 6.7~GHz methanol masers (see Fig.~\ref{hot_cores_contour_map}). Indeed, column 8 of Table~\ref{result-fit-channel-map} shows that within the uncertainties, the peak positions of line density of the three hot cores are consistent with the positions of the class II methanol masers reported in Sgr~B2(N) by \citet{caswell1996}. The 6.7~GHz methanol transition is the strongest and most widespread of the class II methanol masers \citep{menten1993}. It is known to be one of the best tracers of star formation and thought to be associated exclusively with regions forming high-mass stars \citep{minier2003, xu2008}. Indeed for a methanol transition to exhibit maser emission, both suitable physical conditions and a sufficient abundance of methanol are required. \citet{sobolev1997} showed that pumping requires dust temperature $>$ 150 K, high methanol column densities ($>$ 2$\times$10$^{15}$ cm$^{-2}$), and moderate densities ($n_{\rm H}$ $<$ 10$^8$ cm$^{-3}$) to excite the 6.7~GHz methanol maser transition. For these reasons low mass stars are not expected to produce class II methanol masers and indeed such masers have not been detected toward regions forming low-mass stars so far \citep[see, e.g.,][]{pandian2008}. The specific conditions for masers to exist make them powerful probes of high-mass star formation sites and confirm the nature of the new hot cores discovered in Sgr  B2(N). 

Methanol masers can also allow us to assess the evolutionary stages of these new sources as they are thought to trace evolutionary stages  from the IR dark cloud \citep{Pillai2006} to the UCHII phase. \citet{urquhart2014} proposed an evolutionary sequence for high-mass star formation in which methanol masers and HII regions trace two different phases, with the masers probing an earlier stage in the high-mass star formation process. \citet{walsh1997} showed that methanol maser emission is detectable before radio continuum emission, that is before the formation of UCHII regions. The masers are thought to be associated with deeply embedded high-mass protostars not evolved enough to ionize the surrounding gas and produce a detectable HII region. The maser emission remains active after first, a hyper and then an ultra compact HII region has formed around the star for a significant portion of the UCHII phase, during which the methanol molecules are shielded from the central UV radiation by the warm dust in the UCHII region's slowly expanding molecular envelope whose emission also provides its mid-IR pumping photons. Finally, the maser emission stops as the UCHII region expands. This picture is supported by recent radio observations with the greatly increased  sensitivity of the newly expanded Karl G. Jansky Very Large Array (JVLA). \citet{Hu2016} find radio continuum emission in the vicinity of a third of their sample of 372 methanol masers. This is a significantly higher percentage than found by \citet{Walsh1998} with the Australia Compact Array (ATCA), which is due to the fact that the new JVLA images are $\sim 20$ times deeper; ie., have a typical rms noise level $45~\mu$~Jy~beam$^{-1}$ at 4--8 GHz compared to the $\sim 1$~mJy~beam$^{-1}$ (at 8.64 GHz) of the Walsh et al. ATCA data. Most likely, at the JVLA's higher sensitivity, one is able to detect the emission from hypercompact HII regions surrounding the still accreting protostar exciting the masers \citep{Keto2007}. Due to such regions' compactness, their radio emission is very weak. However, an increasing number of such objects is now being detected with the JVLA \citep{Rosero2016, Hu2016}. \citet{vanderwalt2005} estimated the lifetime of class II methanol masers between 2.5~$\times$~10$^4$ and 4.5~$\times$~10$^4$~yr~ (depending on the assumed IMF). After this period the HII region then exists without the maser emission. These last considerations suggest that Sgr~B2(N1) and Sgr~B2(N2) are already more evolved than the new hot cores for which the maser emission still exists. Among them, Sgr B2(N5) is associated with both a UCHII region and a class II methanol maser, which might suggest that it is in a phase between Sgr~B2(N3)/Sgr~B2(N4) and Sgr~B2(N1)/Sgr~B2(N2), when a UCHII region has formed and coexists with maser emission.

           \subsection{Chemical composition of the new hot cores}

Another way to assess the evolutionary stages of the hot cores is to compare their chemical compositions. Sgr~B2(N3), Sgr~B2(N4), and Sgr~B2(N5) have spectral line densities of 31, 11, and 22 lines per GHz above 7$\sigma$, respectively (see Sect.~\ref{line-identification}), much lower than the two main hot cores Sgr~B2(N1) and Sgr~B2(N2). Our LTE model allowed us to identify about 91\%, 89\%, and 93\% of the lines detected above $7\sigma$ that have been assigned to 22, 23, and 25 main species, respectively (see Table~\ref{tab-chemical-composition}). This is much less than the 52 species identified so far toward Sgr B2(N2) based on the EMoCA survey (Belloche, priv. comm.). Although Sgr~B2(N3) has the highest spectral line density among the new hot cores, Sgr~B2(N5) contains the largest number of identified molecules. However fewer isotopologs and vibrationally excited states of these molecules are detected toward Sgr~B2(N5) than Sgr~B2(N3). Sgr~B2(N4)'s spectrum shows a low spectral line density and less species have been identified toward this source. 
Most of the remaining unidentified lines are thought to belong to vibrationally excited states of already known molecules for which the spectroscopic predictions are still missing \citep{belloche2013}, but the presence of new molecules in the ALMA spectra is not excluded.

In order to assess whether the differences reported above are real or sensitivity limited, we need to compare in more details the chemical content of the three new hot cores. In Sect.~\ref{chemical-composition}, we reported the chemical abundances of (complex) organic molecules detected toward the new hot cores plus Sgr~B2(N2) (Fig.~\ref{chemical_composition_plot}). Sgr~B2(N3), Sgr~B2(N4), and Sgr~B2(N5) have a similar chemical composition but differ from Sgr~B2(N2). According to our chemical model presented in Fig. 3 of \citet{bellochescience}, the chemical abundance of \ce{C2H5CN} in the gas phase after sublimation is expected to decrease with time while the abundance of \ce{C2H3CN} increases. The abundance of \ce{CH3CN} is also expected to increase with time after sublimation. Table~\ref{ratios-chemical-abundances} and Fig.~\ref{chemical_composition_plot} show that Sgr~B2(N2) has the lowest ratios of [\ce{C2H3CN}]/[\ce{C2H5CN}] and [\ce{CH3CN}]/[\ce{C2H5CN}] among the four hot cores, which could suggest that it is less evolved than the three new hot cores. According to our chemical model presented in Fig.~5 of \citet{belloche2017}, the chemical abundances of \ce{CH3NCO} and HNCO in the gas phase are also expected to increase with time after sublimation. Sgr~B2(N2) also shows low [\ce{CH3NCO}]/[\ce{C2H5CN}] and [HNCO]/[\ce{C2H5CN}] ratios, consistent with the hypothesis made above. However, all these ratios involve \ce{C2H5CN} in the denominator, which may dominate the general trend discussed above. A deeper analysis involving ratios of various molecules will be necessary to constrain the evolutionary stages of Sgr B2(N)'s hot cores from their chemistry in a more reliable way, especially because the conclusion drawn here that Sgr~B2(N2) appears chemically younger than the three new hot cores is in contradiction with the conclusion drawn in Sect.~\ref{newhotcores} from the associations of the hot cores with methanol masers and/or UCHII regions (see discussion in Sect.~\ref{physical_properties}, however).

One should also note that the chemical timescales of molecules within any single model may not relate directly to the dynamical age of the source, as pointed out by \citet{garrod2008}. This may be exacerbated by the physical differences between the sources, which will also affect the chemical timescales. Only a more explicit and individualized treatment of the time- and space-dependent chemistry in each source is likely to give an accurate explanation for the observed chemical differences.

\begin{table}[!t]
\begin{center}
  \caption{\label{ratios-chemical-abundances} Column density ratios of selected molecules.} 
  \setlength{\tabcolsep}{1.5mm}
  \begin{tabular}{ccccc}
    \hline
Source & $\frac{[\ce{C2H3CN}]}{[\ce{C2H5CN}]}$ &   $\frac{[\ce{CH3CN}]}{[\ce{C2H5CN}]}$  & $\frac{[\rm HNCO]}{[\ce{C2H5CN}]}$  & $\frac{[\ce{CH3NCO}]}{[\ce{C2H5CN}]}$  \\  
    \hline
    \hline 
N2 & 0.06 & 0.33 & 0.30 & 0.03  \\
N3 &  0.22 & 1.92 & 0.57 & 0.24 \\
N4 &  0.13 & 2.80 & 0.21 & 0.67 \\
N5 &  0.13 & 1.31 & 0.32 & 0.13    \\
    \hline
\end{tabular}
\end{center} 
\end{table}

           \subsection{Spatial distribution of the molecules toward the new hot cores}

\begin{figure*}[!t]
\resizebox{\hsize}{!}
{\begin{tabular}{ccc}
 \includegraphics[width=20cm]{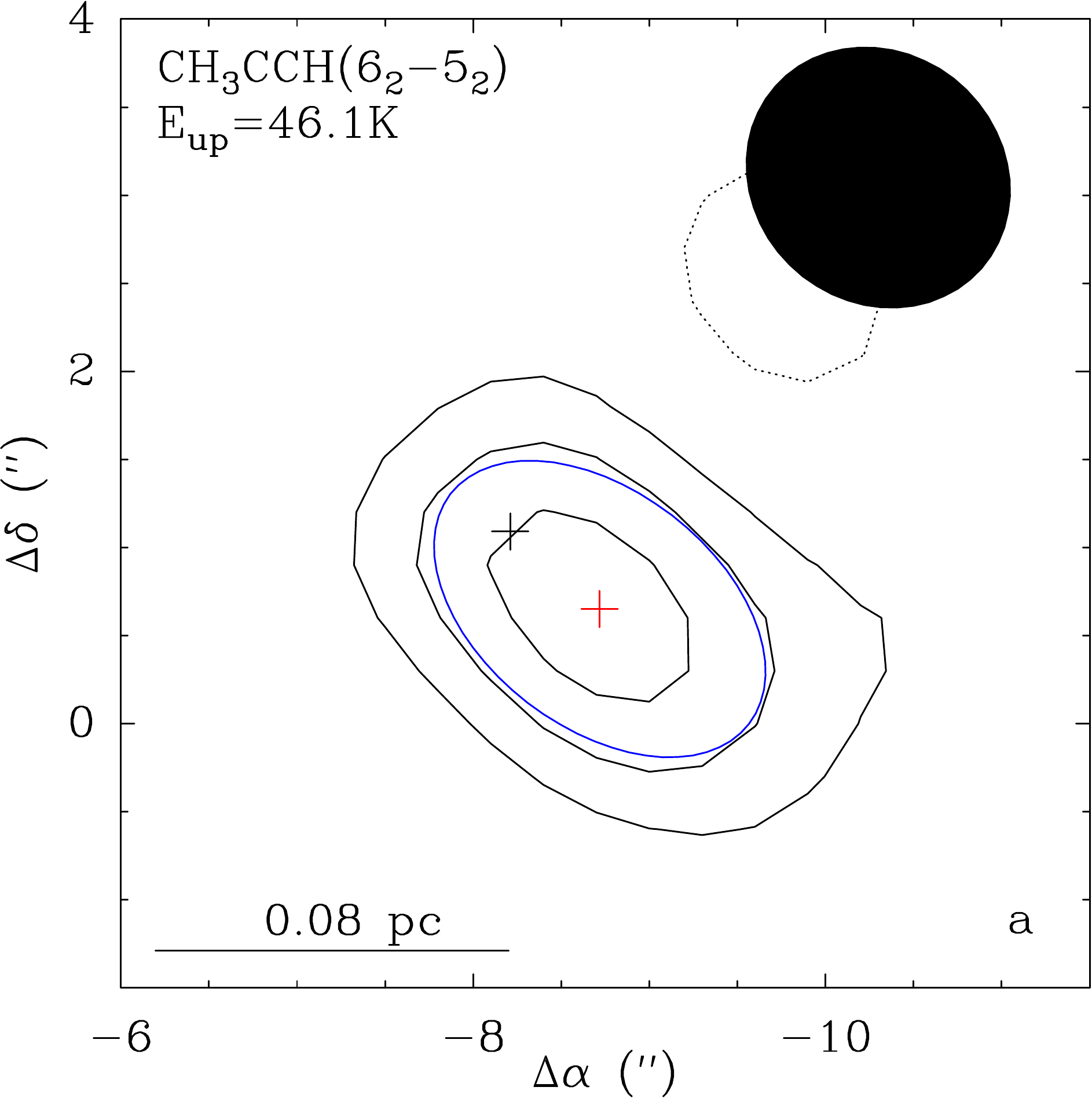} &
 \includegraphics[width=\hsize]{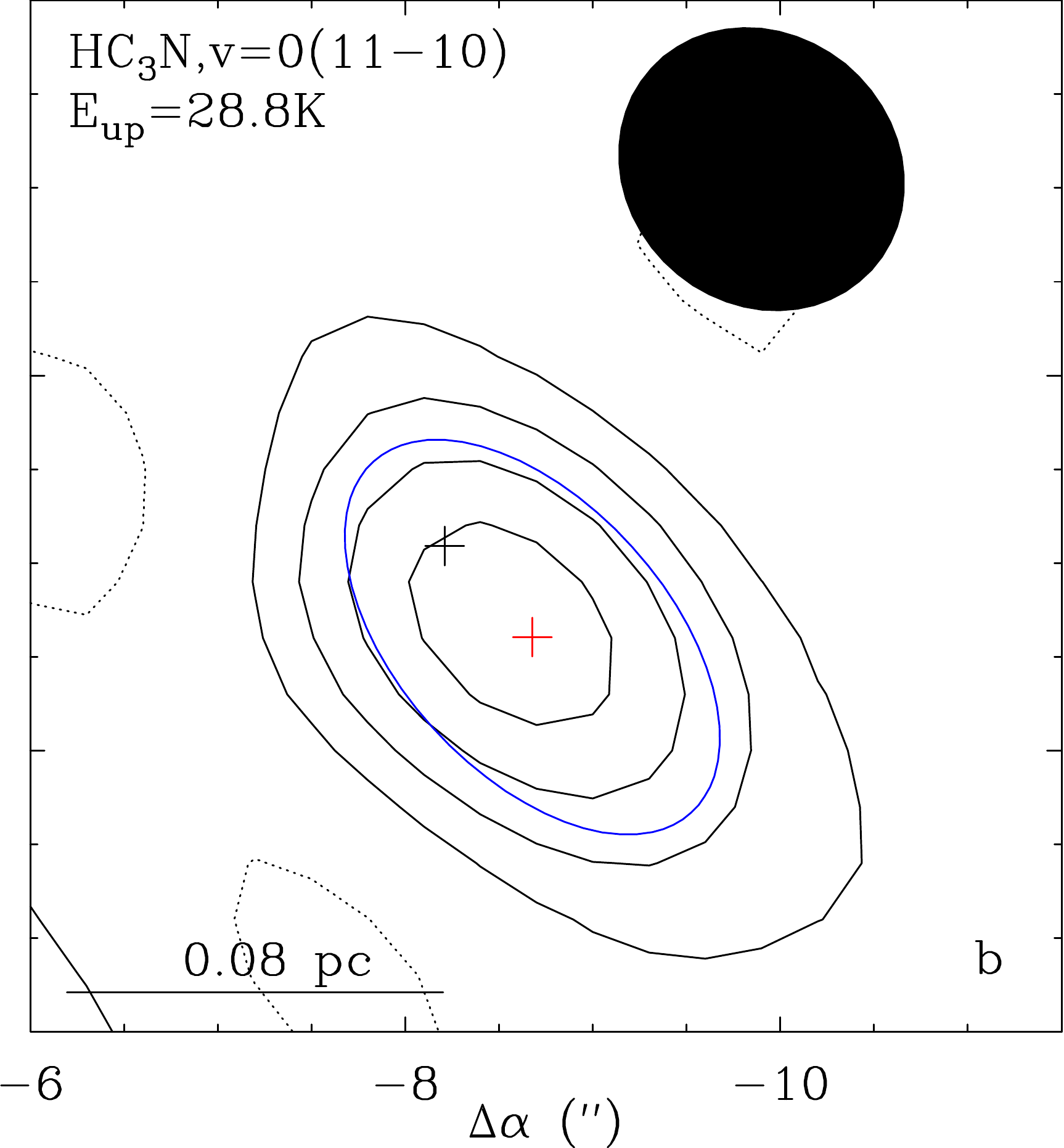} &
 \includegraphics[width=\hsize]{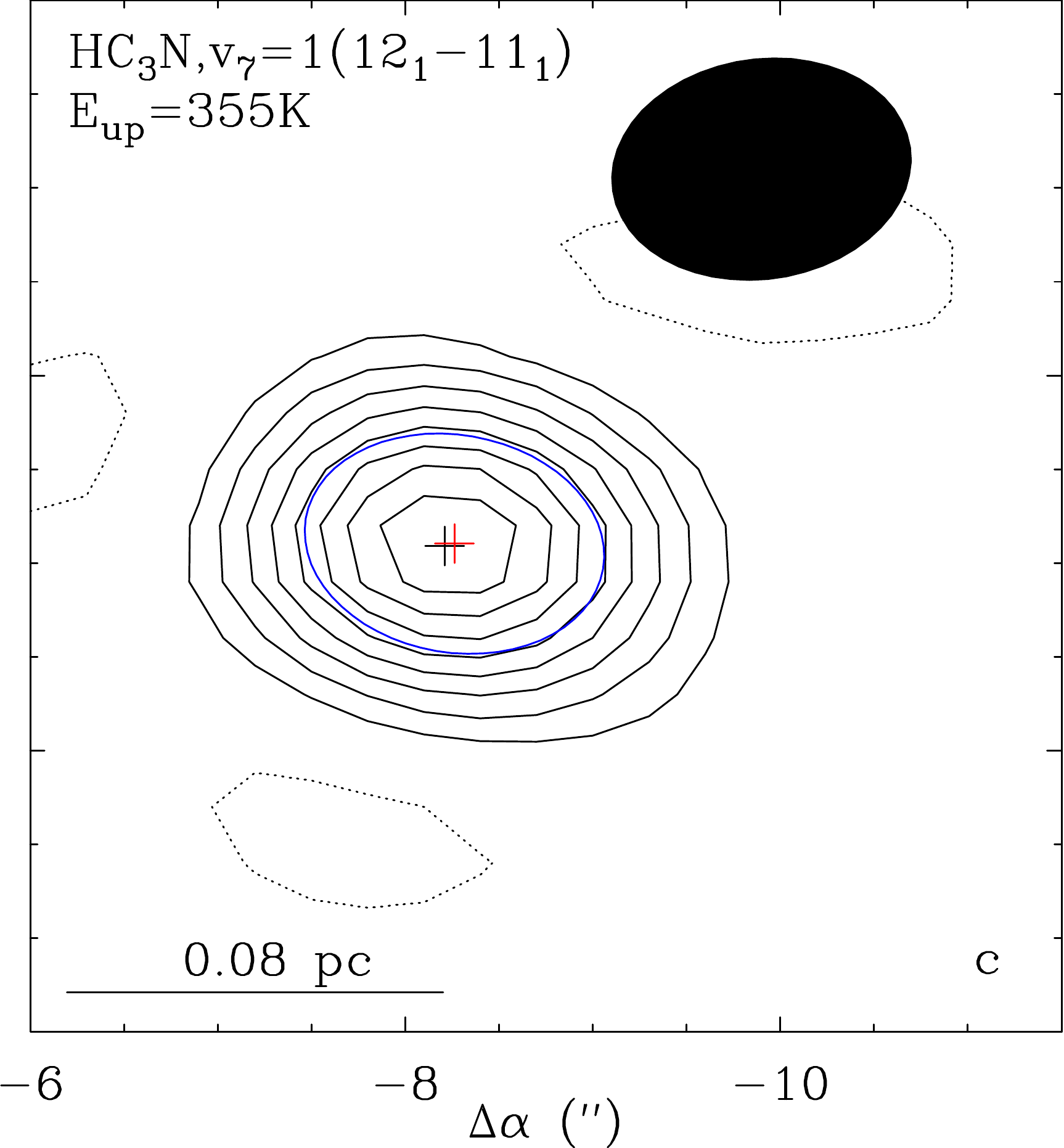} \\
\end{tabular}}
 \caption{\label{fig_discussion_spatial_distribution} Integrated intensity maps of selected transitions detected in Sgr~B2(N3). The contour levels start at 3$\sigma$ (rms $\sim$3~mJ/beam) and increase with a step of $3\sigma$. In each map the blue ellipse shows the result of the 2D-Gaussian fit. The black cross represents the position of the hot core derived from Fig.~\ref{hot_cores_contour_map} and the red cross marks the peak position derived from the fit. The black filled ellipse represents the synthesized beam. The upper level energy of each transition is indicated in temperature unit in each panel.}
\end{figure*}

In Sect.~\ref{spatial-distribution} we investigated the spatial distribution of the molecules detected toward the three new hot cores. Most of the identified species show compact emission with emission peaks within distances of $\sim$0.3$\arcsec$, 0.6$\arcsec$, and 0.4$\arcsec$ of the positions of Sgr~B2(N3), Sgr~B2(N4), and Sgr~B2(N5), respectively (see Fig.~\ref{spatial-distribution-plot}). 
This led us to define the reference position of each hot cores as the position where their spectral line density peaks (Sect.~\ref{search-new-hot-cores}).
Within the uncertainties the emission peak of most species is consistent with this position. Only three molecules, SO, \ce{HC3N}, and \ce{CH3CCH}, show significant offsets between their emission peaks and the position of Sgr~B2(N3). The transitions of SO detected toward Sgr B2(N3) have an emission peak located $\sim$0.5$\arcsec$ South-West of the hot core on average (Fig.~\ref{spatial-distribution-plot}a and top row of Fig.~\ref{kinematic-structure-maps}). Figures~\ref{fig_discussion_spatial_distribution}a,b present the integrated intensity maps of two vibrational ground state transitions of \ce{CH3CCH} and \ce{HC3N}. They both show the same elongated shape as the SO maps. In both cases the peak positions of the emission given by the 2D-Gaussian fit to the map is shifted beyond 0.5$\arcsec$ from the hot core position, due to this extended emission. On the contrary, transitions from within the first vibrationally excited state of \ce{HC3N} have a compact emission that peaks at a position consistent with the position of Sgr B2(N3). 
The $\varv_7$=1 transitions of \ce{HC3N} thus trace the hot core better than the $\varv=0$ transitions. The nature of the extended emission traced by the $\varv=0$ transitions of \ce{HC3N}, SO, and \ce{CH3CCH} is unclear. It may be related to the outflow that is seen along the same direction (Fig.~\ref{kinematic-structure-maps}).

          \subsection{Physical properties of the new hot cores}
\label{physical_properties}

In Sects.~\ref{source-size}, \ref{Trot}, and \ref{kinematic-structure}, we presented basic physical properties of the three new hot cores discovered in Sgr~B2(N). They show similarities with one of the already known hot cores, Sgr~B2(N2). All of them have kinetic temperatures of $\sim$140--180~K. Their spectra show narrow lines with typical linewidth of $\sim$5~km~s$^{-1}$ similar to the values derived by \citet{belloche2016} for Sgr~B2(N2), ranging from 4.7 to 6.5~km s$^{-1}$. Sgr~B2(N4) and Sgr~B2(N5) have mean molecular emission sizes of $\sim$1.0$\arcsec$, which is slightly smaller than the size of Sgr~B2(N2) \citep[$\sim$1.2$\arcsec$, with values ranging between 0.8$\arcsec$ and 1.5$\arcsec$;][]{belloche2016,belloche2017}. Sgr~B2(N3) is more compact, with a molecular emission size of 0.4$\arcsec$.

In Table \ref{best-fit-parameters} we presented the parameters of our best fit models for ten molecules. For each hot core we have decided to adopt the mean angular size derived from the transitions showing resolved emission. As the hot cores appear to be resolved for only few species (especially Sgr B2(N3) for which transitions from only two species could be used to derive the source size, see Table \ref{gaussian-fit}), we decided to use a single source size for each hot core to model all molecules. One has to keep in mind this last consideration while comparing the chemical composition of the hot cores in Section \ref{chemical-composition} because the column densities derived from the model strongly depend on the adopted source size. For instance, we used a source size of 1.0$\arcsec$ to model the spectrum toward Sgr B2 (N5), although the results of the 2D-Gaussian fit to the integrated intensity maps of \ce{C2H5CN} transitions suggest an average emission size of $\sim$0.5$\arcsec$ for this molecule (see Table \ref{gaussian-fit}). The column density of \ce{C2H5CN} in Sgr B2(N5) might thus be underestimated.

In Sect.~\ref{kinematic-structure} we highlighted bipolar structures in the integrated intensity maps of the wings of typical outflow tracers. The North-South velocity gradient observed toward Sgr~B2(N3) in the maps of two SO lines and the OCS(8-7) transition most probably suggests the presence of an outflow (Fig.~\ref{kinematic-structure-maps}). The SO(2$_3$-1$_2$) and CS(2-1) transitions investigated toward Sgr~B2(N5) show a clear bipolar structure also suggesting the presence of an outflow. However, what the other transitions of SO trace toward this source is less clear. Nevertheless our interpretation that Sgr~B2(N3) and Sgr~B2(N5) drive outflows is reinforced by the fact that H$_2$O maser emission is found in the close vicinity of both sources \citep[][ see also Fig.~\ref{appendix_h2o}]{McGrath2004}.

\citet{higuchi2015} have recently reported the presence of a bipolar molecular outflow in the East--West direction in Sgr B2(N1), on the basis of the SiO(2-1) and \ce{SO2}(12$_{\rm4,8}$-13$_{\rm 3,11}$) transitions in the same data set used here. They derived an average dynamical time of $\sim$5 $\times$ 10$^3$ years, similar to our results for Sgr B2(N3) and Sgr B2(N5)($\sim3 \times 10^3$~yr).
The fact that no outflow has been detected toward Sgr B2(N4) does not necessarily mean a real lack of outflow motion but instead it might reflect the youth of the source compared to the others. The outflow structure in this case might be too small to be detected at the resolution of our ALMA survey. 

\citet{codella2004} explored the possibility of using molecular outflows to estimate the age of the associated source. They investigated a large survey of outflows toward UCHII regions. They proposed an evolutionary scheme for regions forming high-mass stars in which class II methanol masers appear before an outflow is detectable, then both coexist in the same phase before the maser switches off as the UCHII region expands. For these reasons, the fact that no outflow has been detected toward Sgr B2(N2) might suggest that it is more evolved than the other cores because a UCHII region is already detected in the source and no class II methanol maser has been reported. However, this hypothesis is in contradiction with the conclusion derived from the comparison of the chemical composition of Sgr~B2(N2) and the three new hot cores. An alternative explanation could be that there are two sources in Sgr~B2(N2). Indeed, there is an offset of 0.43$\arcsec$ ($\sim$3600~au) between the position of peak line density and the position of the UCHII region, K7 (see Fig.~\ref{hot_cores_contour_map} and Table~\ref{result-fit-channel-map}). A similar situation occurs in the star-forming region W51e2 at a distance of 5.4~kpc. This source contains two objects separated by 0.8$\arcsec$, a hypercompact HII region, W51e2-W, and a hot molecular core, W51e2-E \citep[][]{shi2010,ginsburg2017}. At the distance of Sgr~B2(N), these two objects would have a separation of 0.5$\arcsec$, similar to the offset seen between K7 and Sgr~B2(N2).
Observations at higher angular resolution with ALMA will be necessary to test this scenario.

           \subsection{Star formation timescales and evolutionary sequence in Sgr~B2(N)}

In a previous analysis of Sgr~B2(N), \citet{belloche2013} used the UCHII regions reported by \citet{gaume1995} to estimate a star formation rate of 0.028-0.039 $\Msol$ yr$^{-1}$ averaged over 10$^5$ years for the whole Sgr B2 complex. 
We assume a constant star formation rate to estimate the lifetime of Sgr B2(N)'s hot cores. Combining the results of \citet{gaume1995} and \citet{depree2015}, we count ten HII regions within the ALMA primary beam centered between Sgr~B2(N1) and Sgr~B2(N2) (see Fig.~\ref{ALMA_continuum_map}). Eight of these sources are UCHII regions. Only five hot cores are detected in the same area. Considering a lifetime of $\sim$10$^5$ yr for the UCHII regions \citep{peters2010}, we estimate a lifetime of 10$^5$ $\times$ $\frac{5}{8}$ $\sim$6$\times$10$^4$~yr for Sgr~B2(N)'s hot cores. In the same way we derive a statistical lifetime of about 4$\times$10$^4$~yr for the class II 6.7~GHz methanol masers detected in this area, which is, given the low number statistics of the Sgr~B2(N) region, surprisingly consistent with the lifetime derived by \citet{vanderwalt2005} for class II methanol masers (2.5$\times$10$^4$ to 4.5$\times$10$^4$~yr). 

\begin{figure*}[!t]
\begin{center}
 \includegraphics[width=\hsize]{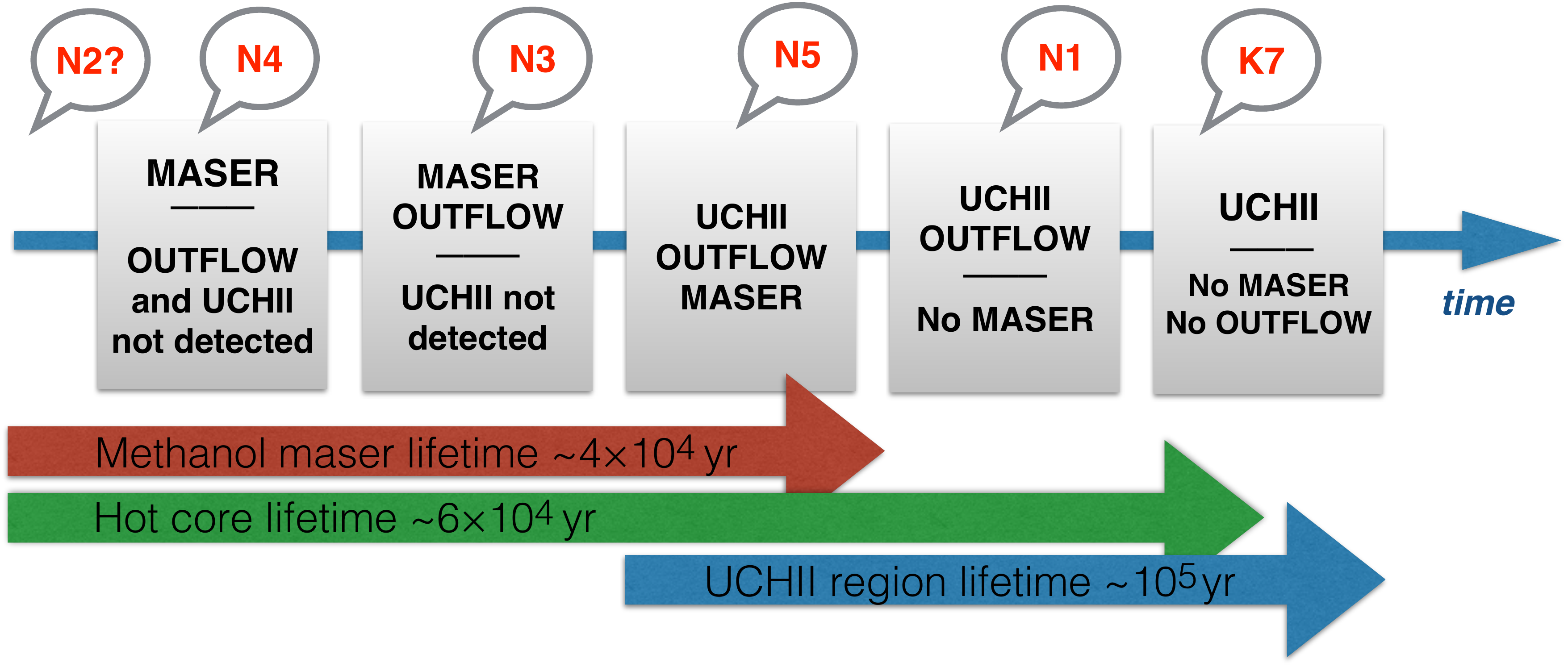}    
 \caption{\label{timescale-plot}  Proposed evolutionary sequence for all the hot cores embedded in Sgr~B2(N) based on their associations with UCHII regions, class II methanol masers, and outflows. K7 is the UCHII region located close to Sgr~B2(N2) and may be a distinct source.}
\end{center}
\end{figure*}

Based on Fig. 5 of \citet{codella2004} and the statistical lifetimes calculated above, we propose in Fig. \ref{timescale-plot} an evolutionary sequence of the hot cores embedded in Sgr B2(N). Among the three new hot cores, Sgr~B2(N5) would be the most evolved one because it is associated with both an outflow and a class II methanol maser, and has already entered the UCHII phase. Sgr~B2(N4) appears like the youngest core because it is only associated with class II methanol maser emission. With an associated methanol maser and a detected outflow but no associated UCHII region, Sgr~B2(N3) would be in-between. As mentioned in Sect.~\ref{chemical-composition}, the status of Sgr~B2(N2) is unclear. Its association with a UCHII region but no methanol maser and no outflow would suggest that it is more evolved than the new hot cores, but its chemical composition seems to suggest the opposite. As discussed in Sect.~\ref{physical_properties}, one reason for this apparent contradiction could be that Sgr~B2(N2) actually contains two distinct sources, one associated with the UCHII region K7, and therefore more evolved than all the other hot cores, and the other one, the hot core, with a line density peak shifted from K7, and still too young to show outflow emission on the scales probed with ALMA or to harbor a UCHII region. 

As shown in Table~\ref{continuum-pics-table}, the five hot cores detected in Sgr~B2(N) have peak column densities that differ by more than one order of magnitude. Therefore, they may form stars of different final masses and our attempt to classify them with a single evolutionary sequence should be considered as tentative only. A deeper analysis of their properties, in particular at higher angular resolution, will be necessary to improve our understanding of their respective evolutionary states.

There is one complication to the evolutionary sequence proposed above: as discussed in Sects. 1.1 and 4.1 of \citet{belloche2008} and shown here in Fig.~\ref{appendix_h2o}, Sgr~B2(N1) coincides with the centroid position of a powerful \ce{H2O} maser compact ($4\arcsec\times2\arcsec$ sized) outflow \citep{Reid1988}. For their (collisional) pumping, H$_2$O masers require temperatures of $\sim$400~K and densities of $\sim 10^9$~cm$^{-3}$, much higher than the values derived in this paper for Sgr~B2(N1). These conditions are met in the post shock regions of fast (J) shocks \citep{Elitzur1989, Hollenbach2013}. The compact H$_2$O maser outflow may originate from a different source than that which drives the UCHII region K2, a situation reminiscent of  the archetypical UCHII region W3(OH), which has powerful OH and methanol masers in its expanding envelope \citep{Menten1992} and is separated by 5$\arcsec$ from the multiple hot core W3(OH)-\ce{H2O} \citep{Wyrowski1999}. Like Sgr~B2(N1), the latter drives a powerful bipolar \ce{H2O} maser outflow \citep{Hachisuka2006}, but shows no methanol maser emission. 

We also mention that several H$_2$O masers are associated with a faint peak of continuum emission in our ALMA data (see Fig.~\ref{appendix_h2o}), which also coincides with a region of moderately enhanced spectral line density located $\sim$3$\arcsec$ West of Sgr~B2(N5) (Fig.~\ref{hot_cores_contour_map}). This region probably harbors an additional (faint) hot core that is not associated with any UCHII region or class II methanol maser.

\section{Conclusions}

We used the 3~mm line survey EMoCA conducted with ALMA in its cylces 0 and 1 to search for new hot cores in the Sgr~B2(N) region, taking advantage of the high sensitivity of these observations. We report the discovery of three new hot cores that we called Sgr~B2(N3), Sgr~B2(N4), and Sgr~B2(N5), located at ($\alpha _{\rm J2000}$=17$^h$47$^m$19.248$^s$,  $\delta _{\rm J2000}$ = -28$^{\rm o}$22$'$14.91$\arcsec$), 
($\alpha _{\rm J2000}$  = 17$^h$47$^m$19.528$^s$, $\delta _{\rm J2000}$ = -28$^{\rm o}$22$'$32.41$\arcsec$), and ($\alpha _{\rm J2000}$  = 17$^h$47$^m$20.047$^s$ , $\delta _{\rm J2000}$ = -28$^{\rm o}$22$'$41.34$\arcsec$) respectively. We analyzed the line survey to characterize their chemical composition and physical structure, and we searched for outflows and associations with UCHII regions or methanol masers. Our main results are summarized as follows: 

\begin{enumerate}

\item{Sgr~B2(N3), Sgr~B2(N4), and Sgr~B2(N5) have spectral line densities above 7$\sigma$ of 31, 11, and 22 lines per GHz respectively, which qualify them as hot cores. About 91\%, 89\%, and 93\% of these lines have been identified and assigned to 22, 23, and 25 main species, respectively.}

\item{The typical linewidth of the three new hot cores is 5~km~s$^{-1}$. They have rotational temperatures of $\sim$140--180~K and the maps of emission lines indicate mean source sizes of 0.4$\arcsec$ for Sgr~B2(N3) and 1.0$\arcsec$ for Sgr~B2(N4) and Sgr~B2(N5).}

\item{Assuming a dust temperature of 150~K similar to the rotational temperatures obtained from the emission lines, we derive a dust emissivity index $\beta \sim 1.2$ by combining ALMA and SMA data. We obtain H$_2$ column densities of $<$5$\times$10$^{23}$, 3$\times$10$^{24}$, and 1$\times$10$^{24}$~cm$^{-2}$ for Sgr~B2(N3), Sgr~B2(N4), and Sgr~B2(N5), respectively, for the sizes listed above.}

\item{We report the detection of outflows in Sgr~B2(N3) and Sgr~B2(N5) based on the analysis of three SO lines, and the OCS(8-7) and CS(2-1) transitions. The outflow dynamical time is $\sim$2.5$\times$10$^3$~yr for both sources. No outflow is detected toward Sgr~B2(N4) and Sgr~B2(N2).}

\item{Each new hot core is associated with a 6.7~GHz class II methanol maser. Sgr~B2(N4) is also associated with a UCHII region.}

\item{We derived the column densities and abundances of ten (complex) organic molecules toward Sgr~B2(N3), Sgr~B2(N4), and Sgr~B2(N5). The three sources share a similar chemical composition, with Sgr~B2(N3) resembling Sgr~B2(N5) a bit more than Sgr~B2(N4). However, they differ from Sgr~B2(N2) and several molecular ratios suggest that Sgr~B2(N2) is chemically less evolved than the three new hot cores.}

\item{Assuming a lifetime of $10^5$ yr for UCHII regions, we derive statistical lifetimes of $4 \times 10^4$ yr for the class II methanol maser phase and 6$ \times 10^4$ yr for the hot core phase in Sgr~B2(N).}

\end{enumerate}
Given that their peak column densities differ by more than one order of magnitude, the five hot cores may form stars of different final masses. Still, their associations with class II methanol masers, outflows, and/or UCHII regions tentatively suggest the following age sequence from the youngest to the oldest source: Sgr~B2(N4), Sgr~B2(N3), Sgr~B2(N5), Sgr~B2(N1). The status of Sgr~B2(N2) is puzzling. Its association with a UCHII region but no outflow and no methanol maser suggests that it should be the oldest source in this sequence. However, this contradicts its youth suggested by its chemical composition. An explanation may be that Sgr~B2(N2) contains two distinct sources, as suggested by the small angular offset separating its embedded UCHII region from the line density peak of the hot core. On-going observations at higher angular resolution with ALMA will help understanding the status of this source.


\begin{acknowledgements}
We thank Sheng-Li Qin and Peter Schilke for providing us the SMA map 
of Sgr~B2(N) in electronic form. This paper makes use of the following ALMA data: 
ADS/JAO.ALMA\#2011.0.00017.S, ADS/JAO.ALMA\#2012.1.00012.S. 
ALMA is a partnership of ESO (representing its member states), NSF (USA), and 
NINS (Japan), together with NRC (Canada), NSC and ASIAA (Taiwan), and KASI 
(Republic of Korea), in cooperation with the Republic of Chile. The Joint ALMA 
Observatory is operated by ESO, AUI/NRAO, and NAOJ. The interferometric data 
are available in the ALMA archive at https://almascience.eso.org/aq/. 
This work has been in part supported by the Deutsche Forschungsgemeinschaft 
(DFG) through the collaborative research grant SFB 956 ``Conditions and Impact 
of Star Formation'', project area B3.
\end{acknowledgements}

\bibliographystyle{aa}
\bibliography{biblioPhD}

\begin{appendix}

\section{Estimating the level of free-free emission}
         \label{appendix-continuum}

\begin{figure*}[!t]
\resizebox{\hsize}{!}
{\begin{tabular}{cc}
   \includegraphics[width=\hsize]{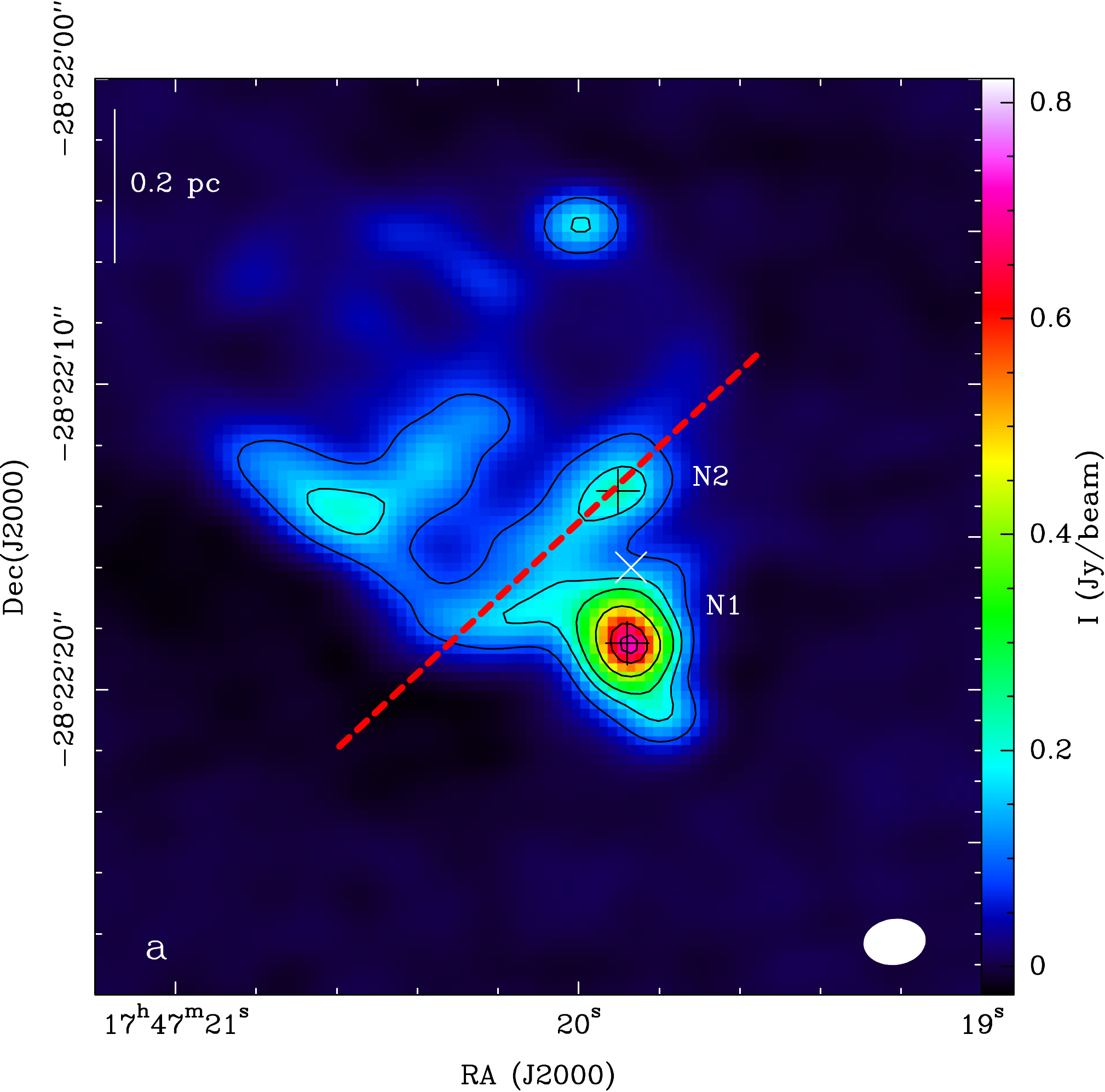} & 
   \includegraphics[width=16.5cm]{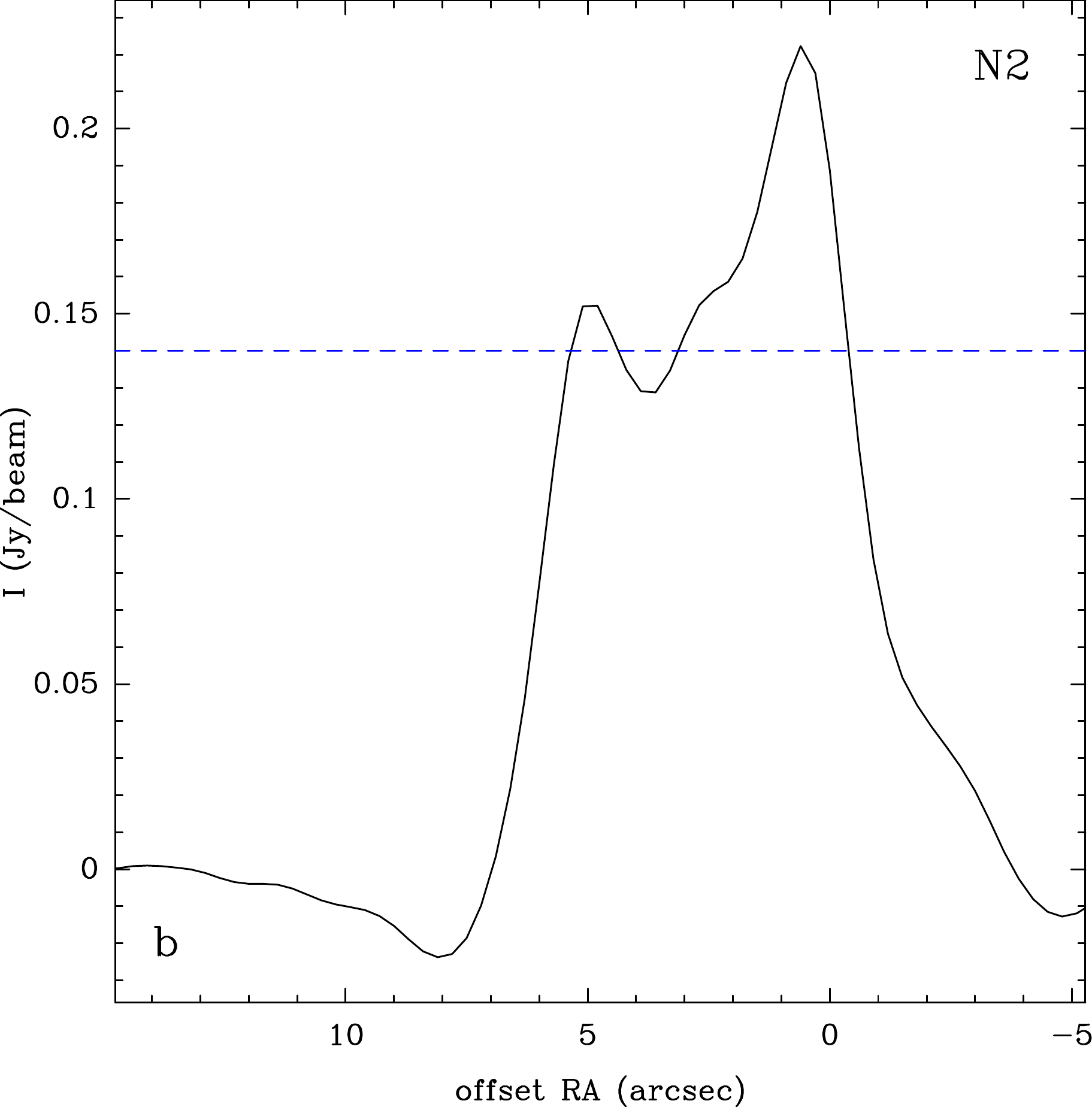} \\
\end{tabular}}
\caption{\label{appendix-continuum-alma-strip}  \textbf{a} Continuum map obtained at 86.5~GHz with ALMA. The contour levels are 20$\sigma$, 40$\sigma$, 60$\sigma$, 100$\sigma$, 140$\sigma$, and 160$\sigma$, with $\sigma$=4.3~mJy. The black crosses mark the peak positions of the continuum emission observed toward Sgr~B2(N1) and Sgr~B2(N2) derived from Fig.~\ref{ALMA_continuum_map}. The white cross represents the phase center. The filled ellipse shows the clean beam (2.04$\arcsec$ $\times$ 1.51$\arcsec$, PA=-83.3$^{\rm o}$). \textbf{b} Profile of the continuum emission along the direction plotted in red in panel \textbf{a}. The blue line marks the estimated level of free-free emission.}   
\end{figure*}

\begin{figure}[!t]
\resizebox{\hsize}{!}
{\begin{tabular}{c}
   \includegraphics[width=\hsize]{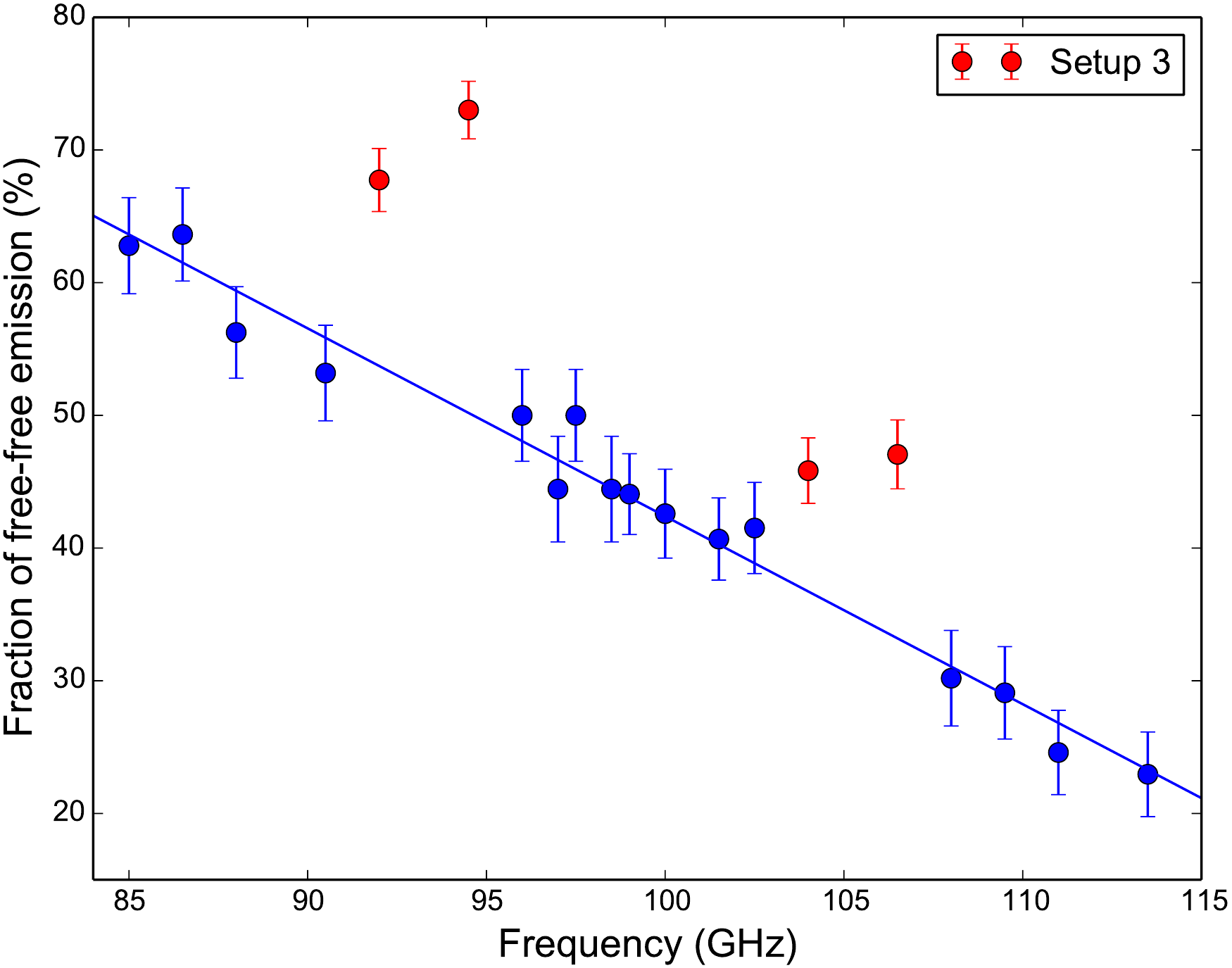} \\   
\end{tabular}}
\caption{\label{appendix-continuum-alma-freefree} Fraction of free-free emission relative to the total flux density measured toward Sgr~B2(N2). This does not include the contribution of K7. The blue line is the weighted linear fit to the results excluding setup 3 that is shown in red.}   
\end{figure}

Here we present the procedure used to estimate the contribution of free-free emission expected toward Sgr~B2(N1), Sgr~B2(N2), and Sgr~B2(N5) in the ALMA continuum maps. As mentioned in Sect.~\ref{continuum-alma}, the overall shape of the extended continuum emission detected with ALMA is similar to the shape of the 1.3~cm free-free emission reported by \citet{gaume1995} which suggests that the extended emission detected with ALMA is dominated by free-free emission. Figure~\ref{appendix-continuum-alma-strip}b shows the profile of the continuum emission along the direction going through Sgr~B2(N2) indicated in red in Fig.~\ref{appendix-continuum-alma-strip}a. Because Sgr~B2(N2) is located in the direction of extended 1.3~cm free-free emission, we decompose the profile shown in Fig.~\ref{appendix-continuum-alma-strip}b into a peak, which we attribute to dust emission associated with Sgr~B2(N2), and a pedestal, which we attribute to the extended free-free emission. We proceed in the same way for the 20 ALMA continuum maps. The results are plotted in Fig.~\ref{appendix-continuum-alma-freefree}. The weighted linear fit to the results excluding setup 3 gives us the correction to subtract from the measured peak density $S^{\rm beam}_{\nu}$ as a function of frequency. For the maps belonging to the frequency range of setup 3 (with the lowest angular resolution), we directly apply the corrections estimated from the plots of the continuum profile. The corrections applied here take into account the contamination from the extended continuum emission detected in the ALMA continuum maps but a correction for the free-free emission arising from the UCHII region K7 that is associated with Sgr~B2(N2) still has to be applied. We estimate the flux expected from K7 based on the flux reported by \citet{depree2015}. They measured an integrated flux of 30~mJy at 44.2~GHz for a source size of 0.08$\arcsec$. We assume that the free-free emission of K7 is optically thin and thus that the measured flux is proportional to $\nu^{\alpha}$, with $\alpha=-0.1$. From this relation we extrapolate a flux of $\sim$28~mJy for the frequency range of the ALMA survey. This value is added to the correction for Sgr~B2(N2) derived  from Fig.~\ref{appendix-continuum-alma-freefree}. 

We proceed in the same way for Sgr~B2(N5), based on the flux reported by \citet{gaume1995} toward Z10.24. They measured an integrated flux of 36~mJy at 22.4~GHz for a source size $<$0.25$\arcsec$. We extrapolate this value to the ALMA frequency range assuming optically thin emission, and we and obtain $\sim$31~mJy. This value is directly subtracted from the values of the peak flux density measured in the ALMA continuum maps toward Sgr~B2(N5).

Finally in the case of Sgr~B2(N1) we use the flux measured by \citet{depree2015} toward K2 and K3 at 44.2~GHz. They measured integrated flux densities of 80~mJy toward K2 for a source size of 0.12$\arcsec$ and 242~mJy toward K3 for a source size of 0.27$\arcsec$. K3 being located 1.3$\arcsec$ North-East of K2, we estimate that only$\sim$15\% of the free-free emission arising from this UCHII region contributes to the total flux density measured in the ALMA synthesized beam of $\sim$1.6$\arcsec$ centered on K2. We thus obtain a total free-free contribution of $\sim$116~mJy that we subtract from the flux densities measured toward Sgr~B2(N1) in the ALMA continuum maps.

\definecolor{lightgray}{gray}{0.85}

\section{Additional tables : continuum analysis}
        \label{appendix-continuum-tables}

\begin{table}[!t]
\begin{center}
  \caption{\label{tab-continuum-n1} Peak flux densities and H$_2$ column densities toward Sgr~B2(N1).}
  \begin{tabular}{cllcc}
    \hline
Freq. & $S_{\nu}^{\rm ALMA}$ \tablefootmark{(a)} & \multicolumn{1}{c}{$S_{\nu}^{\rm SMA}$ \tablefootmark{(b)}} & $N_{\rm H_2}^{\rm ALMA}$          & $N_{\rm H_2}^{\rm SMA}$ \\
(GHz)& \multicolumn{2}{c}{(Jy/beam)}                           & \multicolumn{2}{c}{(10$^{25}$ cm$^{-2}$)} \\
    \hline
    \hline 
 85.0  & 0.569(3)  &  24.15(11) &   1.01(1)  &  0.67(1) \\
 86.5  & 0.594(3)  &  23.69(2)  &   1.05(1)  &  0.69(1) \\
 88.0  & 0.688(4)  &  24.35(3)  &   1.13(1)  &  0.70(1) \\   
 90.5  & 0.789(3)  &  23.08(1)  &   1.38(1)  &  0.77(1) \\
\rowcolor{lightgray} 92.0  & 0.878(5)  &  25.67(2)  &   0.93(1)  &  0.50(1) \\
\rowcolor{lightgray} 94.5  & 0.942(7)  &  25.65(4)  &   0.93(1)  &  0.51(1) \\
 96.0  & 0.810(7)  &  21.57(3)  &   1.31(1)  &  0.82(1) \\
 97.0  & 0.776(5)  &  21.39(4)  &   1.21(1)  &  0.82(1) \\
 97.5  & 0.883(7)  &  20.93(2)  &   1.34(1)  &  0.84(1) \\
 98.5  & 0.809(6)  &  20.91(2)  &   1.27(1)  &  0.85(1) \\
 99.0  & 1.028(9)  &  21.58(14) &   1.57(2)  &  0.83(1) \\
100.0  & 1.022(7)  &  20.78(2)  &   1.68(1)  &  0.90(1) \\
101.5  & 1.047(10)  &  21.11(2)  &   1.55(2)  &  0.86(1) \\
102.5  & 1.051(11)  &  20.41(31) &   1.65(2)  &  0.93(3) \\
\rowcolor{lightgray}104.0  & 1.149(10)  &  23.40(6)  &   1.06(1)  &  0.60(1) \\
\rowcolor{lightgray}106.5  & 1.220(11)  &  22.85(1)  &   1.11(1)  &  0.62(1) \\
108.0  & 1.300(10)  &  18.80(9)  &   1.53(2)  &  0.98(1) \\
109.5  & 1.012(10)  &  18.25(4)  &   1.51(2)  &  1.02(1) \\
111.0  & 1.192(11)  &  19.44(3)  &   1.59(2)  &  0.97(1) \\
113.5  & 1.161(14)  &  18.51(16) &   1.55(2)  &  1.02(2) \\
    \hline
\end{tabular}
\end{center}
\tablefoot{Uncertainties in parentheses are given in units of the last digit. They take into account the error on $S_{\nu}$ given by the Gaussian fitting procedure and the uncertainty on the free-free correction factor. Results obtained from maps belonging to the frequency range covered by setup 3 are highlighted in grey. \tablefoottext{a}{Peak flux density derived from the 2D-Gaussian fit to the continuum map, corrected for the primary beam attenuation and for the free-free contamination.} \tablefoottext{b}{Peak flux density measured on the SMA map obtained at 343~GHz and smoothed to the resolution of the ALMA map.}}
\end{table}

\begin{table}[!t]
\begin{center}
  \caption{\label{tab-continuum-n2} Peak flux densities and H$_2$ column densities toward Sgr~B2(N2).}
  \begin{tabular}{cllcc}
    \hline
Freq & $S_{\nu}^{\rm ALMA}$  & \multicolumn{1}{c}{$S_{\nu}^{\rm SMA}$} & $N_{\rm H_2}^{\rm ALMA}$ & $N_{\rm H_2}^{\rm SMA}$ \\
(GHz)& \multicolumn{2}{c}{(Jy/beam)} & \multicolumn{2}{c}{(10$^{24}$ cm$^{-2}$)} \\
    \hline
    \hline 
85.0  & 0.051(8) &  6.51(2)  & 0.82(13) & 1.35(1) \\ 
86.5  & 0.058(8) &  6.42(2)  & 0.92(13) & 1.38(1) \\ 
88.0  & 0.070(8) &  6.60(2)  & 1.03(12) & 1.39(1) \\ 
90.5  & 0.078(9) &  6.34(1)  & 1.17(13) & 1.52(1) \\ 
\rowcolor{lightgray} 92.0 & 0.073(7) & 6.69(1)  & 0.69(7) & 1.04(1) \\ 
\rowcolor{lightgray} 94.5 & 0.060(7) & 6.69(1)  & 0.54(6) & 1.06(1) \\ 
96.0  & 0.103(16) &  5.97(3)  & 1.43(22) & 1.58(1) \\ 
97.0  & 0.098(9) &  5.97(2)  & 1.32(13) & 1.60(1) \\ 
97.5  & 0.114(9) &  5.87(1)  & 1.59(13) & 1.63(1) \\ 
98.5  & 0.104(10) &  5.87(4)  & 1.41(13) & 1.65(1) \\ 
99.0  & 0.138(9) &  6.02(2)  & 1.76(12) & 1.61(1) \\ 
100.0 & 0.137(10) &  5.88(6)  & 1.85(13) & 1.73(2) \\ 
101.5 & 0.152(9) &  5.93(2)  & 1.87(12) & 1.66(1) \\ 
102.5 & 0.144(10) &  5.80(2)  & 1.86(13) & 1.77(1) \\ 
\rowcolor{lightgray} 104.0 & 0.169(9) & 6.26(2)  & 1.38(8)  &  1.23(1) \\ 
\rowcolor{lightgray} 106.5 & 0.159(9) & 6.16(2)  & 1.27(8)  &  1.28(1) \\ 
108.0 & 0.163(10) &  5.42(1)  & 2.00(13) & 1.85(1) \\ 
109.5 & 0.173(9) &  5.30(1)  & 2.14(13) & 1.91(1) \\ 
111.0 & 0.198(10) &  5.58(2)  & 2.15(11) & 1.84(1) \\ 
113.5 & 0.211(10) &  5.39(1)  & 2.31(11) & 1.92(1) \\
    \hline
\end{tabular}
\end{center}
\tablefoot{Same as Table~\ref{tab-continuum-n1} but for Sgr~B2(N2).}
\end{table}

\begin{table}[!t]
\begin{center}
  \caption{\label{tab-continuum-n3} H$_2$ column density upper limits toward Sgr~B2(N3).}
  \begin{tabular}{ccccc}
    \hline
Freq & rms$^{\rm ALMA}$ \tablefootmark{(a)}  & \multicolumn{1}{c}{rms$^{\rm SMA}$ \tablefootmark{(b)}} & $N_{\rm H_2}^{\rm ALMA}$ \tablefootmark{(c)} & $N_{\rm H_2}^{\rm SMA}$ \tablefootmark{(c)} \\
(GHz)& (mJy/beam) & (Jy/beam) & \multicolumn{2}{c}{(10$^{23}$ cm$^{-2}$)} \\
    \hline
    \hline 
85.0  & 5.3  &  0.12 &  $<$4.4  & $<$1.9 \\
86.5  & 4.3  &  0.19 &  $<$3.5  & $<$1.9 \\
88.0  & 9.6  &  0.12 &  $<$7.2  & $<$2.0 \\
90.5  & 8.3  &  0.19 &  $<$6.6  & $<$2.1 \\
\rowcolor{lightgray}92.0  & 7.4  &  0.23 &  $<$3.7  & $<$1.7 \\
\rowcolor{lightgray}94.5  & 7.5  &  0.23 &  $<$  & $<$1.7 \\
96.0  & 2.7  &  0.17 &  $<$2.0  & $<$2.1 \\
97.0  & 4.4  &  0.17 &  $<$3.1  & $<$2.1 \\
97.5  & 2.6  &  0.17 &  $<$1.9  & $<$2.1 \\
98.5  & 4.3  &  0.16 &  $<$3.0  & $<$2.1 \\
99.0  & 4.8  &  0.17 &  $<$3.2  & $<$2.1 \\
100.0 & 5.0  &  0.16 &  $<$3.5  & $<$2.2 \\
101.5 & 3.7  &  0.17 &  $<$2.4  & $<$2.1 \\
102.5 & 5.3  &  0.16 &  $<$3.6  & $<$2.2 \\
\rowcolor{lightgray}104.0 & 5.0  &  0.20 &  $<$2.1  & $<$1.8 \\
\rowcolor{lightgray}106.5 & 5.0  &  0.19 &  $<$2.1  & $<$1.9 \\
108.0 & 3.0  &  0.15 &  $<$1.9  & $<$2.3 \\
109.5 & 2.5  &  0.14 &  $<$1.6  & $<$2.3 \\
111.0 & 8.2  &  0.15 &  $<$4.7  & $<$2.3 \\
113.5 & 7.6  &  0.14 &  $<$4.4  & $<$2.3 \\
    \hline
\end{tabular}
\end{center}
\tablefoot{Same as Table~\ref{tab-continuum-n1} but for Sgr~B2(N3). \tablefoottext{a}{Noise level measured in the ALMA continuum map with GO NOISE.} \tablefoottext{b}{Noise level measured in the SMA map inside a polygon defined around Sgr~B2(N3)'s position as described in Sect.~\ref{continuum-SMA}.} \tablefoottext{c}{The upper limits correspond to 5$\sigma$.}} 
\end{table}

\begin{table}[!t]
\begin{center}
  \caption{\label{tab-continuum-n4} Peak flux densities and H$_2$ column densities toward Sgr~B2(N4).}
  \setlength{\tabcolsep}{1.0mm}
  \begin{tabular}{clcccc}
    \hline
Freq & \multicolumn{1}{c}{$S_{\nu}^{\rm ALMA}$}  & rms$^{\rm ALMA}$ \tablefootmark{(a)}  & rms$^{\rm SMA}$ \tablefootmark{(b)} & $N_{\rm H_2}^{\rm ALMA}$ \tablefootmark{(c)}  & $N_{\rm H_2}^{\rm SMA}$ \tablefootmark{(d)} \\
(GHz)& \multicolumn{2}{c}{(mJy/beam)} & (Jy/beam) & \multicolumn{2}{c}{(10$^{23}$ cm$^{-2}$)} \\
    \hline
    \hline 
85.0  & \_        & 5.3  & 0.53  & $<$4.38  &  $<$5.21 \\
86.5  & 21.82(60) & 4.3   & 0.51  & 3.43(9)  &  $<$5.26 \\ 
88.0  & \_        & 9.6   & 0.53  & $<$7.28  &  $<$5.28 \\
90.5  & \_        & 8.3   & 0.48  & $<$6.56  &  $<$5.41 \\
\rowcolor{lightgray}92.0  & \_      & 7.4    & 0.64  & $<$3.70  &  $<$4.79 \\
\rowcolor{lightgray}94.5  & \_      & 7.5   & 0.63  & $<$3.49  &  $<$4.82 \\
96.0  & 35.90(12)  & 2.7   & 0.44  & 4.94(2)  &  $<$5.45 \\ 
97.0  & 33.06(50)  & 4.4  & 0.43  & 4.43(7)  &  $<$5.47 \\ 
97.5  & 33.77(50)  & 2.6  & 0.42  & 4.62(7)  &  $<$5.50 \\ 
98.5  & 28.88(76)  & 4.3  & 0.42  & 3.85(10)  &  $<$5.52 \\ 
99.0  & 31.10(76)  & 4.8  & 0.44  & 3.90(10)  &  $<$5.48 \\ 
100.0 & 33.53(64)  & 5.0  & 0.41  & 4.44(8)  &  $<$5.62 \\ 
101.5 & 34.03(64)  & 3.7  & 0.42  & 4.11(8)  &  $<$5.54 \\ 
102.5 & 36.90(103) & 5.3  & 0.40  & 4.67(13)  &  $<$5.65 \\ 
\rowcolor{lightgray}104.0 & 29.66(104) & 5.0  & 0.54  & 2.38(8)  &  $<$5.03 \\ 
\rowcolor{lightgray}106.5 & 38.82(105) & 5.0 & 0.51  & 3.04(8)  &  $<$5.08 \\ 
108.0 & 41.37(66)  & 3.0  & 0.36  & 4.96(8)  &  $<$5.71 \\ 
109.5 & 42.23(13)  & 2.5  & 0.35  & 5.09(2)  &  $<$5.77 \\ 
111.0 & 44.44(54)  & 8.2 & 0.37  & 4.71(6)  &  $<$5.70 \\ 
113.5 & \_         & 7.6  & 0.35  & $<$4.36  &  $<$5.78 \\

    \hline
\end{tabular}
\end{center}
\tablefoot{Same as Table~\ref{tab-continuum-n1} but for Sgr~B2(N4). 
\tablefoottext{a}{Noise level measured in the ALMA continuum map with GO NOISE.}
\tablefoottext{b}{Noise level measured in the SMA map inside a polygon defined around Sgr~B2(N4)'s position as described in Sect.~\ref{continuum-SMA}.}  
\tablefoottext{c}{Where Sgr~B2(N4) is not detected in the ALMA maps, an upper limit to its H$_2$ column density has been calculated at the 5$\sigma$ level.}
\tablefoottext{d}{The upper limits correspond to 5$\sigma$.}}
\end{table}

\begin{table}[!t]
\begin{center}
  \caption{\label{tab-continuum-n5} Peak flux densities and H$_2$ column densities toward Sgr~B2(N5).}
  \begin{tabular}{clccc}
    \hline
Freq & \multicolumn{1}{c}{$S_{\nu}^{\rm ALMA}$}  & rms$^{\rm SMA}$\tablefootmark{(a)}  & $N_{\rm H_2}^{\rm ALMA}$ & $N_{\rm H_2}^{\rm SMA}$ \tablefootmark{(b)} \\
(GHz)& \multicolumn{2}{c}{(Jy/beam)} & \multicolumn{2}{c}{(10$^{24}$ cm$^{-2}$)} \\
    \hline
    \hline 
85.0  & 0.073(1)  &  1.19 &  1.17(1)  &  $<$1.23 \\
86.5  & 0.070(1)  &  1.15 &  1.10(1)  &  $<$1.23 \\
88.0  & 0.063(1)  &  1.18 &  0.93(1)  &  $<$1.23 \\
90.5  & 0.055(1)  &  1.06 &  0.83(1)  &  $<$1.25 \\
\rowcolor{lightgray}92.0  & 0.059(1)  &  1.51 &  0.57(1)  &  $<$1.19 \\
\rowcolor{lightgray}94.5  & 0.054(1)  &  1.49 &  0.48(1)  &  $<$1.19 \\
96.0  & 0.053(1)  &  0.96 &  0.73(1)  &  $<$1.26 \\
97.0  & 0.067(11) &  0.96 &  0.91(16) &  $<$1.26 \\
97.5  & 0.058(1)  &  0.93 &  0.79(1)  &  $<$1.26 \\
98.5  & 0.070(1)  &  0.92 &  0.95(1)  &  $<$1.26 \\
99.0  & 0.046(1)  &  0.96 &  0.58(1)  &  $<$1.26 \\
100.0 & 0.062(1)  &  0.89 &  0.82(2)  &  $<$1.28 \\
101.5 & 0.062(1)  &  0.93 &  0.75(1)  &  $<$1.27 \\
102.5 & 0.056(1)  &  0.87 &  0.71(1)  &  $<$1.28 \\
\rowcolor{lightgray}104.0 & 0.075(1)  &  1.24 &  0.60(1)  &  $<$1.21 \\
\rowcolor{lightgray}106.5 & 0.076(1)  &  1.18 &  0.60(1)  &  $<$1.22 \\
108.0 & 0.048(1)  &  0.78 &  0.58(2)  &  $<$1.29 \\
109.5 & 0.051(1)  &  0.74 &  0.62(1)  &  $<$1.29 \\
111.0 & 0.074(1)  &  0.81 &  0.79(1)  &  $<$1.29 \\
113.5 & 0.075(1)  &  0.75 &  0.80(1)  &  $<$1.30 \\
    \hline
\end{tabular}
\end{center}
\tablefoot{Same as Table~\ref{tab-continuum-n1} but for Sgr~B2(N5). 
\tablefoottext{a}{Noise level measured in the SMA map inside a polygon defined around Sgr~B2(N5)'s position as described in Sect.~\ref{continuum-SMA}.}
\tablefoottext{b}{The upper limits correspond to 5$\sigma$.}}
\end{table}

\section{Additional figures}

\begin{figure}[!t]
\begin{center}
   \includegraphics[width=\hsize]{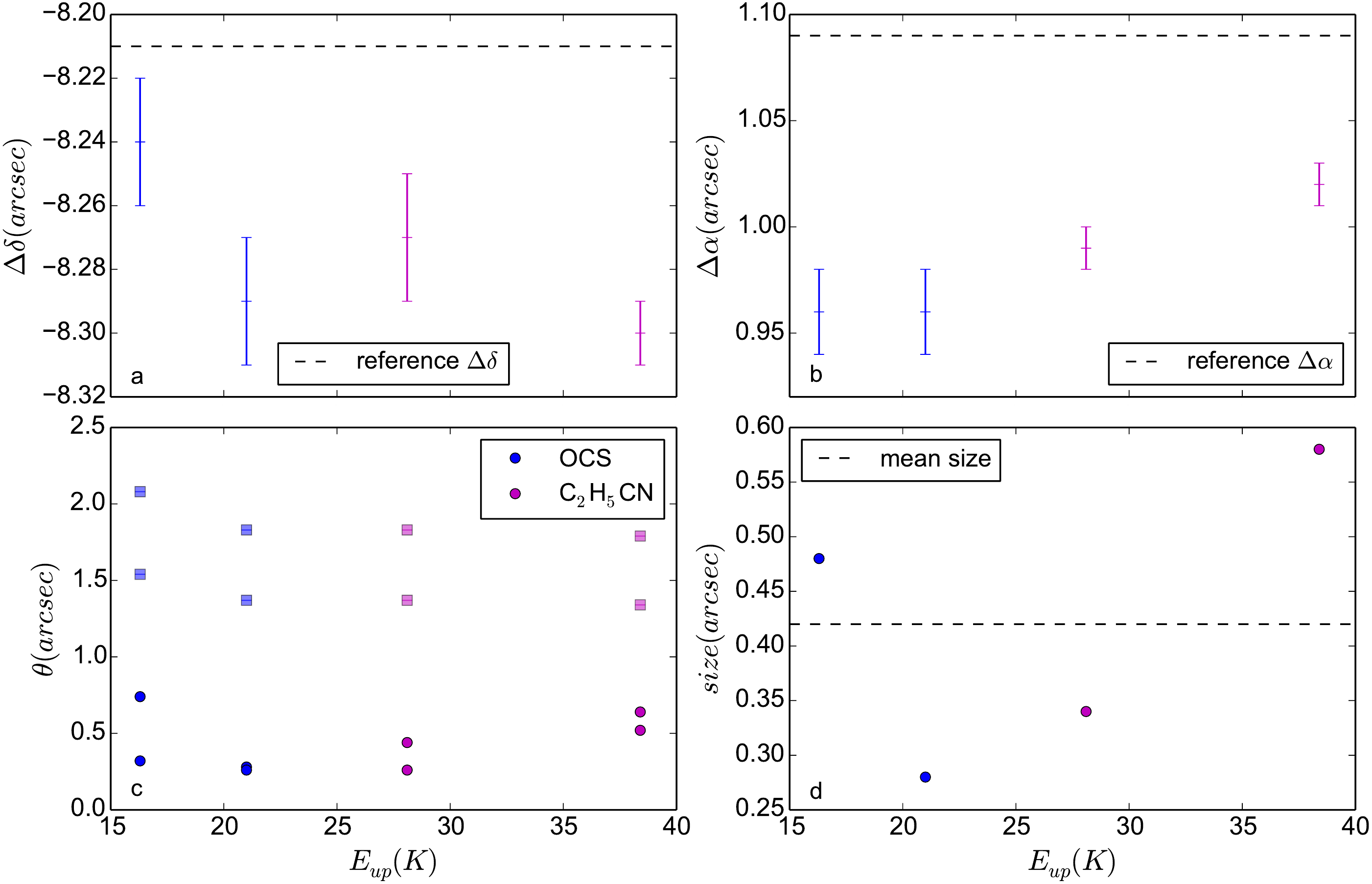} 
\caption{\label{source-size-deconvolved-python-N3} Results of the 2D-Gaussian fits to the integrated intensity maps showing resolved emission toward Sgr~B2(N3). \textbf{a, b} Fitted position. The error bars are the uncertainties given by the Gaussian fits. \textbf{c, d} Fitted size. The dots represent the deconvolved major and minor diameters of the emission (\textbf{c}) and the resulting average source size (\textbf{d}). The squares represent the major and minor axes of the synthesized beam (\textbf{c}). The dashed lines represent the reference position of the hot core (\textbf{a, b}), and the mean deconvolved angular size (\textbf{d}).}
\end{center}
\end{figure}

\begin{figure}[!t]
\begin{center}
   \includegraphics[width=\hsize]{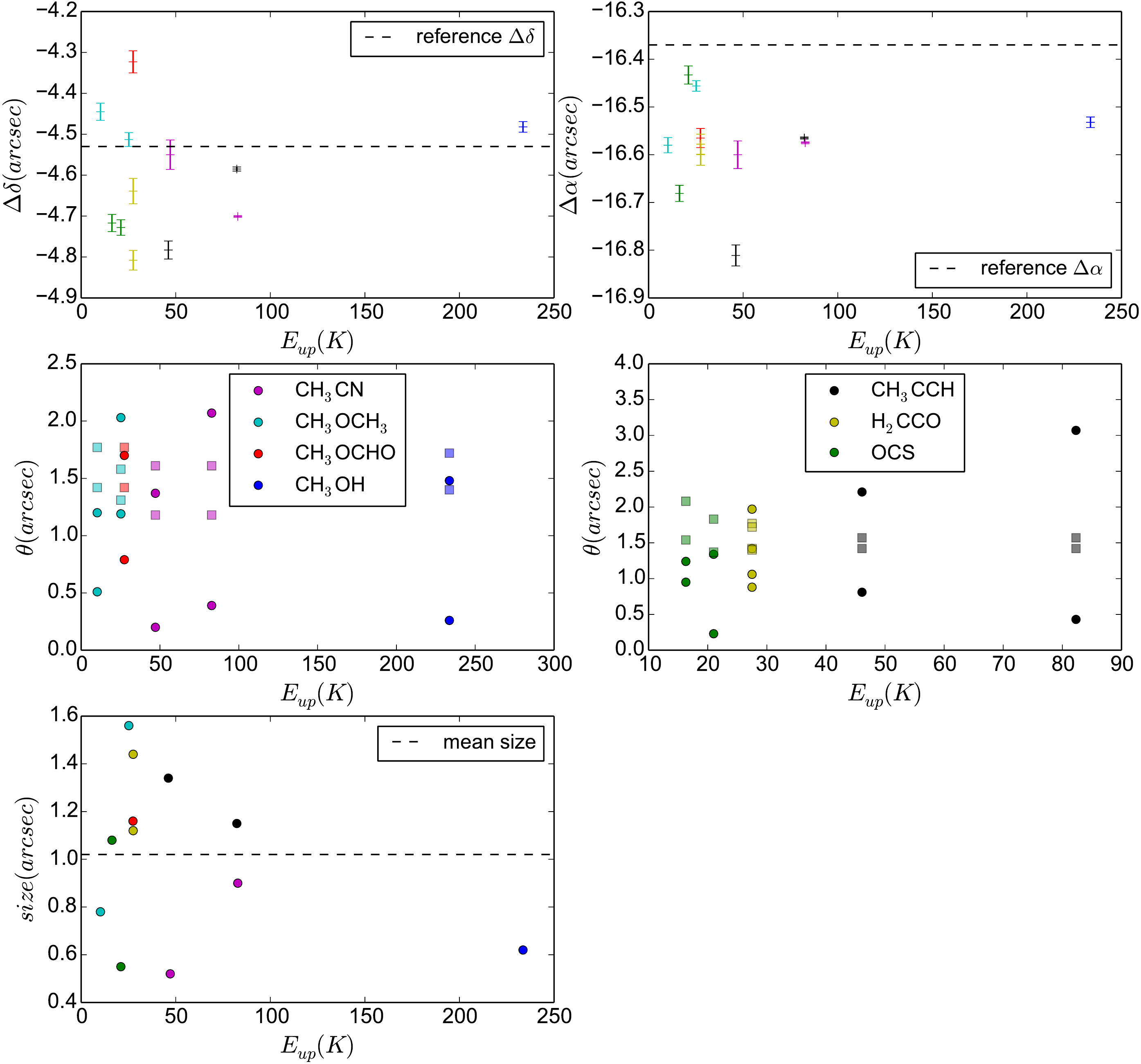} 
\caption{\label{source-size-python-N4} Same as Fig.~\ref{source-size-deconvolved-python-N3} but for Sgr~B2(N4).}
\end{center}
\end{figure}

\begin{figure}[!t]
\begin{center}
   \includegraphics[width=\hsize]{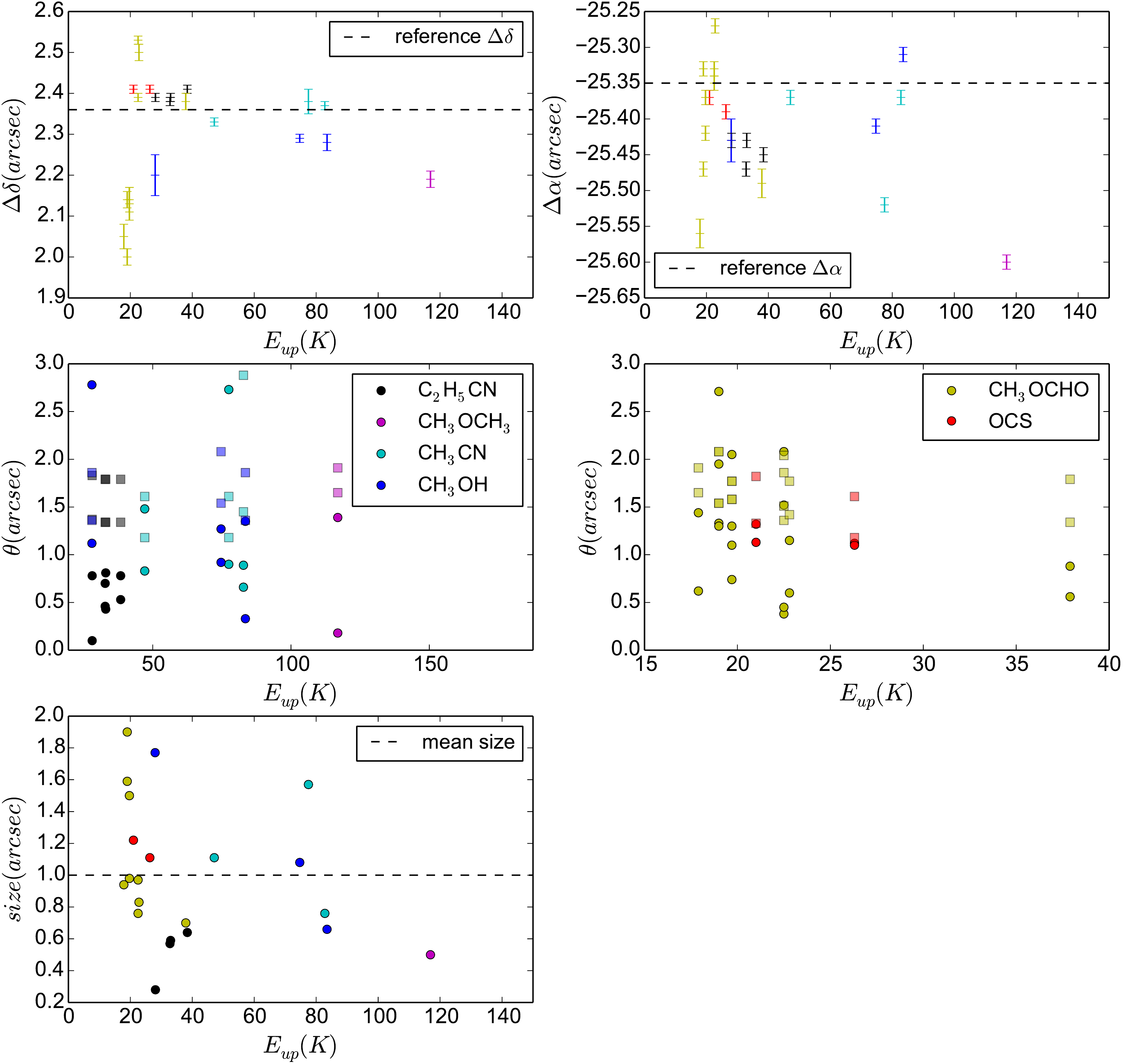} 
\caption{\label{source-size-python-N5} Same as Fig.~\ref{source-size-deconvolved-python-N3} but for Sgr~B2(N5).}
\end{center}
\end{figure}

\begin{figure*}[!t]
\resizebox{\hsize}{!}
{\begin{tabular}{cccc}
   \includegraphics[width=\hsize]{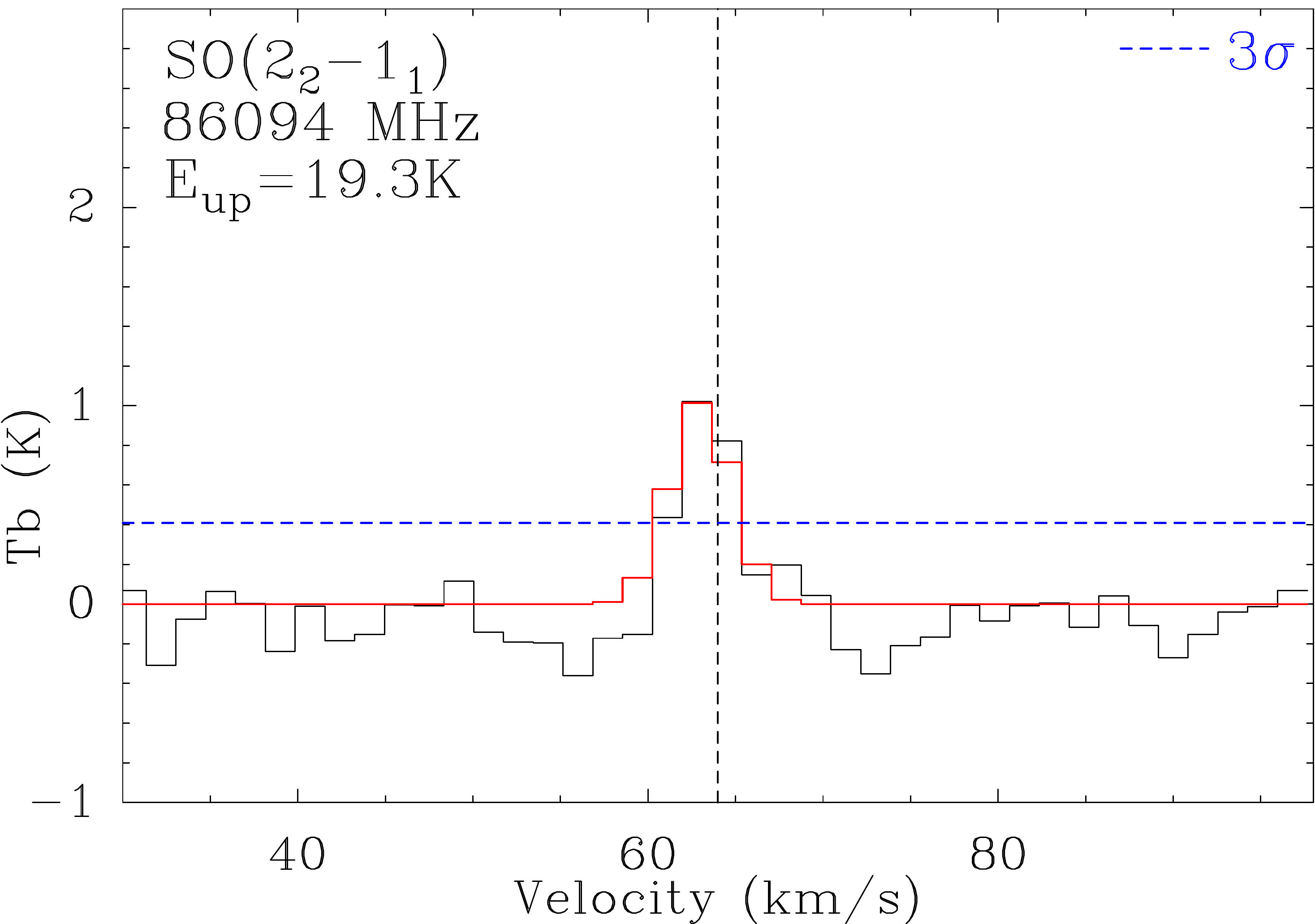} & 
   \includegraphics[width=\hsize]{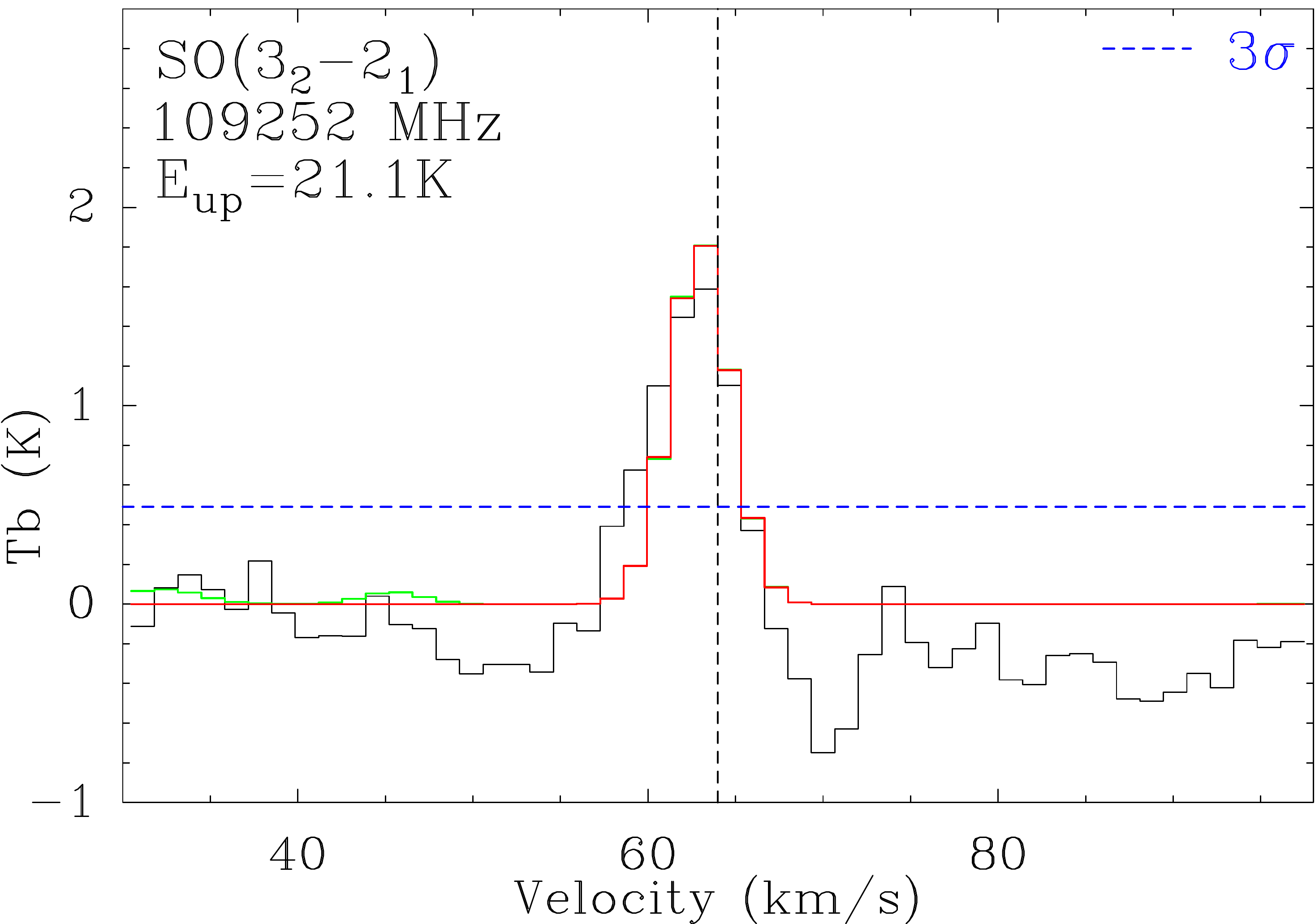} & 
   \includegraphics[width=\hsize]{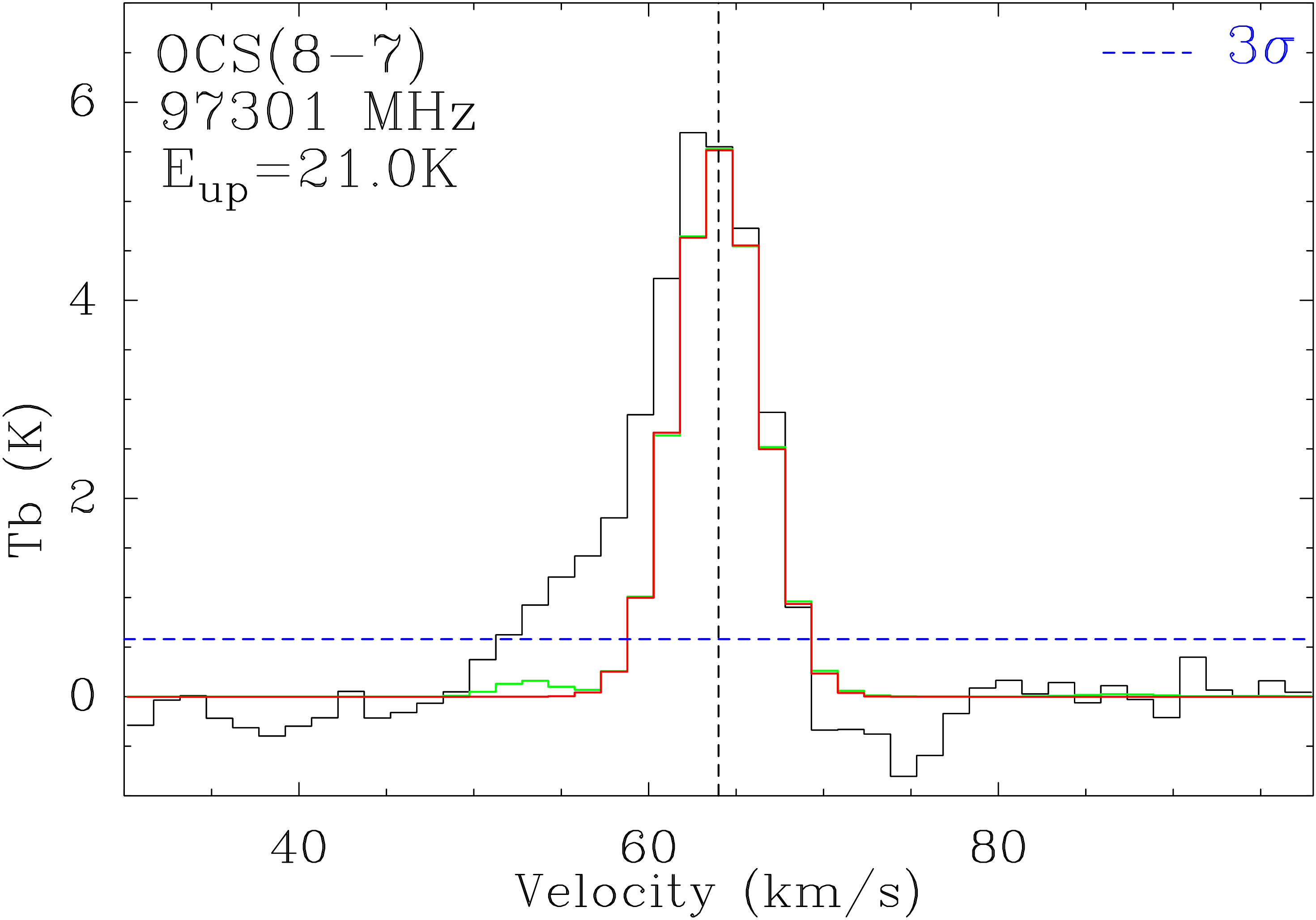} &
   \includegraphics[width=\hsize]{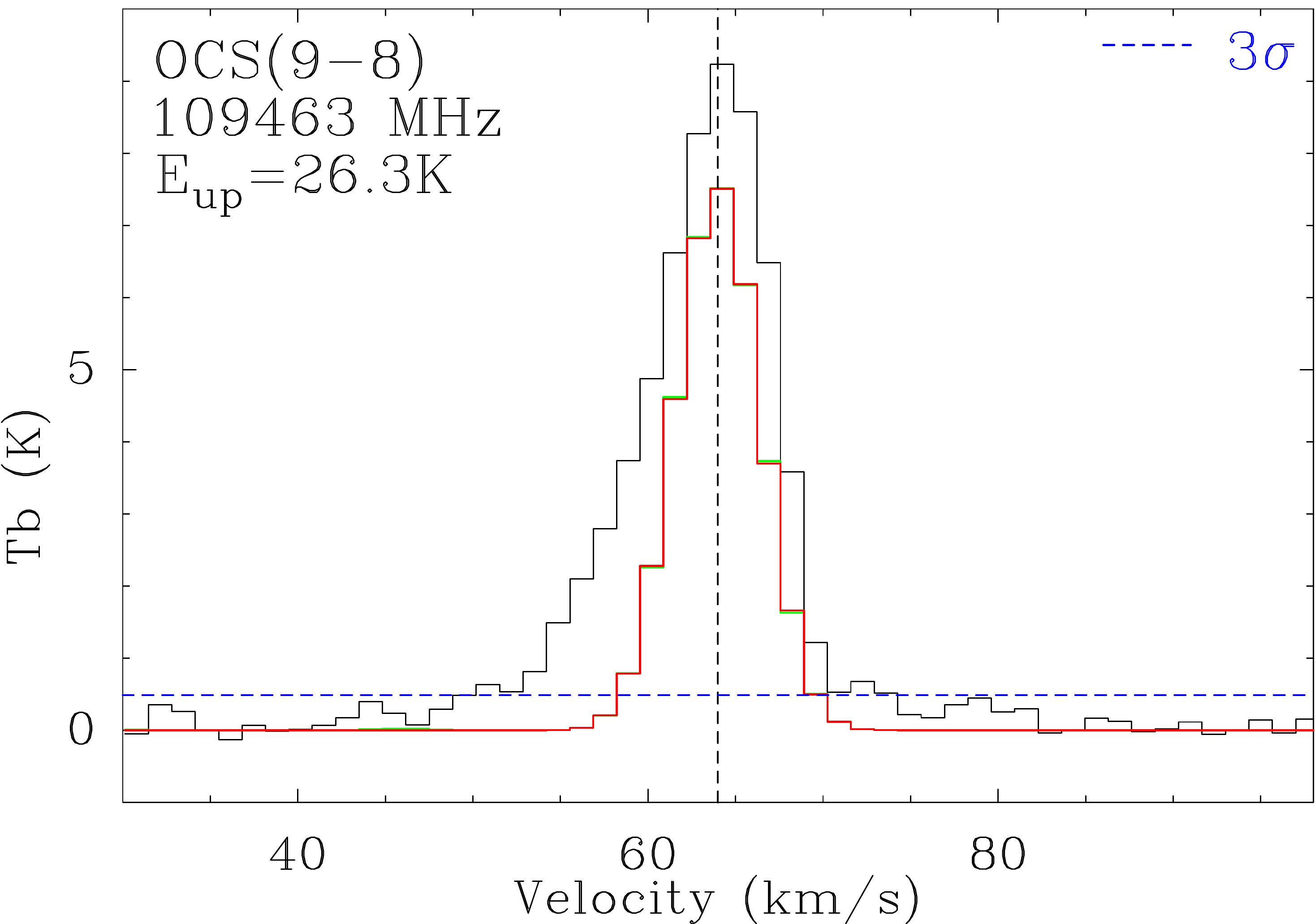} \\
\end{tabular}}
\caption{\label{appendix-kinematic-n4-raies1} Lines of typical outflow tracers detected toward Sgr~B2(N4). The red spectrum represents our LTE model fit to the complete observed spectrum of the molecule. The green spectrum shows the model that contains all the molecules identified so far toward Sgr~B2(N4). The dashed horizontal line shows the $3\sigma$ level. The systemic velocity of the source is marked with the dashed vertical line. The rest frequency and upper level energy (in temperature unit) of each transition are indicated in each panel.}   
\end{figure*}

\begin{figure*}[!t]
\resizebox{\hsize}{!}
{\begin{tabular}{cccc}
   \includegraphics[width=\hsize]{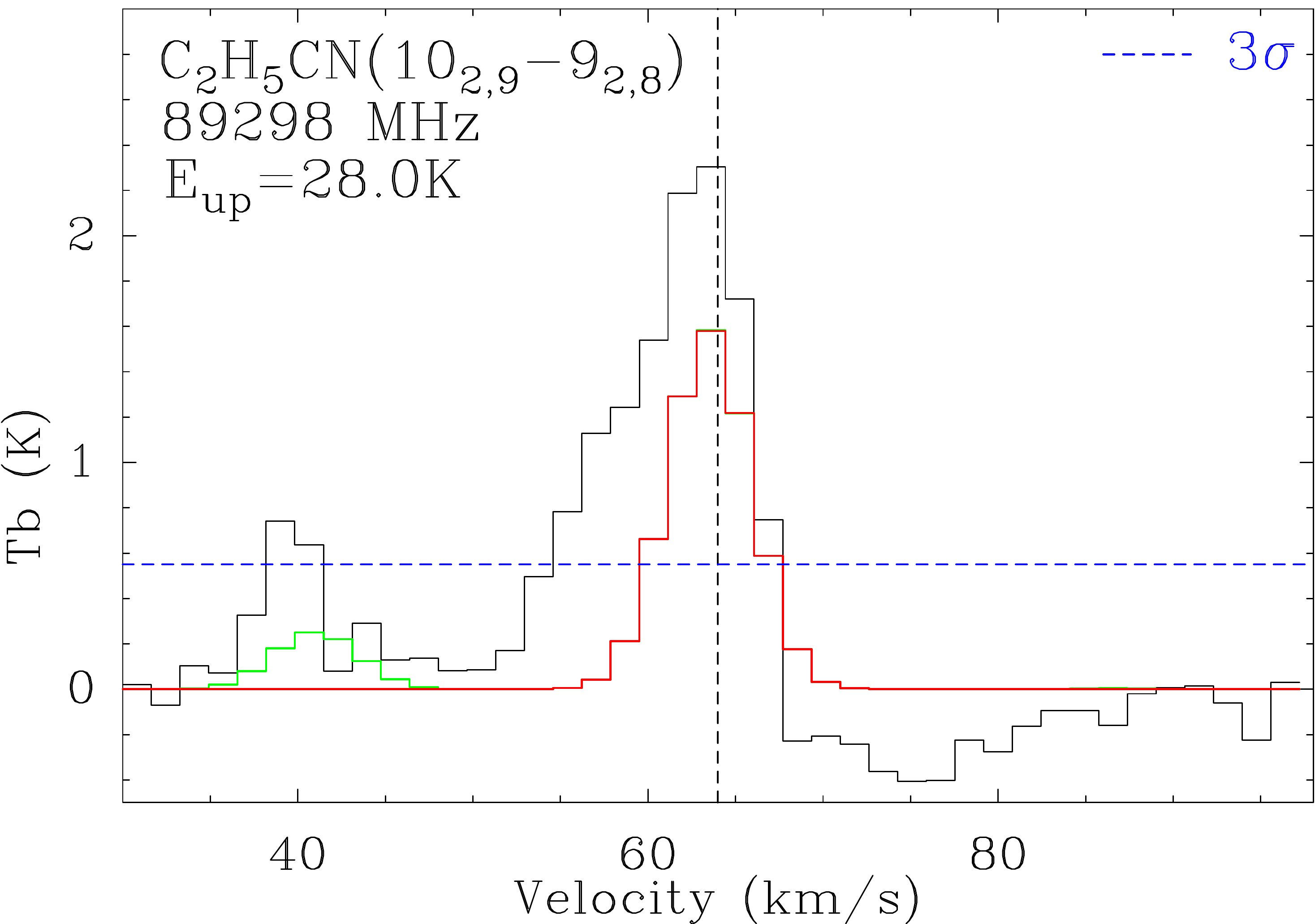} & 
   \includegraphics[width=\hsize]{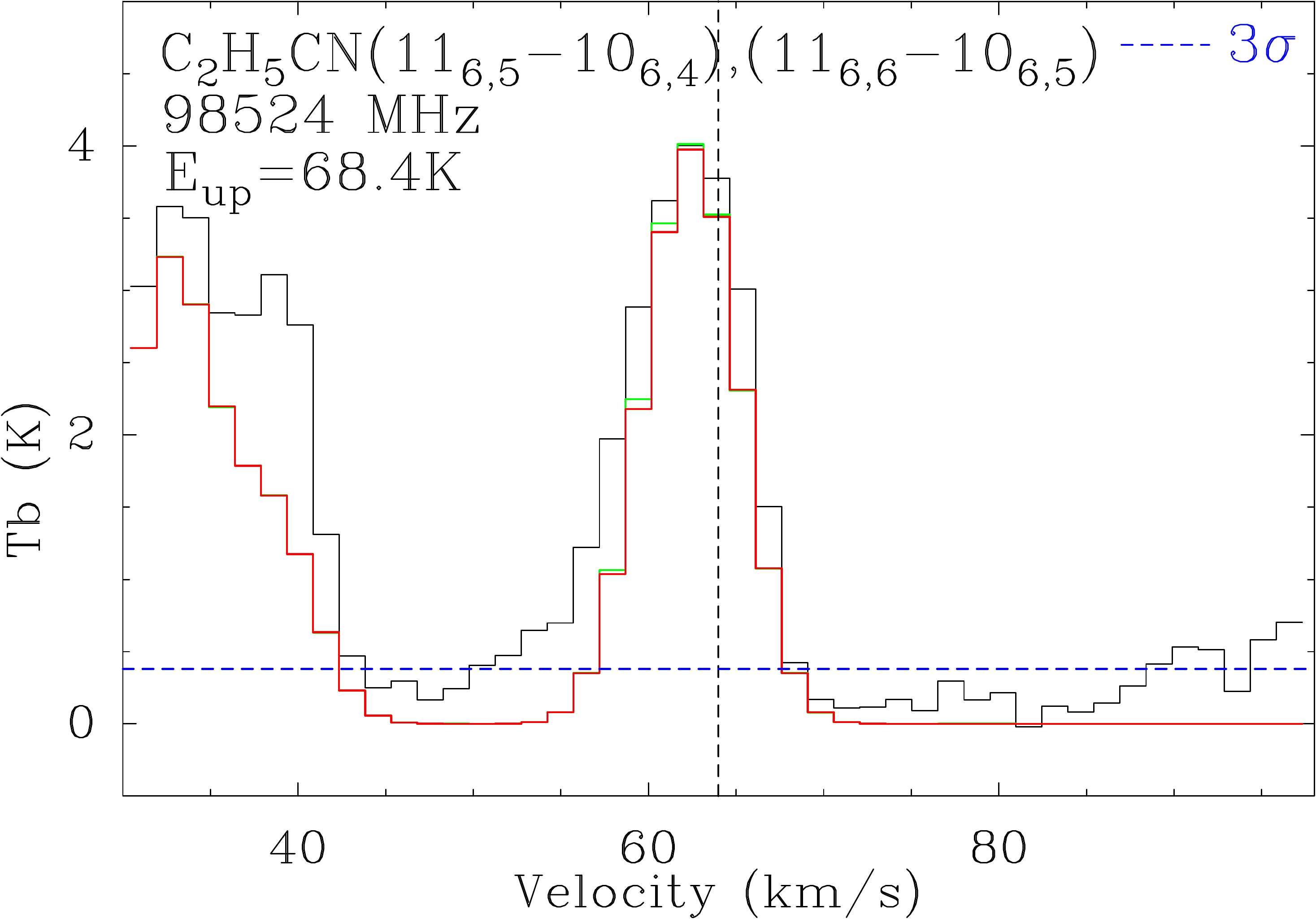} & 
   \includegraphics[width=\hsize]{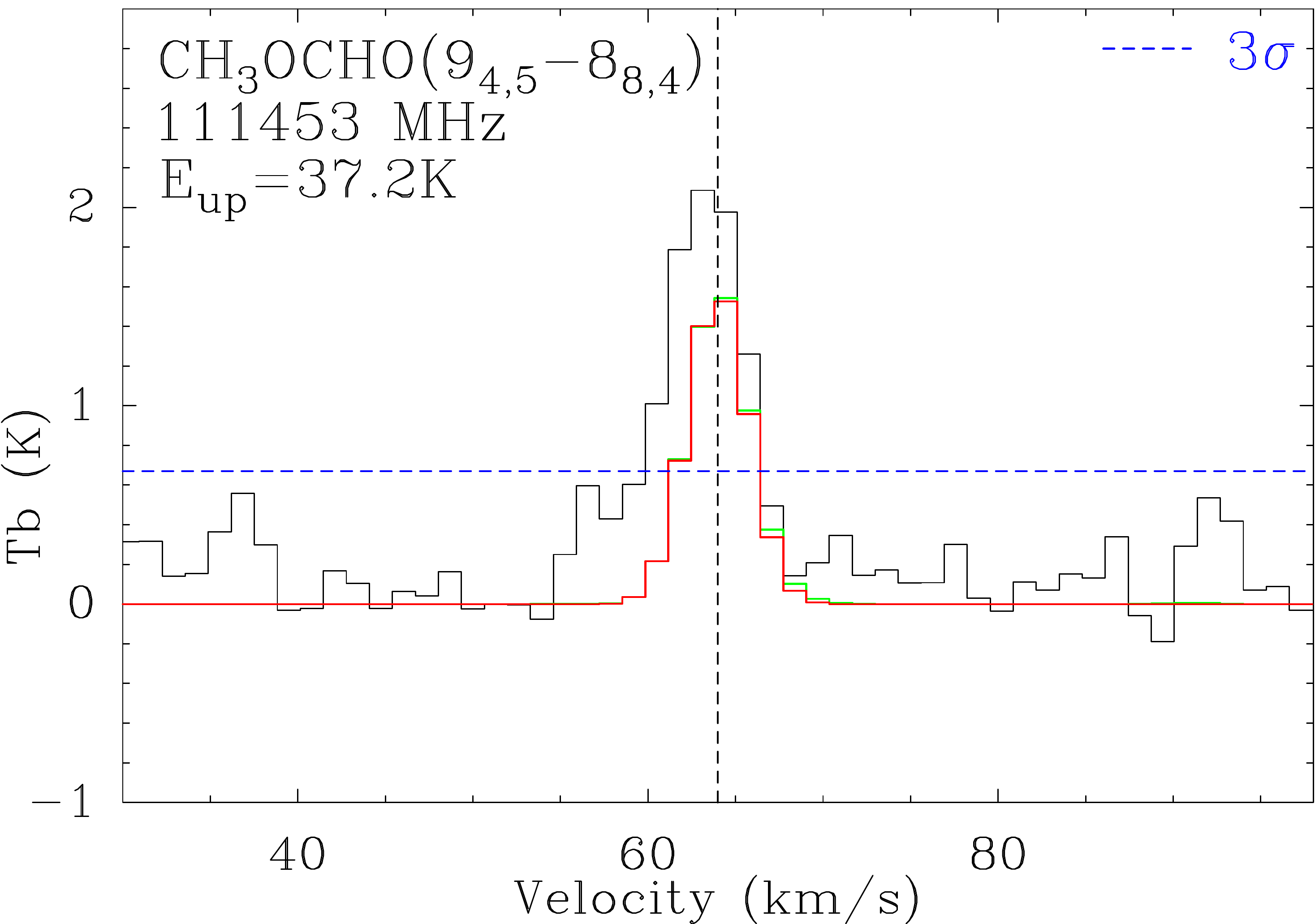} &
   \includegraphics[width=\hsize]{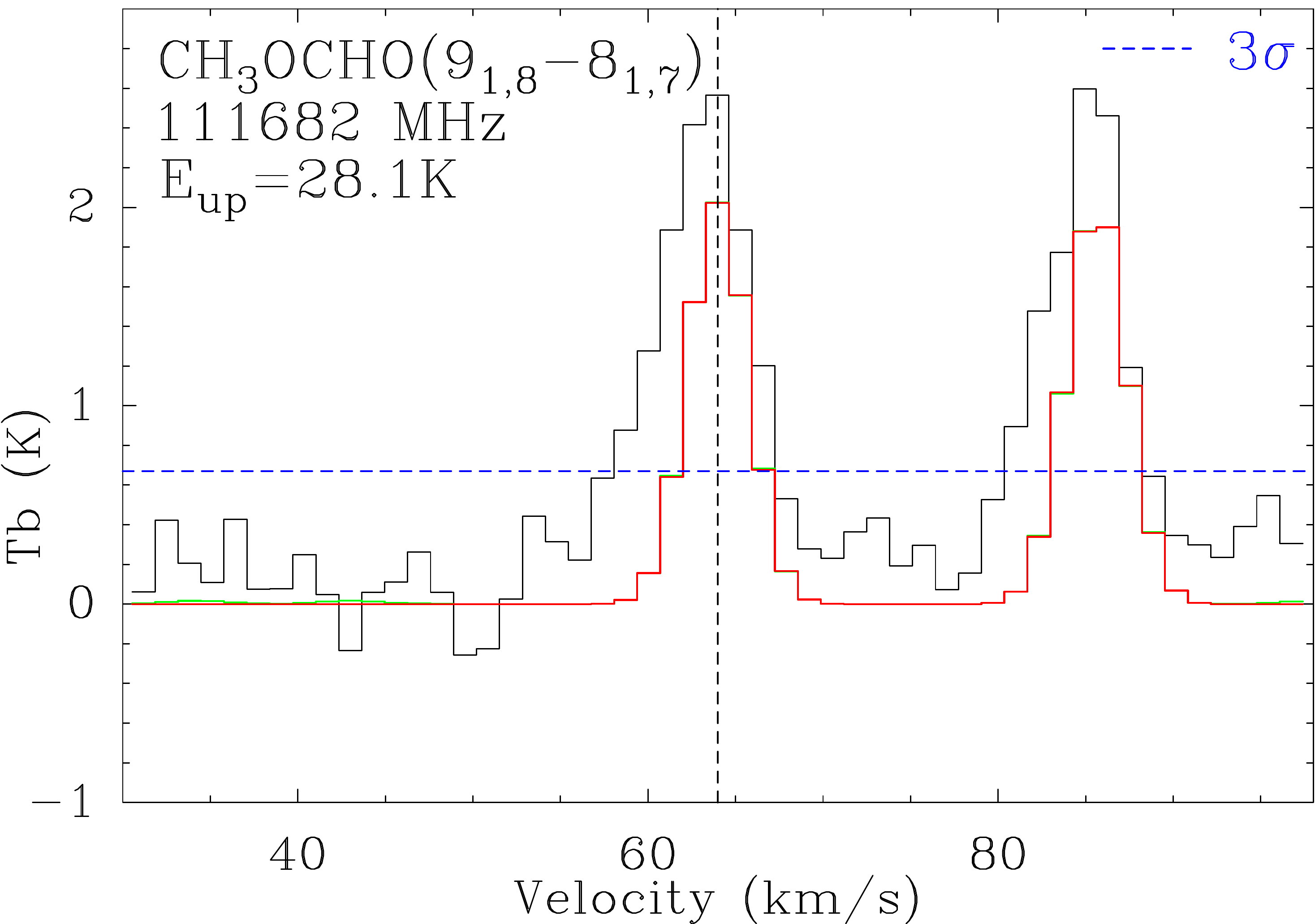} \\
\end{tabular}}
\caption{\label{appendix-kinematic-n4-raies2} Same as  Fig. \ref{appendix-kinematic-n4-raies1} but for species not considered as typical outflow tracers.}
\end{figure*}

\begin{figure}[!t]
\resizebox{\hsize}{!}
{\begin{tabular}{c}
   \includegraphics[width=\hsize]{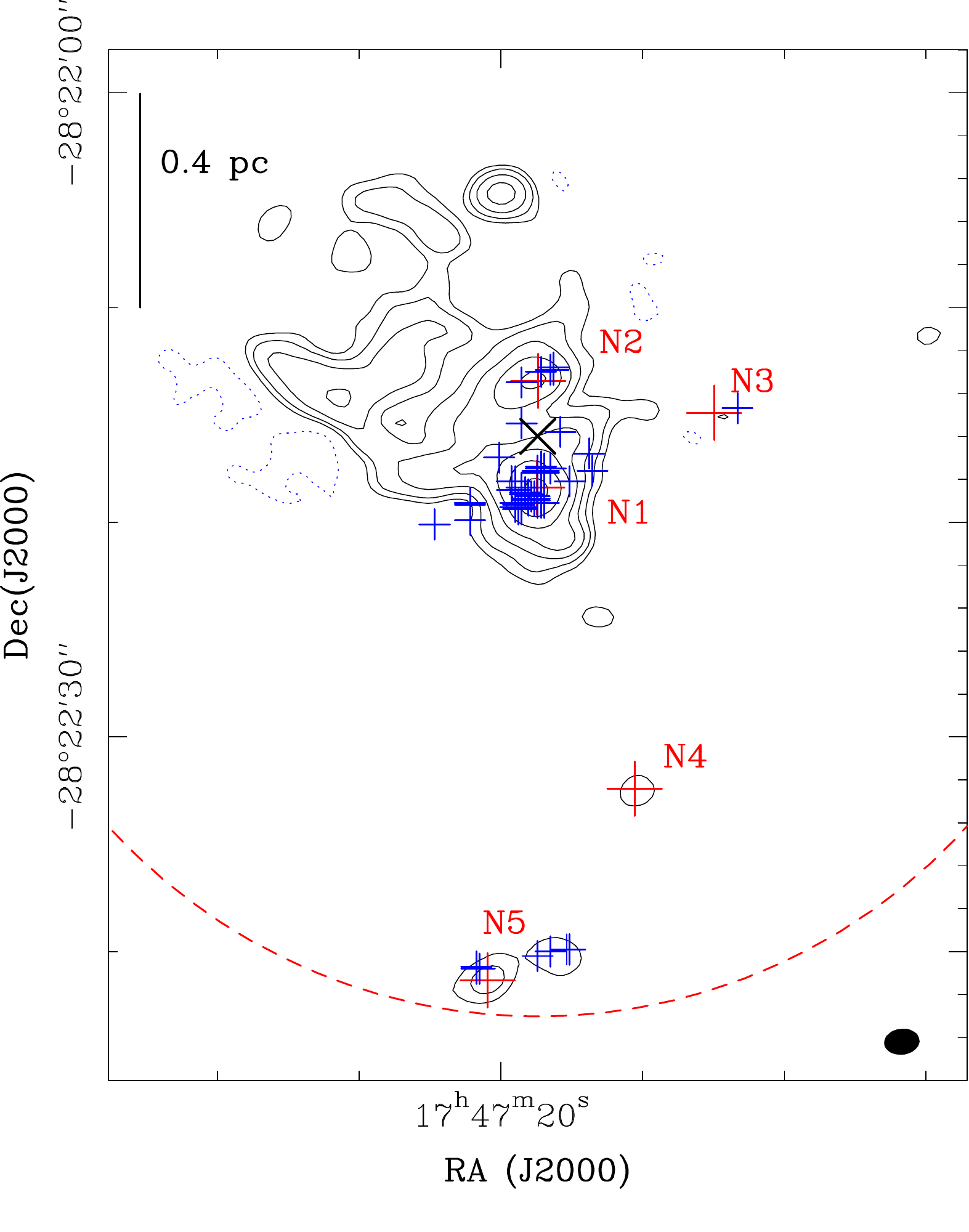} \\
\end{tabular}}
\caption{\label{appendix_h2o} Continuum map of the Sgr~B2(N) region obtained with ALMA at 108~GHz. Contour levels (positive in black solid line and negative in dashed line) start at 5 times the rms noise level, $\sigma$, of 3.0~mJy/beam and double in value up to 320$\sigma$. The filled ellipse shows the synthesized beam (1.65$\arcsec$ $\times$ 1.21$\arcsec$, PA=-83.4$^{\rm o}$). The black cross represents the phase center. The red crosses mark the positions of the five hot cores embedded in Sgr~B2(N). The blue crosses represent \ce{H2O} masers \citep{McGrath2004}. The dotted red circle represents the size (HPBW) of the primary beam of the 12~m antennas at 108~GHz.}
\end{figure}

\end{appendix}

\end{document}